\documentclass[final,5p,times,twocolumn]{elsarticle}

\usepackage{amssymb}
\usepackage{url}

\usepackage{caption,geometry}
\usepackage[colorlinks]{hyperref}
\usepackage{float}
\usepackage[figuresright]{rotating}
\usepackage{amsmath, amssymb}
\usepackage{bm}
\usepackage{url}
\usepackage{rotating}
\usepackage{rotfloat}

\journal{Science Bulletin}

\begin{document}

\begin{frontmatter}



\title{Artificial intelligence for celestial object census: the latest technology meets the oldest science}
\author[inst1]{Baoqiang Lao\corref{cor1}\fnref{}}
\ead{lbq@shao.ac.cn}
\cortext[cor1]{Corresponding authors.}
\affiliation[inst1]{organization={Shanghai Astronomical Observatory, Key Laboratory of Radio Astronomy, CAS},
            addressline={Nandan Road 80},
            city={Shanghai},
            postcode={200030},
            state={},
            country={China}}

\author[inst1]{Tao An\corref{cor1}\fnref{}}
\ead{antao@shao.ac.cn}

\author[inst1,inst2]{Ailing Wang }
\affiliation[inst2]{organization={University of Chinese Academy of Sciences},
            addressline={19A Yuquanlu},
            city={Beijing},
            postcode={100049},
            state={},
            country={China}}

\author[inst1]{Zhijun Xu }
\author[inst1]{Shaoguang Guo}
\author[inst1]{Weijia Lv }
\author[inst1]{Xiaocong Wu }
\author[inst1]{Yingkang Zhang }

\end{frontmatter}


Astronomy is the oldest natural science based on observation, and a census of objects in the sky map to create a catalog is the basis for further research. This effort is achieved through astronomical object detection, also known as ``source finding," which aims to identify individual objects in an astronomical image and then retrieve the properties of those objects to form a catalog.  The completeness, reliability, and accuracy of the resulting catalog has a profound impact on astrophysical research.

We are currently in an era of explosive information growth, where big data is revolutionizing human life, as well as changing the paradigm of scientific research. For example, the Large Synoptic Survey Telescope (LSST)\footnote{LSST to explore the deepest and widest optical sky:
\url{https://www.lsst.org/}} under construction will generate upto 20 terabytes of raw data per night, a scale comparable to that of the Sloan Digital Sky Survey (SDSS)\footnote{SDSS is one of the most successful surveys in the history of astronomy: \url{https://www.sdss.org/}.} in a decade! The Euclid space mission\footnote{Euclid is an ESA medium class astronomy and astrophysics space mission: \url{https://www.euclid-ec.org/}.} is expected to create approximately tens of petabytes of total data. The Square Kilometre Array (SKA)\footnote{SKA is the flagship telescope being built in the field of radio astronomy: \url{https://www.skatelescope.org/}.} is taking the scale of astronomical big data to a new level, generating raw data in a rate of several Tera bits per second in its first phase (10\% of the total scale) and 700 petabytes of scientific data per year \cite{2019SCPMA..6289531A,2020SRCWP}. The challenge for astronomers around the world is how to access and utilize this massive amount of information.

Sky surveys with modern telescopes have led to a dramatic increase in image size and quality, presenting enormous challenges as well as opportunities for new discoveries. As an example, the Australian SKA Pathfinder (ASKAP) all-sky survey is expected to detect 70 million radio galaxies \cite{2011PASA...28..215N}, and the classification and morphology of these radio sources provide key information for understanding the formation and evolution of the Universe. However, it is impossible to identify such a vast amount of objects by visual inspection, and classifying the extracted sources is even challenging. To tackle these challenges, algorithms for automatic source finding and classification need to be developed \cite{2021MNRAS.500.3821B}.

Early source finding algorithms were integrated in data processing software packages. To process large astronomical data, a number of stand-alone source finding software packages have been developed and offered higher reliability and accuracy than the old ones. Artificial intelligence (AI) technology has been widely used in industries, such as geological monitoring, robotics, autonomous driving, face recognition, medical image analysis, etc. Compared to standard (non-AI) source finders in radio astronomy, AI-based automated methods, especially deep learning (DL) methods, aided by the acceleration of graphics processing unit (GPU) devices, offer pronounced advantages in terms of operating speed. Moreover, machine learning can also analyze data without our instructions, i.e., it can identify unexpected patterns, e.g., identifying more types of galaxies. This new discovery capability will certainly improve our understanding of the Universe.

Convolutional neural networks (CNNs) are the best performing image recognition classifiers in academia and industry. Recently, many DL-based astronomical source detectors or classifiers have been developed (\cite{2019MNRAS.482.1211W,2021MNRAS.tmp..358B} and references therein), and they are divided into single-stage and two-stage algorithms: the former is faster but less accurate and suitable for fast detection; the later is more accurate but slower. A suitable CNN architecture needs to be chosen according to the task requirements, with a trade-off between recognition precision and computational cost.

Among the region-based CNN (R-CNN) family, Faster R-CNN has the advantage of faster training speed and easier data annotation.
\textsc{ClaRAN} \cite{2019MNRAS.482.1211W} v0.1 is a detector built on top of Faster R-CNN, and it can locate and associate the radio source components with $\sim$90\% precision, making it one of the highest-precision CNN-based classifiers. However, the performance of \textsc{ClaRAN} v0.1 is limited by the backbone network it uses, Visual Geometry Group Network (VGGNet). When the number of network layers of VGGNet increases, the model complexity increases and the recognition performance decreases accordingly; in addition, increasing parameters significantly increases the computational complexity, training time, and GPU memory usage.
Another shortcoming of \textsc{ClaRAN} v0.1 is that it can only classify radio sources by peak and component, and lacks relevance for extended sources.
In fact, most of the commonly used source finders can only identify compact point-like sources, rather than directly identifying and classifying extended sources, which is done by visual inspection in post-processing.

To overcome these shortcomings, we construct a new source finder, named \textsc{HeTu}\footnote{\textsc{HeTu} is named after two mysterious patterns handed down from ancient China, which contains profound cosmic astrology.}, based on an improved CNN model and a new classification method.
\textsc{HeTu} uses a combined network structure with Residual Network (ResNet) and Feature Pyramid Networks (FPN) as the backbone network.  It exploits the advantage of ResNet in balancing recognition precision and computational cost and the advantage of FPN in multi-feature object detection. As a result, \textsc{HeTu} not only increases the network depth, but also provides multi-scale feature maps without causing a significant decrease in running speed.
In this study, we used two different layers (50 and 101 layers) of ResNet, the generated models are called \textsc{HeTu}-50 and \textsc{HeTu}-101, respectively. The backbone network ResNet50-FPN is also used in \textsc{ClaRAN} v0.2 \cite{2021MNRAS.500.3821B}. The workflow of \textsc{HeTu} is depicted in Fig.~\ref{fig:resnet_fpn}
and the \textsc{HeTu} network is discussed in details in \ref{sec:2.3} and \ref{sec:2.4}.

We run three experiments to verify the performance of \textsc{HeTu}: (1) training experiment; (2) testing experiment; (3) predicting experiment. All experiments were conducted on the China SKA Regional Centre Prototype \cite{2019NatAs...3.1030A} (\ref{sec:3}).

In the training experiment, we used the same datasets and the same source classification scheme as \textsc{ClaRAN}~\cite{2019MNRAS.482.1211W} in order to compare the results.
The metric of the mean Average Precision (\textit{mAP}) \cite{2016arXiv161203144L} increases from 78.4\% for \textsc{ClaRAN} to 86.7\% for \textsc{HeTu}-50 and to 87.6\% for \textsc{HeTu}-101 (Table \ref{tab:AP_D1}), indicating a significant improvement in the recognition performance of \textsc{HeTu} compared to \textsc{ClaRAN}. \textit{mAP}s obtained from \textsc{HeTu} are also much higher than those derived from the ResNet models alone (Tables~\ref{tab:AP_D3}--\ref{tab:AP_D4}), validating the higher performance of the combined ResNet-FPN network.
The deeper \textsc{HeTu}-101 network increases the precision by 0.9\% over \textsc{HeTu}-50, therefore, \textsc{HeTu}-101 is used for both testing and predicting experiments.
We also found that \textsc{HeTu}'s performance is not strongly dependent on the dataset used and it is therefore widely adaptable.
\textsc{HeTu} supports parallel execution using multiple GPU devices, and in our training experiment the training speed is 2.5 times faster than without parallelism.

In the testing experiment (\ref{sec:testing}), we used a different source classification scheme from \textsc{ClaRAN}. \textsc{HeTu} automatically locates radio sources in the images and at the same time assigns them to one of the four classes according to their morphology: compact point-like sources (CS), Fanaroff-Riley type I (FRI) sources characterized with a central core and prominent two-sided jets which are weaker further from the core, Fanaroff-Riley type II (FRII) sources characterized by two prominent terminal components with symmetric shapes, and core-jet (CJ) sources showing a bright core component at one end of an elongated weaker jet feature. This classification scheme encompasses most of the radio sources with practical astrophysical meaning. We re-labelled all images of the training dataset by visual recognition according to the new classification scheme.
To avoid overfitting due to the imbalance of the different classes, we used the data augmentation technique to enlarging the FRI, FRII and CJ samples.
It took 4.9 hours to train the workflow of \textsc{HeTu}-101 over 40,000 steps for the re-labelled augmented dataset on 8 GPU devices.
The processing time is about 5.4 milliseconds per image, two orders of magnitude faster than the visual recognition.
$mAP$ is 94.2\% for the re-labelled augmented dataset, 4.3\% higher than the unaugmented dataset (Table~\ref{tab:AP_D1new1}).
The average precisions ($AP$s) for some source classes are as high as 0.994 (CS) and 0.981 (FRII). After augmentation, increasing the network depth did not greatly improve the recognition performance. Moreover, the total loss curves show that the training model for the augmented dataset is stable for all classes (Fig.~\ref{fig:P-R}).

Based on the successful establishment of the training set and CNN model from the training experiment, we applied \textsc{HeTu} to the practical astronomical data processing
(the predicting experiment, see details in \ref{sec:predicting}).
We used \textsc{HeTu} for source detection and classification on the images from the all-sky survey GLEAM \cite{2017MNRAS.464.1146H} observed with the SKA-low precursor telescope MWA \cite{2013PASA...30....7T}, and compared the results with those obtained with the traditional source finding software \textsc{Aegean} \cite{2018PASA...35...11H}.
\textsc{HeTu} detection (and classification) speed is 100 milliseconds per image, 21 times faster than \textsc{Aegean}.
If only the identification task is performed without classification (Gaussian fitting), \textsc{HeTu}'s runtime is even $\sim$2.5 times faster.
We cross-matched the sources detected by \textsc{HeTu} and \textsc{Aegean} with a search radius of 30 arcsec.
The cross-matching fraction varies when different detection thresholds are adopted (Table~\ref{tab:resnet+fpn_aegean}).
For example, when the detection threshold is $6\sigma$, the cross-matched CS objects account for 94.5\% of the \textsc{HeTu}-detected CS sources and 94.3\% for the \textsc{Aegean}-detected CS sources. If the detection threshold is set to $5\sigma$, the cross-match rates change to 96.9\% and 89.2\% for the \textsc{HeTu} and \textsc{Aegean} CS catalogs. A lower detection threshold results in more weaker sources detected, but at the cost of introducing more fake sources. A large fraction of the un-matched sources are found at the image edges (\textit{e.g.}, Fig.~\ref{fig:nomatch_all_aegean}
), and they are discarded by \textsc{HeTu} since \textsc{HeTu} considers them morphologically incomplete.
At lower thresholds, \textsc{Aegean} detects fake sources associated with sidelobes of very bright sources, which are not identified by \textsc{HeTu}.
The predicting experiment shows that \textsc{HeTu} not only has high recognition precision, but also has excellent ability in identifying weak sources (Fig.~\ref{fig:agean_resnet_fpn_fr2}).

\textsc{HeTu} is able to classify the detected sources into relevant classes while recognizing them; in contrast, \textsc{Aegean} only identifies the components of a source and can not directly determine whether there is a connection between adjacent components, leading to the classification of extended sources to be done in an offline manner by visual inspection.
After associating the \textsc{HeTu}-detected extended sources with the brightest component of the corresponding \textsc{Aegean}-detected sources, we found that: the cross-match rate is 100\% for FRII, 97.4\% for FRI and 97.6\% for CJ classes (Table S8), respectively, indicating that \textsc{HeTu} performs very well in identifying extended sources.

The ongoing and upcoming large radio continuum survey projects using the SKA pathfinder telescopes\footnote{see latest advances of the SKA pathfinder telescopes at 2021 SKA Science Meeting: \url{https://www.skatelescope.org/skascicon21/}} and SKA itself will produce a tremendous amount of images. Automated and accurate source finding and classification tools are particularly important to support these large sky surveys and to mine the data archive. Future predicting experiments will be performed to further improve \textsc{HeTu}'s recognition performance and speed, to support larger-scale images and to focus more on extended sources.

Neural networks have a deeper understanding of data than expected, but require large data sets for training (learning), and the vast Universe provides neural networks with a naturally enormous amount of data, and AI will undoubtedly have a profound impact on astronomy. However, it is important to note that AI can only perform certain tasks well if there is a large, correctly labelled dataset to learn from, and the trained model performs a single type of tasks. In other words, AI is not an ``all-around champion". But even so, the speed and efficiency of AI is increasingly shaping our understanding of the natural world. The network framework of \textsc{HeTu} is used  not only for astronomical source identification and classification but also in other fields such as medical CT image analysis (e.g., automated tumor detection).

\section*{Conflict of interest}
The authors declare that they have no conflict of interest.

\section*{Author contributions}
TA initiated the project. BL performed the experiments. BL and TA wrote the paper. AW contributed to result check and discussion. ZX, SG, WL, XW, and YZ contributed to the experiments.

\appendix

\section{Supplementary materials} \label{sec:appendix}


Large surveys using modern telescopes are producing images that are increasing exponentially in size and quality. Identifying objects in the generated images by visual recognition is time-consuming and labor-intensive, while classifying the extracted radio sources is even more challenging. To address these challenges, we develop a deep learning-based radio source detector, named \textsc{HeTu}, which is capable of rapidly identifying and classifying radio sources in an automated manner for both compact and extended radio sources. \textsc{HeTu} is based on a combination of a residual network (ResNet) and feature pyramid network (FPN). We classify radio sources into four classes based on their morphology. The training images are manually labeled and data augmentation methods are applied to solve the data imbalance between the different classes. \textsc{HeTu} automatically locates the radio sources in the images and assigns them to one of the four classes. The experiment on the testing dataset shows an average operation time of 5.4 millisecond per image and precision of 99.4\% for compact point-like sources and 98.1\% for double-lobe sources. We applied \textsc{HeTu} to the images obtained from the GaLactic and the Galactic Extragalactic All-Object Murchison Wide-field Array (GLEAM) survey project. More than 96.9\% of the \textsc{HeTu}-detected compact sources are matched compared to the source finding software used in the GLEAM. We also detected and classified 2,298 extended sources (including Fanaroff-Riley type I and II sources, and core-jet sources) above $5\sigma$. The cross-matching rates of extended sources are higher than 97\%, showing excellent performance of \textsc{HeTu} in identifying extended radio sources. \textsc{HeTu} provides an efficient tool for radio source finding and classification and can be applied to other scientific fields.

\section{Background}
We are currently in an era of explosive information growth, where big data is profoundly changing economic development, social progress and daily 
life, as well as changing the paradigm of scientific research. Astronomy is one of the oldest natural sciences based on observations, and astronomical data are growing at an unprecedented rate as large modern astronomical facilities survey the sky in ever-increasing depth and scope. The Large Synoptic Survey Telescope (LSST)~\cite{2019ApJ...873..111I} under construction will generate upto 20 terabytes of raw data per night, a scale comparable to that of the Sloan Digital Sky Survey (SDSS)~\cite{2000AJ....120.1579Y} in a decade! The Square Kilometre Array (SKA)~\cite{2009IEEEP..97.1482D} is taking the scale of astronomical big data to a new level, generating raw data in a rate of several Tera bits per second in its first phase (constructing 10\% of the total scale) and $\sim$700 petabytes of pre-calibrated scientific data per year \cite{2019SCPMA..6289531A,2020SRCWP}. The challenge for astronomers around the world is how to access and make use of this massive amount of information.

Establish catalogs is the basis for conducting further astronomical research. This effort is achieved through astronomical object detection, also known as ``source finding'', which is a fundamental step in the standard procedures for astronomical data processing and analysis.
The purpose of source finding is to identify individual astronomical objects in the survey images and then return the physical parameters (\textit{e.g.}, position, flux density, size, etc) of these objects to form a catalogue. Based on the created catalogue, the sky models can be built and further calibration can be performed. The study of radio source classes and morphology can address key scientific questions about the formation and evolution of the galaxies and Universe \cite{2015aska.confE..81M}. Therefore, the completeness, reliability and precision of the resulting catalogues have a profound impact on astrophysical studies. 

Early source finding algorithms were integrated into the data processing packages. These algorithms still play an important role in survey campaigns today, for example, the task Search and Destroy (SAD) in Astronomical Image Processing System (AIPS) 
\cite{2003ASSL..285..109G}, was used to create the catalogue of both the Faint Images of the Radio Sky at Twenty-Centimeters (FIRST) survey \cite{1997ApJ...475..479W} and National Radio Astronomy Observatory (NRAO) VLA Sky Survey (NVSS) \cite{1998AJ....115.1693C}; the tasks IMSAD and SFIND in the MIRIAD package \cite{1995ASPC...77..433S,2002AJ....123.1086H}
were used for the surveys of the Australia Telescope Compact Array (ATCA)  \cite{2000A&AS..146...41P,2015MNRAS.448.3731R}.  

With the upgrading of radio telescopes and the need for new survey projects, many stand-alone source finding software packages have been developed \cite{2015PASA...32...37H}, such as Source Extractor (\textsc{SExtractor}) \cite{1996A&AS..117..393B}, \textsc{DUCHAMP} \cite{2012MNRAS.421.3242W}, \textsc{BLOBCAT} \cite{2012MNRAS.425..979H}, \textsc{Aegean} \cite{2012MNRAS.422.1812H}, Python Blob Detection and Source Finder (\textsc{PyBDSF}) formerly known as Python Blob Detection and Source Measurement (\textsc{PyBDSM}) \cite{2015ascl.soft02007M}, etc. These packages offer higher reliability and precision than the old ones and are widely used in modern radio surveys. However, they are mainly focused on compact sources with the exception of \textsc{PyBDSM} which has good detection performance for both compact and extended sources.

The existing radio telescopes are carrying out increasingly wide-field, deep, and high-resolution radio surveys. Examples include the Evolutionary Map of the Universe (EMU) survey of the Australian Square Kilometre Array Pathfinder (ASKAP)  \cite{2011PASA...28..215N}, and the GaLactic and Extragalactic All-Sky MWA (GLEAM) survey of the Murchison Widefield Array (MWA) \cite{2015PASA...32...25W,2017MNRAS.464.1146H}. 
EMU is expected to detect about 70 million radio sources, which is 25 times the number of radio sources detected by the largest survey NVSS in the past \cite{1998AJ....115.1693C}. In addition, the EMU's sensitivity and angular resolution are 45 times deeper and five times better than the NVSS, respectively. 
In the next step, the massive amount of image data generated by the survey projects places higher demands on the source 
finding software, which should be more complete and faster (near real-time) \cite{2015PASA...32...37H}. To further improve the operation speed and sample completeness, the current source finding software packages have been upgraded, such as \textsc{Aegean} 2.0 \cite{2018PASA...35...11H}. And new source finding software packages are under development, such as CAESAR Source Finder \cite{2019PASA...36...37R} and ProFound \cite{2019MNRAS.487.3971H}, which will provide better performance on both compact sources and extended sources. CAESAR supports distributed processing of large radio images on a single computing node or on multiple computing nodes. Radio sources detected by ProFound will be labelled as segmented maps, showing the pixel attribution of the detected sources, and ProFound can give more accurate flux density measurements for extended sources than component-based source finding algorithms or software packages.  

Artificial intelligence (AI) technology has been widely used in industries, such as geological monitoring, robotics, autonomous driving, face recognition, medical image analysis, etc.
Compared to standard non-AI source finding algorithms or software packages, AI-based automated methods, especially deep learning methods, aided by the acceleration with Graphics Processing Unit (GPU) devices, offer pronounced advantages in running speed and thus have promising applications. Moreover, the standard source finders can only identify islands or components of a radio source, but are not able to link the separated components for direct classification, which is done by humans in post processing by visual inspection. This may result in incomplete information about the morphology of the diffuse extended sources obtained. 

Convolutional neural networks (CNNs) are the best performing classifiers for image recognition in academic and industrial communities. CNN is a special type of neural network that learns which features are important to extract from an image and then uses these learned features for classification. To our knowledge, \textsc{TOOTHLEss} was the first CNN developed for the radio galaxy classification based on their morphology \cite{2017ApJS..230...20A}.  Recently, many deep learning-based radio sources classifiers 
\cite{2018MNRAS.480.2085A}, \cite{2020MNRAS.498.5620S}, \cite{2019Galax...8....3L}, \cite{2021MNRAS.501.4579B}, \cite{2021arXiv210208252S}, \cite{2021arXiv210201007B} or detectors \cite{2019MNRAS.482.1211W}, \cite{2019MNRAS.484.2793V} have been developed.

In computer vision,  the radio source classifiers are related to the classification task, while the detectors belong to the object detection task. Classification is concerned with the overall characteristics of the object, and it gives a description of the content of the whole image in which there is a single object. Detection, on the other hand, is concerned with specific objects and returns information about the classification and localization of multiple objects. An image obtained from an actual observation may contain multiple sources, and one needs to localize each detected radio source with a bounding box and then determine the morphological class of each source. Therefore the detection task algorithm in deep learning is more suitable for astronomical source finding. 

Object detection is one of the key problems in computer vision and its purpose is to determine whether there are instances of a given class in a given image. If so, the attributes of each object, including label name, localization, coverage and bounding box, are returned. Deep learning-based object detection algorithms are divided into single-stage algorithms and two-stage algorithms \cite{2019arXiv190505055Z}. Single-stage algorithms perform region proposal and classification throughout the process and give detection results directly, such as the single-shot multibox detector (SSD) \cite{2015arXiv151202325L} and the you-only-look-once (YOLO) \cite{2015arXiv150602640R}, \cite{2016arXiv161208242R}, \cite{2018arXiv180402767R}, \cite{2020arXiv200410934B}. In contrast, two-stage algorithms first generate region proposals, and then classify and recognize objects based on the region proposals, such as the region-based convolutional neural network (R-CNN) family: R-CNN \cite{2014arXiv1409.5403G}, Spatial Pyramid Pooling Pooling Netwo (SPP-Net) \cite{2015arXiv150201852H}, Fast R-CNN \cite{2015arXiv150408083G}, Faster Region-based Convolutional Neural Network (R-CNN) \cite{2015arXiv150601497R}, Mask R-CNN \cite{2017arXiv170306870H}. Single-stage algorithms are faster but less accurate and suitable for fast detection, while two-stage algorithms have higher precision but slower speed. There is a trade-off between recognition precision and computational cost \cite{2021MNRAS.tmp..358B}, and a suitable CNN architecture needs to be selected according to the actual requirements of the task.

With the development of the R-CNN family, its performance and precision continue to improve. In particular, Mask R-CNN not only returns a bounding box for each detected object, but also marks whether each pixel in the bounding box belongs to that object.
However, the dataset annotation file of Mask R-CNN is completely different from other R-CNNs, and it needs to contain not only the class name and bounding box, but also the mask of each object, which requires a lot of human eye recognition to achieve the data annotation. When the training set is very large, modeling before running Mast CNN becomes a time-consuming challenge. 
In contrast, Faster R-CNN has slightly lower performance but faster training speed and easier data annotation.  
\textsc{ClaRAN} is a detector built on Faster R-CNN, and it can locate and associate the radio source components with $\sim$90\% precision in each image  \cite{2019MNRAS.482.1211W}. \textsc{ClaRAN} is one of the highest-precision CNN-based classifiers, so the present study is mainly based on the pioneering work of \textsc{ClaRAN}. The performance of the published \textsc{ClaRAN} version 0.1 (v0.1) is limited by the old backbone network used \footnote{\url{https://github.com/chenwuperth/rgz\_rcnn}}; moreover, it can only classify radio sources by the number of peaks and components, and such classification lacks astrophysical relevance for most extended sources.

We developed a radio source finding and classification tool called \textsc{HeTu} based on Faster R-CNN but using a more advanced backbone network compared to \textsc{ClaRAN} v0.1. 
The development of \textsc{HeTu} is described in Sec.~\ref{sec:method}. An existing labelled dataset is used to train \textsc{HeTu} in Sec.~\ref{sec:training}. The recognition performance based on a re-labelled dataset is presented in Sec.~\ref{sec:testing}.
In Sec.~\ref{sec:predicting}, \textsc{HeTu} is applied to source finding and classification of the GLEAM images and the detection results from \textsc{HeTu} are compared with those  from \textsc{Aegean}.
The main results are summarized in Sec.~\ref{sec:conclusion}.

\section{Methods and Data}\label{sec:method}

Architecture is undoubtedly a key factor affecting the performance of CNNs.
The network structure of Faster R-CNN mainly consists of feature extraction, Region Proposal Network (RPN), classification and regression \cite{2015arXiv150601497R}. Feature extraction is performed by convolving the input image with a convolutional layer network composed of CNNs. Such a convolutional layer network is called the backbone network in deep learning and determines the eventual performance of the Faster R-CNN.

\subsection{ResNet}\label{sec:2.1}

The Visual Geometry Group Network (VGGNet) is a deep CNN developed by the Computer Vision Group at University of Oxford and Google DeepMind company \cite{2014arXiv1409.1556S}. VGGNet investigates the relationship between the depth of a CNN and its performance. It successfully constructed deep CNNs with 16 layers (called VGGNet-16) and 19 layers (VGGNet-19) and demonstrated that increasing the depth of the network can improve the final performance of the network to some extent. VGGNet has been widely used for image feature extraction in object detection, such as \textsc{ClaRAN} v0.1 \cite{2019MNRAS.482.1211W}. However, it should be noted that while there is a weak correlation between the model complexity and recognition performance, increasing the number of parameters will increase computational complexity, training time, and GPU memory usage. The number of parameters of VGGNet increases dramatically due to the rapid expansion of parameters and the construction of deep structures. The network has to continuously perform gradient propagation during the back propagation. When the number of network layers increases, the gradient decays during the layer-by-layer propagation. As a result, the weights of the previous network layers cannot be adjusted effectively. To address this limitation of the VGGNet, Microsoft Research Institute proposed the Residual Network (ResNet), which successfully trained a 152-layer neural network with fewer parameters than VGGNet  \cite{2015arXiv151203385H}. Due to the seamless integration of the ResNet with other network structures, the training of the neural network can be greatly accelerated and the precision of the trained model can be significantly improved.

\subsection{Feature Pyramid Networks}\label{sec:2.2}

In the case of high resolution, the shallow neural networks obtain smaller receptive fields and learn more detailed features in the image.
On the contrary, the deep neural networks have lower resolution, acquire feature maps with larger receptive fields and learn more semantic features. In wide-field radio images, the size of radio sources varies from a few arcseconds to a few degrees, and each source has different characteristics: small-size and simple-structure sources can be identified with shallow feature maps, while complex and extended sources need to be distinguished with deep feature maps. However, the original Faster R-CNN and most of its variants use a single high-level feature map for object detection. This may result in that small-sized objects with less pixel information can be easily missed by the down-sampling process. For example, using a single high-level feature map network for a point-like source or compact source can limit its detection precision and performance. To address this problem, Lin et al. \cite{2016arXiv161203144L}  proposed a feature pyramid network (FPN) structure suitable for multi-scale object detection, which can solve the problem of multi-scale variation in object detection with little memory consumption and computational effort.

\subsection{ResNet-FPN network} \label{sec:2.3} 

Using a combined network structure with ResNet and FPN (hereafter ResNet-FPN) as the backbone network takes advantage of ResNet in balancing recognition precision and computational cost and FPN in multi-feature object detection. ResNet-FPN not only increases the network depth, but also provides multi-scale feature maps without causing a significant decrease in running speed.
ResNet-FPN  consists  of  two  networks:  a  bottom-up  network  and  a  top-down,  laterally  connected network.

\textit{Bottom-up network.} The bottom-up network uses a ResNet network, which is processed no differently from the original ResNet. Modern CNN networks are generally divided into different stages according to the size of the feature map, and the difference of the feature map between each stage is scaled by a factor of 2. In the bottom-up network, each stage corresponds to a feature pyramid level, and the last layer of each stage is selected as the feature of the corresponding level in ResNet-FPN. For ResNet, it has 5 layers of features, denoted as C1, C2, C3, C4, and C5 \cite{2015arXiv151203385H}. C1 is the last layer of head convolution (conv1) in ResNet. C2, C3, C4, and C5 are the last residual block layer features of conv2, conv3, conv4 and conv5. Due to the large amount of memory occupied by C1, C2--C5 were finally selected as features of ResNet-FPN \cite{2015arXiv151203385H}, \cite{2016arXiv161203144L}. The step sizes of these feature layers relative to the input processed image are 4, 8, 16 and 32, respectively.

{\it Top-down and laterally connected networks}.
Four sets of feature maps are obtained through the bottom-up network. Shallow feature maps such as C2 contain more texture information, while deep feature maps such as C5 involve more semantic information. To combine these four groups of feature maps with different characteristics, a top-down and laterally connected strategy is used in ResNet-FPN. C2--C5 obtained by ResNet experience different down-sampling times, so the obtained feature maps have different sizes. To improve the computational efficiency, we first convolved the C2--C5 feature maps with a convolution layer of size $\mathrm n\_chan\times1\times1$ to reduce the dimensionality of the channels, where $\mathrm n\_chan$ is 256. After processing by this convolution layer, they have the same number of channels, but different sizes. Then, the top-level small feature maps are up-sampled to make them the same size as the feature maps of the previous stage, for example, the height and width of C5 will be the same size as C4 after up-sampling. The upsampling method is implemented by an un-pooling algorithm that un-pools the input with a fixed matrix for kronecker product \cite{2013arXiv1311.2901Z}. The feature map is updated by adding the up-sampled features and the previously convolved features. The whole process is updated from the top to the bottom layers of the network, so it is called the top-down network. The units of addition operations are called lateral connections. The advantage of this network structure is that it utilizes both the strong semantic features of the upper layer (more suitable for classification) and the high-resolution information of the lower layer (better for localization).

After updating the feature maps, these feature maps are convolved with a convolutional layer of size $\mathrm n\_chan\times3\times3$ to obtain the final feature maps (P2, P3, P4 and P5). This convolution operation is performed to reduce the aliasing effect of up-sampling. The feature map P6 is used for the 5th anchor scale in the RPN, which is generated by sub-sampling P5 with a step size of 2, using  two-dimensional Max Pooling layers.

\begin{figure*}
\centering
\includegraphics[scale=1]{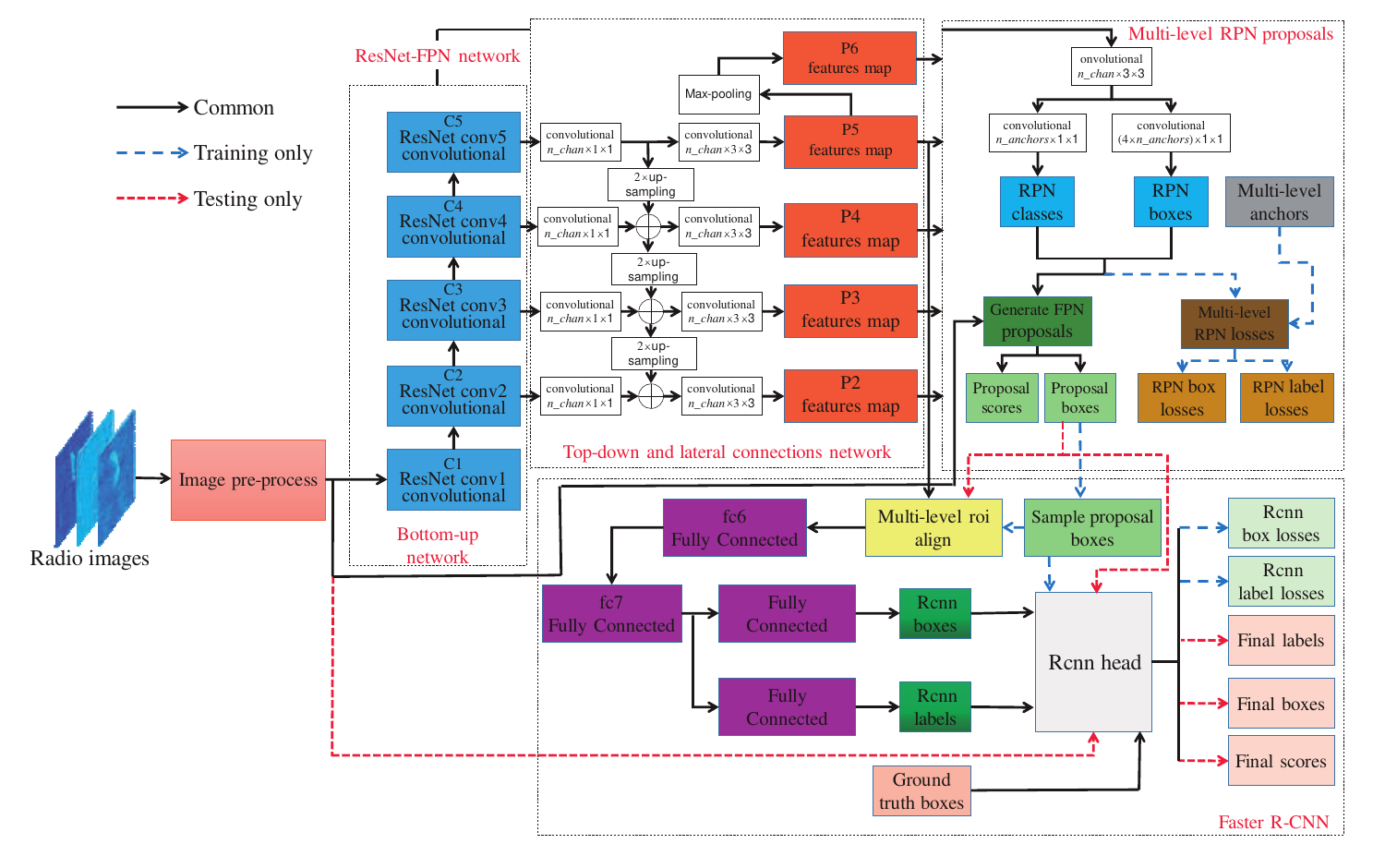}
    \caption{The workflow of \textsc{HeTu}. It consists of four steps: image pre-processing, ResNet-FPN network, multi-level RPN proposal network, and classification and box regression network (Faster R-CNN). Image pre-processing is performed for image resizing and image data normalization. ResNet-FPN network is used to generate multi-scale feature maps (P2 to P6) from the processed images. The multi-scale feature maps are then imported into a multi-level RPN proposal network to generate proposal scores and proposal bounding boxes one by one. Finally, multi-scale features maps, proposal results, processed images and ground truth boxes are subjected to classification and box regression using a Faster R-CNN network. The  label names, scores and boxes of detected sources are exported.
    }
    \label{fig:resnet_fpn}
\end{figure*}

\subsection{\textsc{HeTu} -- an improved Faster R-CNN}\label{sec:2.4}

We developed a source detector named \textsc{HeTu}\footnote{\url{https://github.com/lao19881213/rgz\_resnet\_fpn}} based on ResNet-FPN.  
The workflow of \textsc{HeTu} is shown in Fig.~\ref{fig:resnet_fpn}, which includes four steps: image pre-processing, ResNet-FPN network, multi-level RPN proposals network, and classification and box regression network (Faster R-CNN network). First, the input radio images are pre-processed; second, the processed images are sent to the ResNet-FPN network to generate multi-scale features maps (P2, P3, P4, P5 and P6). Then, the multi-scale features maps are fed into the multi-level RPN network to obtain the proposal scores and the proposal bounding boxes, respectively. Finally, the processed images, features maps and proposals results are fed into the Faster R-CNN network for final radio source classification and precise localization.

\subsubsection{Image Pre-processing}\label{sec:img-prep}

In the image pre-processing step, two operations are performed for each input image in the streaming format: image resizing and image data normalization. In the image resizing operation, the shortest edge in the input image is resized to a certain value (SHORT\_EDGE\_SIZE) while the longest edge is avoided to exceed MAX\_SIZE using bi-linear interpolation. In the workflow, based on the experience of Wu et al \cite{2019MNRAS.482.1211W}, the SHORT\_EDGE\_SIZE and MAX\_SIZE are set to 600 and 1,333 pixels, respectively. Thus 
the height and width of the input image are adjusted to 600 pixel $\times$ 600 pixel. 
The normalization makes the input image data have the same distribution, which prevents the gradients from disappearing and makes the network converge faster. After image resizing, the pixel values of each image data are subtracted to a certain mean value for each dimension (channel, height, width), and then the pixel values of the resulting data are divided by a given standard deviation value for each dimension.  

\subsubsection{ResNet-FPN Network}

For more details on the ResNet-FPN network, please refer to Sec.~\ref{sec:2.3}.
ResNet has three different layers: 50 (hereafter ResNet-50), 101 (ResNet-101), and 152 (ResNet-152). We note that \textsc{ClaRAN} v0.2 replaces VGGNet with ResNet-50-FPN. In this work, we used both ResNet-50 and ResNet-101, and the corresponding models generated by \textsc{HeTu} are called \textsc{HeTu}-50 and \textsc{HeTu}-101, respectively.
The training results shown in Sec.~\ref{sec:training} are obtained from the experiment on \textsc{HeTu}-101. The learning rate curve, total loss, weight decay curves using \textsc{HeTu}-50 are similar to those of \textsc{HeTu}-101, so we don't repeat the presentation.

\subsubsection{Multi-level RPN Proposals Network}

Based on the ResNet-FPN design, RPN replaces the original single-scale feature map with an FPN model for multi-scale processing \cite{2016arXiv161203144L}. Each layer of the feature pyramid is processed using the same network header. The output is called RPN class and RPN frame, which includes one convolutional layer of size $\mathrm{ n\_chan\times3\times3}$ and two convolutional layers of size $1\times1$. The number of channels in the latter two convolutional layers is $\mathrm{n\_anchors}$ and $\mathrm{4\times n\_anchors}$, respectively. Since the network header is a dense sliding window for all positions of all feature pyramid layers, there is no need to set multi-scale anchors for specific feature pyramid layers. Therefore, $\mathrm{n\_anchors}$ is equal to the number of anchor ratios. P2, P3, P4, P5, and P6 correspond to anchor ratios of $32^2$, $64^2$, $128^2$, $256^2$, and $512^2$ pixels, respectively. Astronomical objects cannot be square in morphology, so three ratios are used: 0.5, 1 and 2. For a common pathway, the RPN class and RPN box results are entered into the ``Generate RPN Proposal" process, which generates proposal scores and proposal boxes using the NMS (Non-Maximum Suppression) algorithm \cite{2015arXiv150601497R}. The RPN class and RPN box results are also input to the multi-level anchors for multi-level RPN loss processing to obtain the RPN box loss and RPN label loss for the training pathway only.

\subsubsection{Classification and box regression network}

We use Faster R-CNN as the classification and recognition network. First, the four feature maps (P2, P3, P4, and P5) from the ResNet-FPN network and the proposal box from the RPN network were imported into the ``multi-level region of interest (roi) align", and the feature maps were cropped and resized to have the same resolution size of $14$ pixel $\times$14 pixel. For the testing route only, the proposal boxes are directly used as the output of the RPN network. And for the training route, the proposal boxes are a number of box samples from all proposals. Then, the process is similar to the original Faster R-CNN, and the aligned feature maps are placed into two fully connected layers (fc6 and fc7) with a total of 1024 channels. Next, classification and box regression are performed using these two fully connected layers, and the detected label names and bounding boxes, called Rcnn labels and Rcnn boxes, are exported. Finally, the Rcnn labels, Rcnn boxes, proposed boxes, processed images and ground truth boxes are processed by the `Rcnn header' function to generate the final labels, boxes and scores of the detected sources in the test dataset, and the box loss and label loss (\textit{i.e.}, Rcnn box loss and Rcnn label loss) of the Faster R-CNN are computed separately. 

\section{Experiments}\label{sec:3}

We run three experiments to verify the performance of \textsc{HeTu}: (1) training experiment; (2) testing experiment; (3) predicting experiment.


All experiments in this paper were conducted on a  single GPU node of the China SKA Regional Centre Prototype (CSRC-P) system \cite{2019NatAs...3.1030A}. As shown in Table~\ref{tab:GPU}, this node is equipped with 36 CPU cores operating at 2.3 GHz, eight NVIDIA V100 SXM2 GPUs with 32 GB high-bandwidth memory, 7.68 GB local Solid State Disk (SSD) storage and 512 GB main memory. These 8 GPUs have a theoretical computing power of over 80 TFLOPS with double precision.  

   \begin{table*}
   \centering
      \caption[]{The specification of the GPU node on CSRC-P}\label{tab:GPU}
       \begin{tabular}{llll}
        \hline\hline
              Type &	Quantity	& Specification &	Capacity \\
        \hline
                CPU	& 2 & 	Intel(R) Xeon(R) Gold 6140, 2.30 GHz 	& 36 cores \\
                Memory	& 16	& DDR4-32GB-2666-ECC	& 512 GB \\
                Hard disk &	4	& 1.92TB-SSD-2.5	& 7.68 TB \\
                GPU & 	8	& Nvidia Tesla V100 SXM2 32 GB	& 80 TFLOPS (double precision)\\
        \hline
        \end{tabular}
   \end{table*}

The proposed workflow described in Sec.~\ref{sec:2.4} is based on the TensorFlow framework. We used a data parallelism strategy to accelerate the training of the synchronous update. In this strategy, the whole model parameters ($i.e.$, weights) were replicated across multiple GPU devices. It created one tower on each GPU device within its own variable scope. Each tower independently processed a different sub-batch (local batch) of data, which was split from the original data batch, and output the gradient of the weights with respect to the loss of the model over the local batch. As described in Table \ref{tab:GPU}, 8 GPU devices were used, so the batch size was 8. Each gradient update was summed across the towers using ``ALL-Reduce" collective communication primitives through Nvidia Collective multi-GPU Communication Library (NCCL) \cite{hua2017mgupgma}. For the same dataset, the training experiments using eight GPU devices run approximately 2.5 times faster than using a single GPU. 

\subsection{Training and validation experiment}\label{sec:training}

The first experiment is used to train \textsc{HeTu} and validate its performance. 

\subsubsection{Training}\label{sec:3.1.1}

The training dataset used for the verification experiment is the same as those used in the Wu et al. \cite{2019MNRAS.482.1211W} for the sake of comparison with our method. The dataset is derived from the citizen science project Radio Galaxy Zoon Data Release 1 (DR1) (\cite{2015MNRAS.453.2326B}, Wong et al., in preparation). The radio source data include the 1.4-GHz FIRST radio images \cite{1995ApJ...450..559B} in both FITS and PNG formats and mid-infrared images at 3.4 $\mu$m from the Wide-field Infrared Survey Explorer (WISE, \cite{2014yCat.2328....0C}) in PNG format. The \textsc{ClaRAN} dataset contains $10,744$ subjects, of which $6,141$ subjects are used for training and $4,603$ subjects for testing. Each subject is a PNG image with a size of $132\times132\times3$ (pixel$\times$pixel$\times$channel). The size of each pixel is 1.375 arcsec, so the field of view of each image is 3 arcmin$\times$3 arcmin. One image (subject) may contain more than one source. The class name of the first dataset is defined as $iC\_jP$, where $i$ is the number of components ($C$), and $j$ is the number of flux density peaks ($P$) \cite{2019MNRAS.482.1211W}. $C$ refers to the discrete independent radio source component identified above $4\sigma$ noise level, and $P$ represents the distinguishable peak identified in each class of objects. To ensure that the visual distinguishability of the radio source morphology and the balanced of the training datasets, finally, \textsc{ClaRAN} used four datasets ($i.e.$ D1, D2, D3, D4) and six source classes ($i.e.$ $1C\_1P$, $1C\_2P$, $1C\_3P$, $2C\_2P$, $2C\_3P$, $3C\_3P$). The number of radio sources in each class is summarized in Table~\ref{tab:claran_data}.

\begin{table}
\centering
      \caption[]{Number of radio sources for each class in the first dataset \cite{2019MNRAS.482.1211W}}
         \label{tab:claran_data}
       \begin{tabular}{llll}
        \hline\hline
             Class name & Training & Testing\\
        \hline
                $1C\_1P$  & 3,516 & 1,782\\
                $1C\_2P$  & 810 & 521\\
                $1C\_3P$  & 728& 684\\
                $2C\_2P$  & 647 & 604\\
                $2C\_3P$  & 608 & 599\\
                $3C\_3P$  & 666 & 668\\ \hline
                Total  & 6367 & 4858\\
            \hline
        \end{tabular}
        \end{table}

\begin{figure}[H]
\centering
\includegraphics[scale=1]{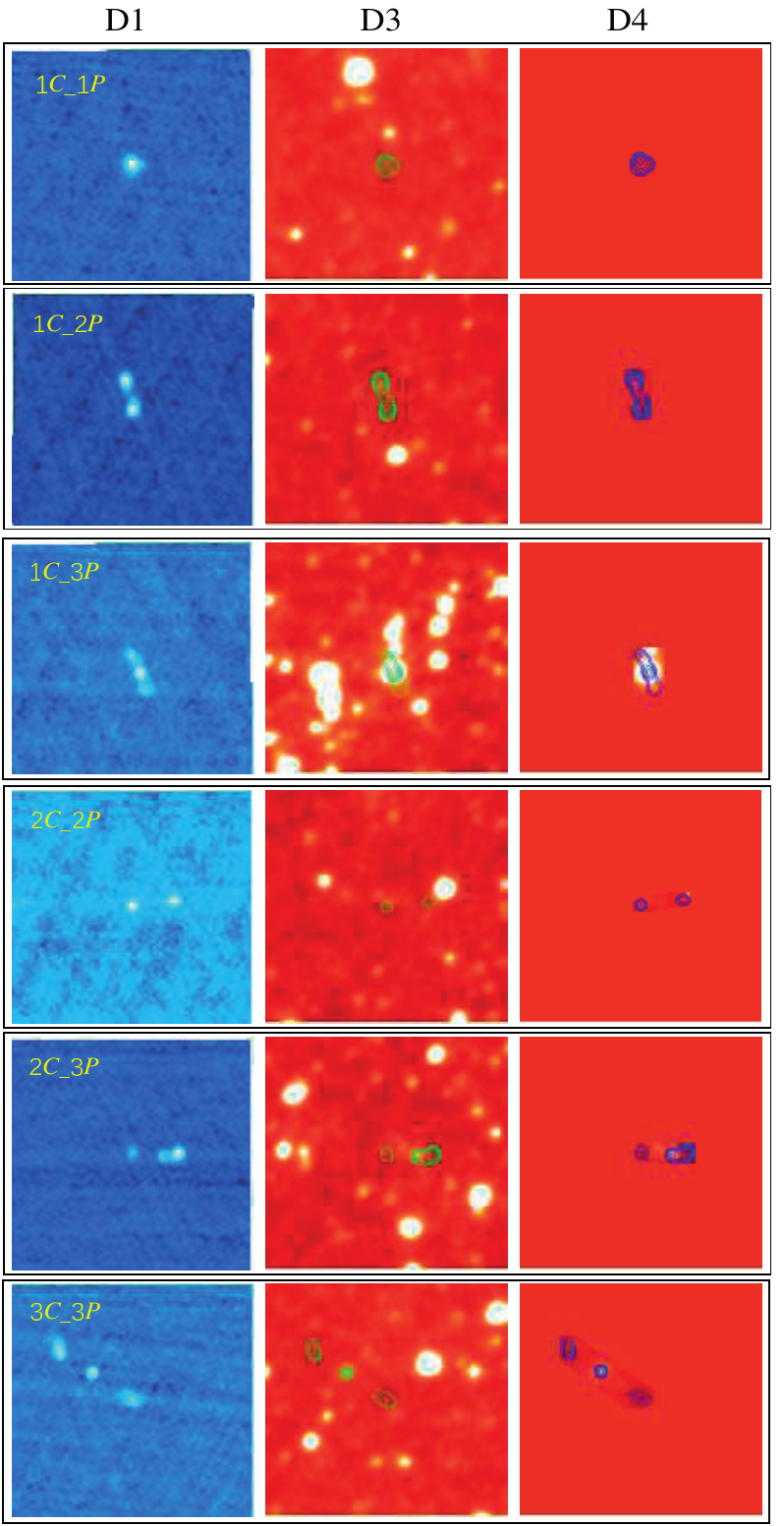}
\caption{Example images of each class in the D1, D3 and D4 datasets, respectively. These images are described in more detail in Sec.~\ref{sec:training}.}
\label{fig:example_claran}
\end{figure}

In this paper, we only utilized the D1, D3, and D4 datasets, as shown in Fig.~\ref{fig:example_claran}. D1 are a set of three-channel (RGB) images generated from the original FIRST FITS files at the log-min-max scale and `cool'-colour index maps. D3 is a composite image set generated by overlaying the FIRST image (contour map) on top of the corresponding WISE image (color-scale map) with a $5\sigma$ threshold. D4 images are derived from D3 images by plotting a convex hull overall radio contours of D3 and then labelling all pixels outside the convex hull with the average pixel value of each channel on all images. 

A linear warm-up strategy was used to adjust the learning rate. The learning rate ($LR$) function is shown in Eq.~\ref{eq:LR}. Since the weights of the model were randomly initialized at the beginning of the training, a large learning rate would introduce the instability of the model. The liner warm-up strategy used a smaller initial learning rate ($LR_{\rm init}$), and the learning rate increased linearly from the beginning of the first global step or iteration to the end of warm-up steps. 
When the trained model was stable and the number of the global step or iteration ($G_{\rm step}$) was equal to the number of warm-up steps ($WP_{\rm steps}$), the learning rate was modified to a normal value (base learning rate, $LR_{\rm base}$). The learning rate was then decayed using a step-based exponential decay strategy, where the learning rate was decayed over multiple specified decay steps ($D_{\rm steps}$) with a decay coefficient ($D_{\rm coef}$). As the learning rate changed dynamically, the training process became smoother and it was easier to reach a local optimum. Besides, we adopted a momentum optimizer of 0.9 and $L2$ regularization (or weight decay) of 0.0001 to accelerate the network convergence and prevent overfitting of the model. 

\begin{equation}
LR = \left\{ \begin{gathered}
  LR_{\rm init}(1 + \frac{{{G_{\rm step}}}}{{{E_{\rm steps}}}})\quad  \quad \quad {\mkern 1mu} {\mkern 1mu} {\mkern 1mu} {G_{\rm step}} < WP_{\rm steps} \hfill ,\\
  LR_{\rm base}{({D_{\rm coef}})^{{G_{\rm step}}/{D_{\rm steps}}}}\quad \;\;\,{G_{step}} \geqslant W{P_{steps}} \hfill ,\\ 
\end{gathered}  \right.
\label{eq:LR}
\end{equation}
where, the symbol $/$ indicates division and rounding, $E_{\rm steps}$ is the number steps of per epoch.

\begin{figure}
\vspace{0.5cm}
\centering
\includegraphics[width=0.45\textwidth]{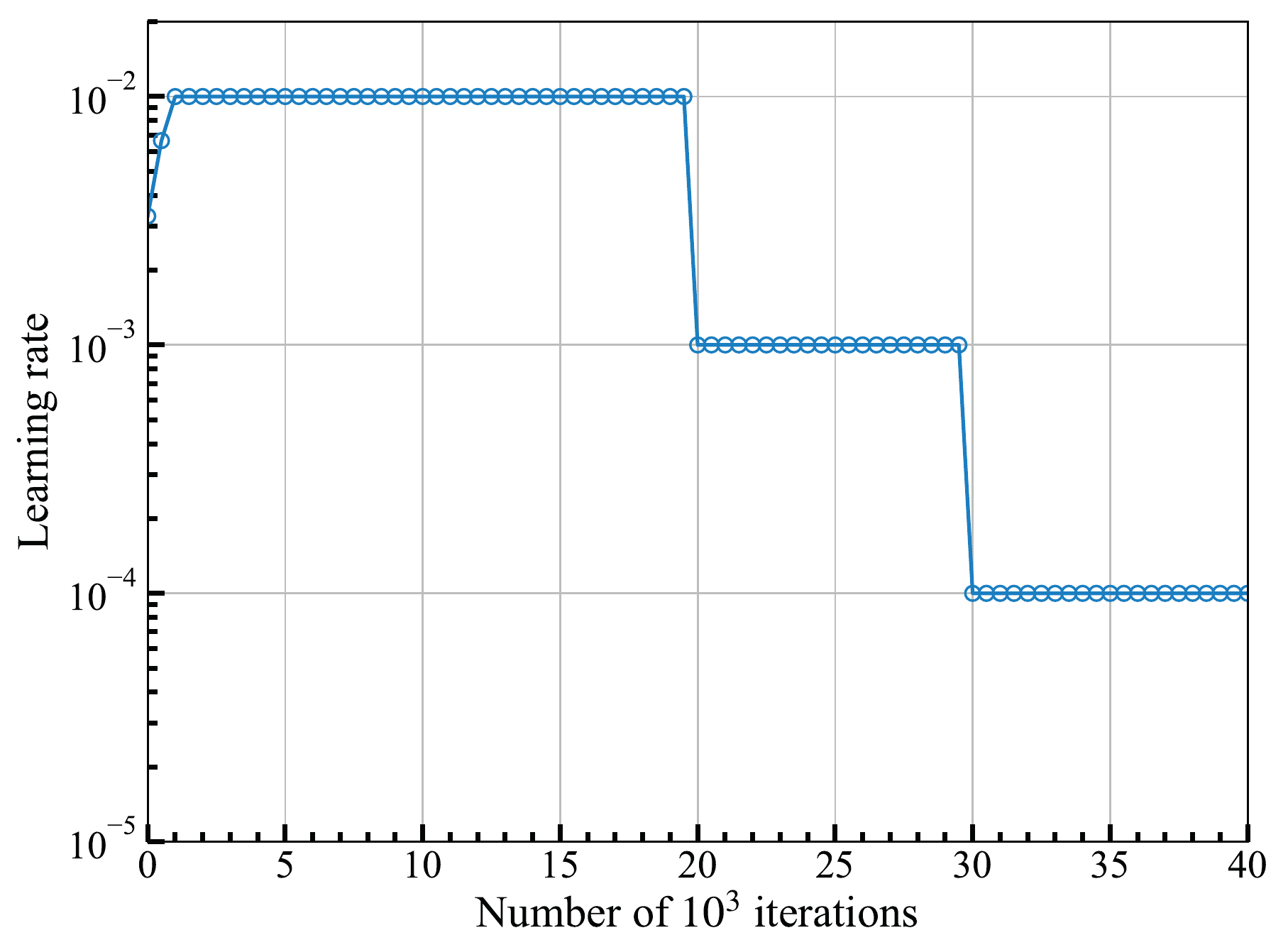}
\caption{The learning rate curve in training. The learning rate increases linearly with the number of iterations when the number is less than 1,000. When the number of iterations equals 1,000, the learning rate is set to 0.01. The learning rate decays to 0.001 after 20,000 iterations, and further decays to 0.0001 after 30,000 iterations.} 
\label{fig:LR-relab}
\end{figure}

In training, $LR_{\rm init}$ was set to 0.0033, $WP_{\rm steps}$ to 1,000, $LR_{\rm base}$ to 0.01, $E_{\rm steps}$ to 500, $D_{\rm coef}$ to 0.1, and $D_{\rm steps}$ to a list set of [20,000, 30,000, 40,000]. Fig.~\ref{fig:LR-relab} shows the learning rate curve in training. At global step 1, the learning rate was initially 0.0033. Then, the learning rate increased linearly according to the $LR$ function. When the global step was equal to $WP_{\rm steps}$, the learning rate was set to 0.01,  after which the learning rate decayed to 0.001 and 0.0001 after 20,000 and 30,000 steps, respectively. Using \textsc{HeTu}-50 and \textsc{HeTu}-101 networks, the training speeds are 0.42 seconds and 0.49 seconds per step on D1, D3, and D4 datasets. 
Thus, a workflow instructed to execute 40,000 steps required 4.7 to 5.4 hours of training time on 8 GPU devices. 

The efficiency and effectiveness of the training workflow is mainly determined by the loss function ($L_{\rm total}$), which is the sum of RPN loss, Faster R-CNN loss and weight decay cost ($L_{\rm wd\_cost}$). Both RPN loss and Faster R-CNN loss include label classification loss ($L_{\rm rpn\_label}$ and $L_{\rm rcnn\_label}$) and bounding box loss ($L_{\rm rpn\_box}$ and $L_{\rm rcnn\_box}$). The $L_{\rm total}$ is calculated as: 
\begin{equation}
L_{\rm total}=L_{\rm rcnn\_box} + L_{\rm rcnn\_label} +L_{\rm rpn\_box} + L_{\rm rpn\_label} +L_{\rm wd\_cost}.
\label{eq:loss}
\end{equation}

\begin{figure}
\vspace{0.3cm}
\centering
\includegraphics[width=0.45\textwidth]{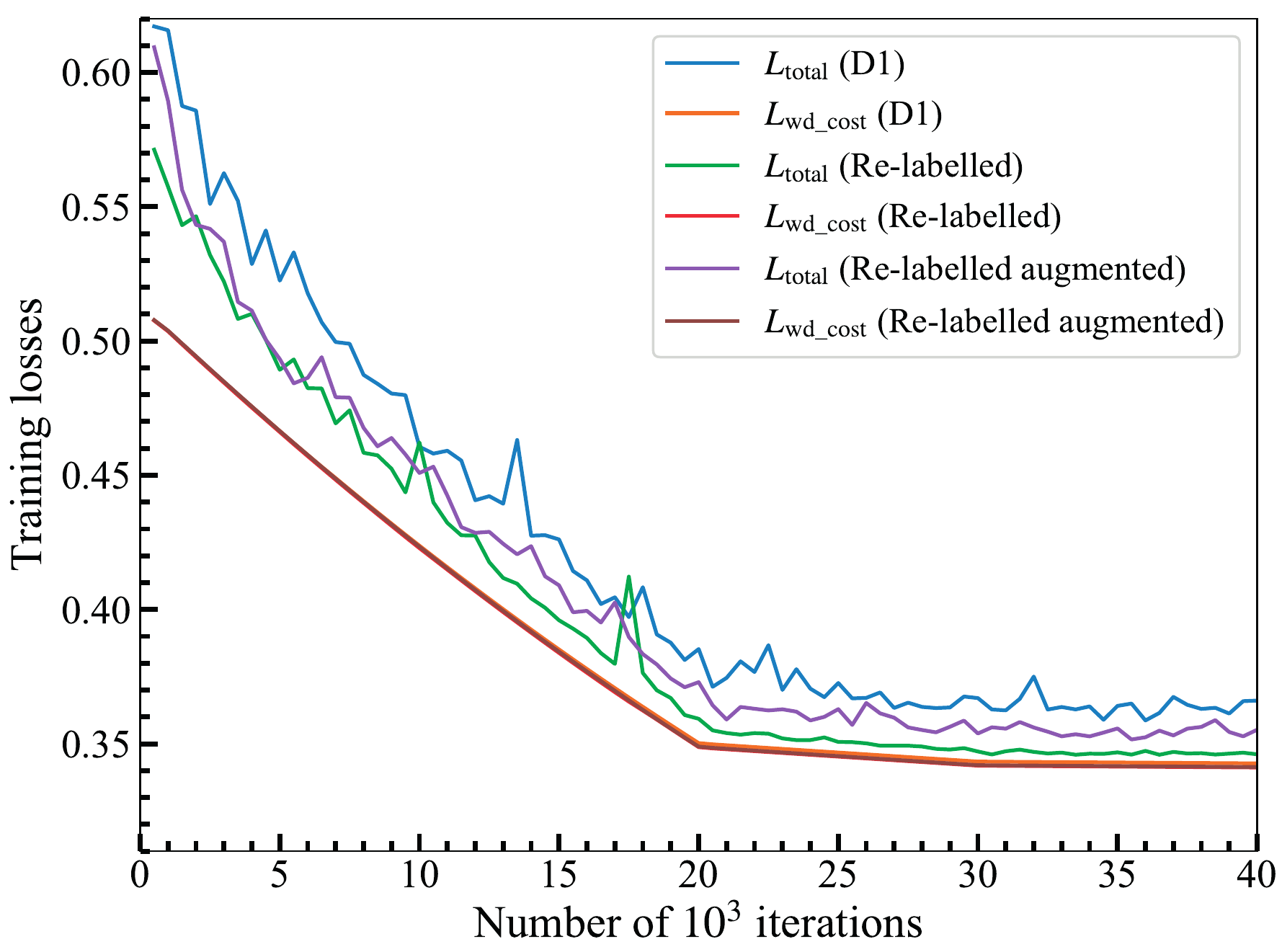}
\caption{Total loss and weight decay cost ($Y$-axis) curves for the D1 and re-labelled datasets trained using \textsc{HeTu}-101 network. $X$-axis denotes the number of iterations. The light blue line, green line and purple line are the total loss curves of the D1, re-labelled and its augmented datasets, respectively; and the orange line, pink line and brown line at the bottom are the corresponding weight decay cost curves, respectively. The losses are averaged every 500 iterations (or epoch) during the training process for visualization. The three weight decay cost curves are so close in shape that they are squeezed together and cannot be separated.} 
\label{fig:total_loss_D1_relabelled}
\end{figure}

The purpose of training is to reduce the training loss and obtain a better model.
The light blue and orange lines in Fig.~\ref{fig:total_loss_D1_relabelled} represent the total and weight decay cost curves for the \textsc{HeTu}-101 model on the D1 dataset. 
The total loss and weight decay cost curves derived from D3 and D4 datasets are similar to Fig.~\ref{fig:total_loss_D1_relabelled} and therefore are not shown in this paper. 
In Fig.~\ref{fig:total_loss_D1_relabelled}, the total training loss of D1 dataset decreases from 0.62 to 0.36 and the weight decay cost decreases from 0.51 to 0.34. Due to the imbalance in the amount of D1 dataset, the total loss curves fluctuates but decreases in the first 20,000 iterations, and fluctuates more when the iteration is around 13,500. In contrast, the weight decay cost curves drop slowly over the first 20,000 iterations. Both curves seem to be stabilized in their decline after 20,000 to 40,000 iterations, indicating that the model has reached its learning capacity with the current network architecture and data sets. This also suggests that most of the loss contribution comes from the weight decay cost.

\begin{figure}
\centering
\includegraphics[width=0.45\textwidth
]{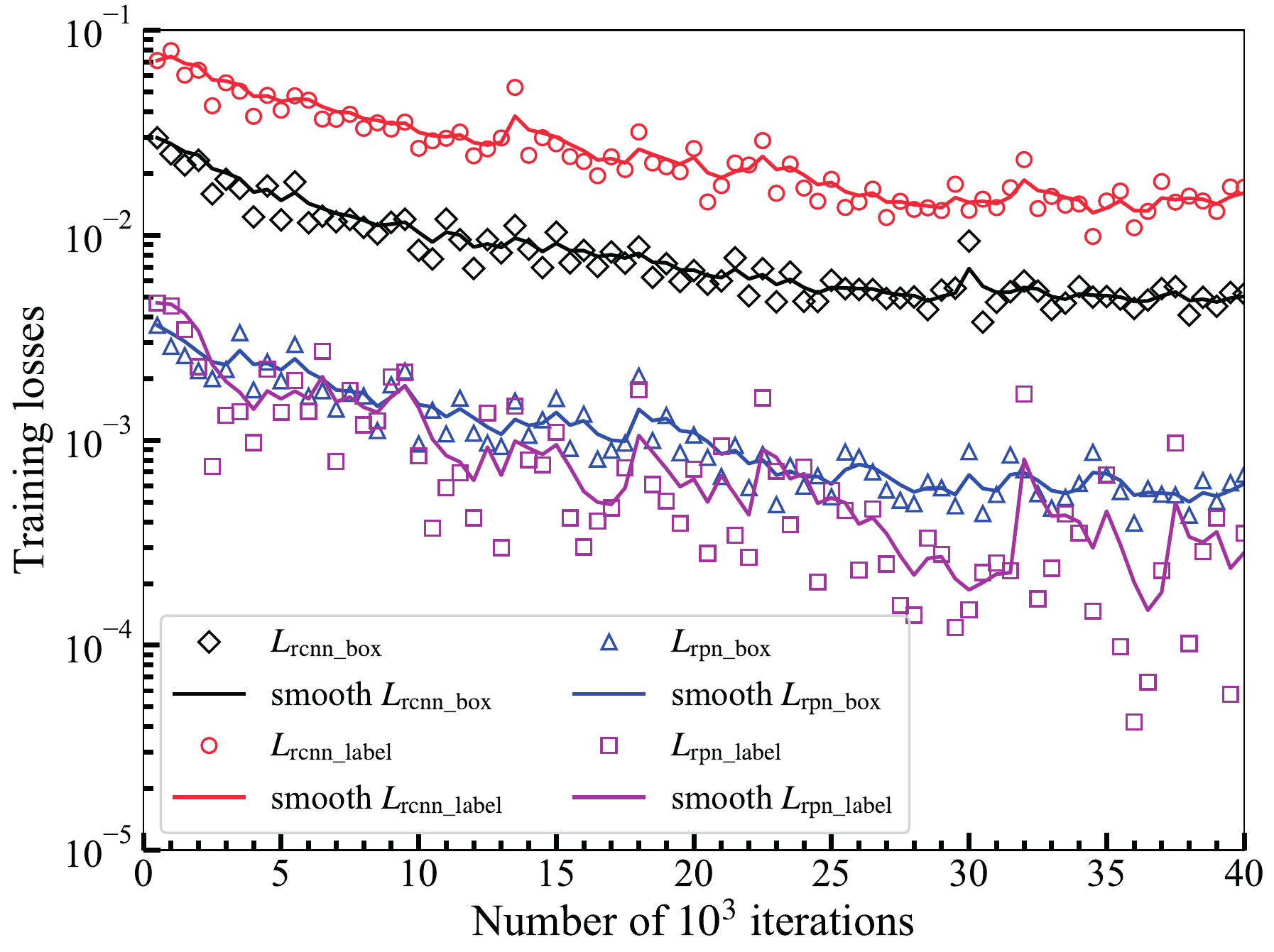}
\caption{RPN and Faster R-CNN loss ($Y$-axis) curves of using \textsc{HeTu}-101 network for training D1 datasets by the number of iterations ($X$-axis). The red line and scatter hollow circle markers are the smoothed and measured Faster R-CNN label classification loss ($L_{\rm rcnn\_label}$), the black line and scatter plus markers are smoothed and measured the Faster R-CNN bounding box loss ($L_{\rm rcnn\_box}$), the blue line and scatter hollow upper triangle markers are the smoothed and measured RPN label classification loss ($L_{\rm rpn\_label}$), the magenta line and scatter hollow squares markers are the smoothed and measured RPN bounding box loss ($L_{\rm rpn\_box}$).}
\label{fig:rpn_rcnn_loss_D1}
\end{figure}

Fig.~\ref{fig:rpn_rcnn_loss_D1} shows four additional loss curves for the \textsc{HeTu}-101 model on the D1 dataset. $L_{\rm rcnn\_box}$ loss decreases from 0.030 to 0.0038, $L_{\rm rcnn\_label}$ decreases from 0.0796 to 0.0099, indicating an approximately 8-fold reduction in loss after Faster R-CNN training. $L_{\rm rcnn\_box}$ loss decreases from 0.0036 to 0.00039, $L_{\rm rpn\_label}$ from 0.0047 to 0.000042, indicating that $L_{\rm rcnn\_box}$ is reduced 9 times and $L_{\rm rpn\_label}$ is reduced 112 times after the training converges. These curves show that the learning ability of the model is well validated on the current datasets.

\subsubsection{Evaluation}\label{sec:3.1.2}

We use the Average Precision ($AP$) and the mean Average Precision ($mAP$) as evaluation metrics for the recognition performance.  $mAP$ was first defined in the PASCAL Visual Objects Classes (VOC) challenge \cite{Everingham15}. Nowadays, it is widely used as an evaluation metric in object detection. 

$AP$ is a function of precision ($P_{rec}$) and recall ($R$), which is usually obtained by calculating the area of the 
graph formed by the $P_{rec}-R$ curve and the $x$ axis. $P_{rec}$ is the percentage of detected sources that are indeed sources based on the ground truth values. $R$ is the percentage of correctly detected ground truth sources. Given a class $c \in 1...{N_c}$, let $L_c$ represent a list of radio sources detected by proposed methods as class $c$ sources from all images in the testing data set. The precision $P_{rec}$ of the class $c$ source can be calculated by     

\begin{equation}
P_{rec}(k) = \frac{{\sum\nolimits_{i = 1}^k {TP(i)} }}{{\sum\nolimits_{i = 1}^k {(TP(i)+FP(i))} }},\quad 1 \leqslant k \leqslant {|L_c|},
\label{eq:3}
\end{equation}
where $|L_c|$ is the length of $L_c$. $TP(i)$ is an indicator function which is equal to $1$ for the Intersection Over Union ($IOU$) greater than $0.5$, and $0$ otherwise. $FP(i)$ is an indicator function which is equal to 1 for $IOU$ less than 0.5 and 0 otherwise. $IOU$ is given by the overlapping area between the predicted bounding box (${B_p}$) and the ground truth bounding box ($B_{gt}$) divided by the area of the union between them:
\begin{equation}
IOU = \frac{{{area}{\kern 1pt} ({B_p}\; \cap \,\;{B_{gt}})}}{{{area}{\kern 1pt} ({B_p}\; \cup \,\;{B_{gt}})}},
\label{eq:4}
\end{equation}
where the symbol $\cup$  represents the union of ${B_p}$ and $B_{gt}$, $\cap$ the intersection of them.

The recall $R$ is given by 
\begin{equation}
R(k) = \frac{{\sum\nolimits_{i = 1}^k {TP(i)} }}{{G{T_c}}},\quad 1 \leqslant k \leqslant {|L_c|},
\label{eq:5}
\end{equation}
where $G{T_c}$ is the total number of ground truth of class $c$ sources in the testing set.

For $AP$, instead of using the $P_{rec}-R$ curve directly, we use the smoothed $P_{rec}-R$ curve. For each point on the $P_{rec}-R$ curve, the value of $P_{rec}$ takes the maximum value of $P_{rec}$ to the right of that point, i.e.,  the envelope ${P_{\rm envelope}}$ of $P_{rec}$ is calculated. The formula for ${P_{\rm envelope}}$ is 

\begin{equation}
{P_{{\text{envelope}}}}(n - 1) = \max \left( {{P_{\rm new}}(n - 1),\;{P_{\rm new}}(n)} \right){\kern 1pt} ,|{L_c}| + 1 \leqslant n \leqslant 1,
\label{eq:6}
\end{equation}
where ${P_{new}}$ is given by appending the sentinel value 0 at both the front-end and back-end of $P_{rec}$.

The $AP$ of class $c$ can be obtained by calculating the sum of the discrete areas formed by the ${P_{\rm envelope}}$ 
\begin{equation}
A{P_c} = \sum\nolimits_{m = 0}^{|{L_c}|} {({R_{\rm new}}(m + 1) - {R_{\rm new}}(m)){P_{
\rm envelope}}(m + 1)},
\label{eq:7}
\end{equation}
where $R_{\rm new}$ is given by appending sentinel values 0 and 1 at the front-end and back-end of $R$, respectively. 

Finally, the $mAP$ of $N_c$ classes is calculated as
\begin{equation}
mAP = \frac{{\sum\nolimits_{c = 1}^{{N_c}} {A{P_c}} }}{{{N_c}}}.
\label{eq:8}
\end{equation}

Eq.~\ref{eq:7} and Eq.~\ref{eq:8} are applied to evaluate the D1, D3, and D4 datasets using their own trained models, as illustrated in Fig.~\ref{fig:example_claran}. The testing speeds using the trained \textsc{HeTu}-50 and \textsc{HeTu}-101 model were 4.9 milli-second (ms) and 5.9 ms per image, respectively. Tables~\ref{tab:AP_D1}, \ref{tab:AP_D3}, and \ref{tab:AP_D4} show the results of $AP$ and $mAP$ for different methods. 

For the D1 datasets,  $mAP$ increased from 78.4\% for \textsc{ClaRAN} v0.1, to 86.7\% for \textsc{HeTu}-50 and to 87.6\% for \textsc{HeTu}-101.
$mAP$ increased from 82.5\% for ResNet-50 to 86.7\% for \textsc{HeTu}-50, and from 83.9\% for ResNet-101 to 87.6\% for \textsc{HeTu}-101.
In particular,  $AP$s for all classes were greatly improved using the trained ResNet-FPN (\textit{i.e.},  \textsc{HeTu}-50 and  \textsc{HeTu}-101) models compare to only using the trained ResNet models (ResNet-50 and ResNet-101)\footnote{\url{https://github.com/lao19881213/rgz\_resnet}}. Especially the $AP$s of compact radio sources have been greatly improved, 
for example, the maximum increase of $AP$ for class $1C\_1P$ was 0.062, and the maximum increase of 0.101 for $1C\_2P$. These values indicate that the recognition performance using \textsc{HeTu} is significantly better than \textsc{ClaRAN} v0.1, and also higher than the un-trained ResNet models as well.

$AP$ and $mAP$ show similar improvements on the D3 and D4 datasets, but their percentage improvements are slightly lower than that of the D1 dataset, indicating that the trained ResNet-FPN model (\textit{i.e.}, \textsc{HeTu}) has better scalability on D1.
When \textsc{HeTu} uses the same trained ResNet-FPN model, the derived $mAP$s for these three datasets are almost the same (Tables~\ref{tab:AP_D1}--\ref{tab:AP_D4}): 87.6\%, 88.2\%, and 88.0\% for  D1, D3 and D4 datasets, respectively. 
This indicates that \textsc{HeTu} is only slightly affected by the visualization of the fusion method. In other words, the performance of \textsc{HeTu} does not strongly depend on the dataset used, and changing the dataset has little effect on the $mAP$. 

\textsc{ClaRAN} v0.2 replaces the VGGNet with the more advanced ResNet-50-FPN, which has a similar backbone network to that of \textsc{HeTu}-50. The $mAP$ values\footnote{\url{https://github.com/chenwuperth/claran}} derived from \textsc{ClaRAN} v0.2 on the D1 and D3 datasets are 85.9\% and 86.1\%, respectively. As a comparison, we can see that the recognition precision of \textsc{HeTu}-50 is slightly (0.8 percentage) higher than that of \textsc{ClaRAN} v0.2.

   \begin{table*}
   \centering
      \caption{$AP$ and $mAP$ results of evaluation on D1 datasets}
      \label{tab:AP_D1}
      \begin{tabular}{l|llll|l}
      \hline\hline
              Class name & ResNet-50 &	ResNet-101	&	\textsc{HeTu}-50	& \textsc{HeTu}-101&	\textsc{ClaRAN} v0.1 \cite{2019MNRAS.482.1211W}  \\
       \hline
        $1C\_1P$ &	0.8904&	0.9180&		0.9519&	0.9621&	0.8580 \\
        $1C\_2P$&	0.6583&	0.6782&	    0.7597&	0.7636&	0.6882\\
        $1C\_3P$&	0.8920&	0.9004&		0.9126&	0.9186&	0.8816\\
        $2C\_2P$&	0.7836& 0.8358&		0.8398&	0.8674&	0.7014\\
        $2C\_3P$&	0.8013&	0.7809&		0.8037&	0.8057&	0.7099\\
        $3C\_3P$&	0.9269&	0.9230&		0.9322&	0.9413	&0.8636\\\hline
        $mAP$	&82.5\%	&83.9\%	&86.7\%	&87.6\%	&78.4\% \\
        \hline
        \end{tabular} 
   \end{table*}
   
   \begin{table*}
   \centering
        \caption[]{$AP$ and $mAP$ results of evaluation on D3 datasets}
        \label{tab:AP_D3}
        \begin{tabular}{l|llll|l}
        \hline\hline
              Class name & ResNet-50 &	ResNet-101	&	\textsc{HeTu}-50	& \textsc{HeTu}-101&	\textsc{ClaRAN} v0.1 \cite{2019MNRAS.482.1211W}  \\
        \hline
            $1C\_1P$ &	0.8971&	0.9456	&0.9523	&0.9582	&0.8485\\
            $1C\_2P$&	0.6312	&0.6817		&0.7452&	0.7476&	0.6746\\
            $1C\_3P$&   0.8709	&0.8911	&	0.8925	&0.9026&	0.8876\\
            $2C\_2P$&	0.8321	&0.8773	&	0.8708	&0.8784&	0.7983\\
            $2C\_3P$&	0.8190	&0.8337		&0.8386&	0.8374&	0.8047\\
            $3C\_3P$&	0.9728	&0.9770 &	0.9670&	0.9678&	0.9424\\ \hline
            $mAP$ &83.7\% &86.8\%&	87.8\%&88.2\%&82.6\% \\
        \hline
        \end{tabular} 
   \end{table*}
   
   \begin{table*}
   \centering
      \caption[]{$AP$ and $mAP$ results of evaluation on D4 datasets}
      \label{tab:AP_D4}
        \begin{tabular}{l|llll|l}
        \hline\hline
              Class name & ResNet-50 &	ResNet-101	&	\textsc{HeTu}-50	& \textsc{HeTu}-101&	\textsc{ClaRAN} v0.1 \cite{2019MNRAS.482.1211W}  \\
        \hline
            $1C\_1P$ &	0.9026&	0.9108&		0.9552&	0.9579&	0.8784\\
            $1C\_2P$&	0.6748&	0.7118&		0.7313&	0.7425&	0.7074\\
            $1C\_3P$&   0.8970&	0.9064&		0.8881&	0.9024&	0.8941\\
            $2C\_2P$&	0.8451&	0.8656&		0.8704&	0.8770&	0.8200\\
            $2C\_3P$&	0.8155&	0.8248&		0.8300&	0.8301&	0.7916\\
            $3C\_3P$&	0.9688&	0.9747&		0.9675&	0.9683&	0.9269 \\ \hline
            $mAP$ &85.1\% &86.6\% &	87.4\% &	88.0\%	&83.6\% \\
        \hline
        \end{tabular} 
   \end{table*}        

\subsection{Testing experiment and evaluation}\label{sec:testing}

The testing experiment makes use of re-labelled augmented dataset. 

\subsubsection{Re-labelled dataset}\label{sec:3.2.1}

The classification scheme used in the training experiment (see Sec.~\ref{sec:training}), also adopted in \textsc{ClaRAN}  \cite{2019MNRAS.482.1211W} , relies solely on the morphological characteristics. With the exception of the compact sources (\textit{i.e.}, class $1C\_1P$), all other classifications in \textsc{ClaRAN} lack a real corresponding classification with astrophysical relevance. Therefore, such classifications may not be the best solution for radio astronomy. 
Typically, radio sources can be classified into compact sources (CS) and extended sources according to their morphology. 
Extended sources are usually divided into Fanaroff and Riley  class I (FRI) and class II (FRII) \cite{1974MNRAS.167P..31F}. Some previous studies \cite{2017ApJS..230...20A,2019ApJS..240...34M,2021MNRAS.tmp..358B} classified the radio sources into a total of six classes: compact sources, FRI, FRII, bent-tailed galaxies, X-shaped galaxies, and ringlike galaxies. Bent-tailed galaxies are by nature FRI galaxies whose jets are bent by the ram pressure of the intra-galactic medium during the motion of the radio sources in the cluster environment. In addition to the six classes mentioned above, there is another class of sources with a prominent compact component and a weaker elongated feature. They are mostly associated with radio-loud quasars, whose core and jets are subject to the Doppler boosting effect and are observed only a one-sided jet, with the opposite-side jet being too weak to be detected. This class of sources is not a small group and their radio emission characteristics are distinctively different from those of FRI/II galaxies, but they have not been considered separately in previous radio source classifications. In this work, we define them as core-jet (CJ) sources. The irregular X-shaped and ringlike galaxies are not well represented in  the total radio galaxy population, and their physical nature is still under debate. Therefore, we exclude X-shaped and ringlike classifications.
In summary, we consider four morphological classes in our study: CS, FRI, FRII and CJ.

\begin{figure*}
\centering
\includegraphics[width=0.9\textwidth]{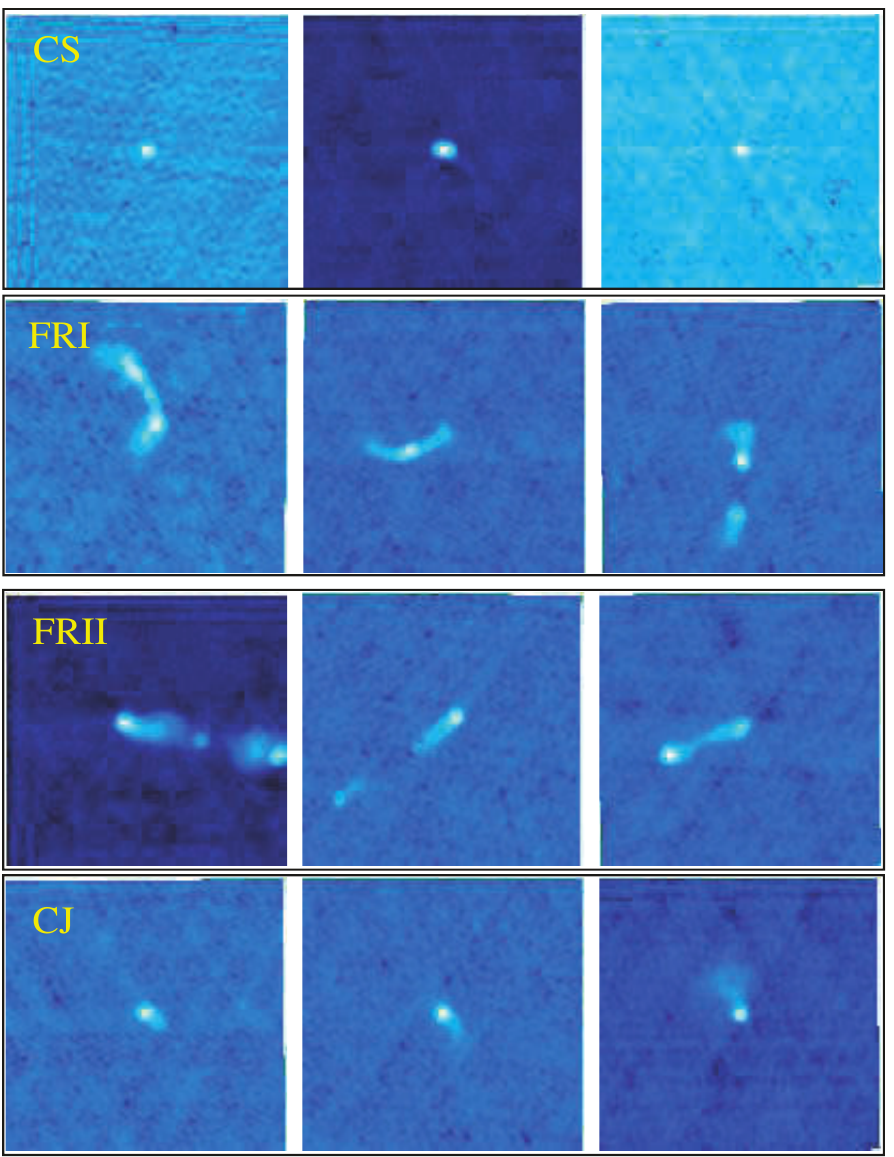}
\caption{Example images for each class in re-labelled datasets, showing 3 images per class. More details are discussed in Sec.~\ref{sec:3.2.1}.
CS is characterized with a single component or a point source and corresponds to the $1C\_1P$ class in the \textsc{ClaRAN}. An FRI is a three-component or multiple-component extended source with a central bright core and two (or more) peaks located on the opposite sides of the core; FRI sources are re-labelled from parts of the $1C\_3P$, $2C\_2P$, $2C\_3P$ and $3C\_3P$ sources in \textsc{ClaRAN}. An FRII is characterized by two prominent terminal components and an extended feature between them. Some FRIIs have a visible central core, but not all. FRIIs are re-labelled from parts of $1C\_2P$, $1C\_3P$, $2C\_2P$, $2C\_3P$ and $3C\_3P$ sources with two peaks or three peaks. A core-jet (CJ) has a prominent compact component (core) at one end of an elongated weaker feature. CJs may contain multiple peaks and are re-labelled from some $1C\_1P$ and $1C\_2P$ sources. 
}
\label{fig:example_relabelled}
\end{figure*}

We re-labelled all images of the D1 dataset of \textsc{ClaRAN} \cite{2019MNRAS.482.1211W} by visual recognition according to our new classification scheme. The re-labelled dataset contains a total of 4,381 PNG-format images. Compared to the original dataset, we have discarded some images with lower quality or with complex source structures, while only kept the high-quality images with high-confidence classifications.  Fig.~\ref{fig:example_relabelled} shows example images of the four classes. Although we used a subset of the D1 dataset as Wu et al. \cite{2019MNRAS.482.1211W} did, there is no direct one-to-one correspondence between our classified sources and theirs because we used different classification schemes and we discarded some images. Similarly, in contrast to Ma et al. \cite{2019ApJS..240...34M}, both we and they have re-labelled the FIRST images by visual inspection using different schemes (note that the D1 dataset stems from the FIRST image data), so it is not possible to directly compare the two training sets. The details of the re-labelled dataset  are presented in Table~\ref{tab:relabelled}. 

\subsubsection{Data Augmentation}\label{sec:3.2.2}

Deep neural networks generally require a large amount of training data to obtain the optimal model. When the training set is relatively small, overfitting often occurs. The overfitting problem can be solved by enlarging the training sample. However, in many cases, the actual number of observed images is fixed and limited, so it is difficult to obtain the required large amount of data. To solve this problem, data augmentation is invoked to increase the diversity and the sample size of training samples, improve the robustness of the model and avoid overfitting. 

Data augmentation is a technique that artificially expands the training datasets by allowing a limited amount of data to produce more comparable data. It is an effective approach to overcome the lack of training data and is currently widely used in various fields of deep learning \cite{2019BasR...31.1017S}. Deep neural networks generally require a large amount of training data to obtain a more desirable model. However, it is usually not possible to get large amounts of data in many scenarios. To solve this problem, data augmentation can be used to increase the diversity of training samples, improve the robustness of the model and avoid over fitting. By changing the training samples randomly, the model's dependence on certain attributes can be reduced, thus improving the generalization energy of the model. There are two modes of data augmentation: offline augmentation and online augmentation. Offline augmentation is the direct processing of image data and allows linear augmentation of specified categories to balance the data volume in each category.
Online augmentation is the augmentation of the batch data in the program prior to training and is very suitable for automated training. However, some datasets cannot be linearly enhanced and this method is always used for large datasets.

\begin{figure*}
\centering
\includegraphics[width=0.8\textwidth]{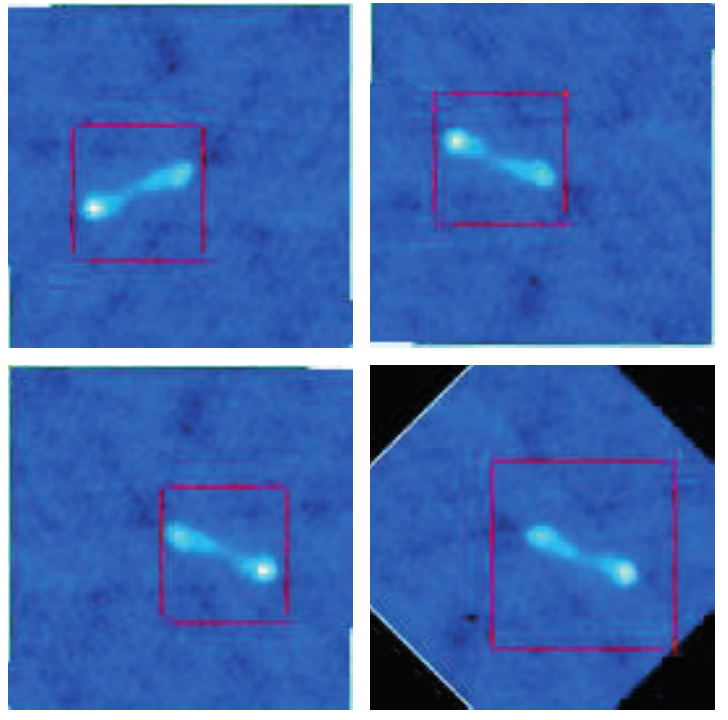}
\caption{An example of image augmentation. The top-left panel is the original image, with the size and localization of the source shown as abounding rectangular box. The top-right panel is flipped up and down on the original image. The bottom-left panel is flipped left and right on the original image. The bottom-right panel is a $45^\circ$ rotation on the original image. The size and localization of the bounding box also change with the image flipping and rotation.} %
\label{fig:example-aug}
\end{figure*}

We use offline methods to enhance the training images of re-labelled datasets. Image data augmentation typically involves flipping, rotation, scaling, cropping, translation and adding noise. Given the specificity and authenticity of radio images, only image flipping and rotation are adopted for our datasets. Becker et al. \cite{2021MNRAS.tmp..358B} considered rotational invariance and used an augmentation method of rotating the images every 15$^\circ$. Instead, we rotated the images by random angles. By randomly changing the training samples, the dependence of the model dependence on certain attributes can be reduced, thus improving the generalization of the model. An example of image augmentation is shown in Fig.~\ref{fig:example-aug}.  
The original datasets contain the PNG images and the annotation files, which record the localization and class label corresponding to each image. While the images are augmented, the localization and size of the bounding box ($4$ points coordinate values of the bounding box) are transformed accordingly to generate the new annotation file. In Fig.~\ref{fig:example-aug}, the top-left panel shows the original image with a red bounding box from its annotations file. The other three show the enhanced images with its new bounding box.

In the re-labelled datasets, there were serious imbalances in the number of four classes, where the number of FRI and core-jet sources was about $11-13$ times less than the number of CS sources. After many training and testing, we found that the following augmentation scheme has the best performance. FRI source images underwent about $15$ times data augmentation overall images, each time the images were randomly performed in one of these three ways: flipping top-down, flipping left-right and rotating randomly in the range from $-180^\circ$ to $+180^\circ$. Core-jet source images were performed about $20$ times, with each data augmentation performed in the same way as FRI. FRII source images were performed flipping top-down and flipping left-right overall images. CS sources images performed flipping top-down overall images. More details of augmented datasets as shown in the third column of Table~\ref{tab:relabelled}.

\begin{table}
\centering
      \caption[]{Number of sources for each class of re-labelled datasets for training and testing}\label{tab:relabelled}
       \begin{tabular}{llll}
        \hline\hline
             Class name & Training & Training (augmented) & testing \\
        \hline
            CS  & 1,720 & 3,949 & 1,127\\
            FRI  & 157 & 2,512 & 106\\
            FRII  & 825 & 2,475 &  553\\
            Core-jet  & 132 & 2,628& 87\\ \hline
            Total & 2,834 & 11,564& 1,873\\
        \hline
        \end{tabular}
\end{table}

\subsubsection{Results and Evaluation} \label{sec:3.2.3}

The experiments are based on the proposed workflow described in Sec.~\ref{sec:2.4}. We used the warm-up and the data parallelism strategies described in Sec.~\ref{sec:training} to optimize and accelerate the training. Using the same initial learning rate $LR_{\rm init}$ and parameters as the D1 dataset in Sec.~\ref{sec:2.4}, the learning rate curve for the re-labelled datasets is identical to Fig.~\ref{fig:LR-relab}. 
Since the amount of the re-labelled augmented dataset is much larger than that without augmentation, parallel operation of the training of these datasets can better utilize the capability of multiple GPUs and significantly reduce the training time. 
Using \textsc{HeTu}-50 and \textsc{HeTu}-101 networks, the average training speeds are 0.45 seconds and 0.49 seconds per step for the re-labelled dataset, 0.41 seconds and 0.45 seconds for the augmented dataset. 
Executing the workflow of more than 40,000 steps, the re-labelled and augmentation dataset required 4.6 and 4.9 hours of training time on 8 GPU devices, which are 0.1 hours and 0.5 hours shorter than those on the D1, D3 and D4 datasets.

In Fig.~\ref{fig:total_loss_D1_relabelled}, green-color and purple-color lines show the total cost curves for \textsc{HeTu}-101 on the re-labelled datasets, and pink and brown line show the corresponding weight decay curves. The weight decay curves almost overlap with the curves for the D1 dataset. The overall trend of the total training loss on the re-labelled datasets is similar to that on the D1 dataset, but decreases from 0.610 to 0.355 for re-labelled un-augmented dataset, and from 0.570 to 0.346 for the augmented dataset. The total loss curves on the augmented dataset is relatively stable and smooth, and the curves on the un-augmented datasets have large fluctuations at 10,000 and 17,500 iterations. This indicates that the augmentation method makes the data samples more diverse and thus the training model more stable.

\begin{figure}
\centering
\includegraphics[width=0.45\textwidth]{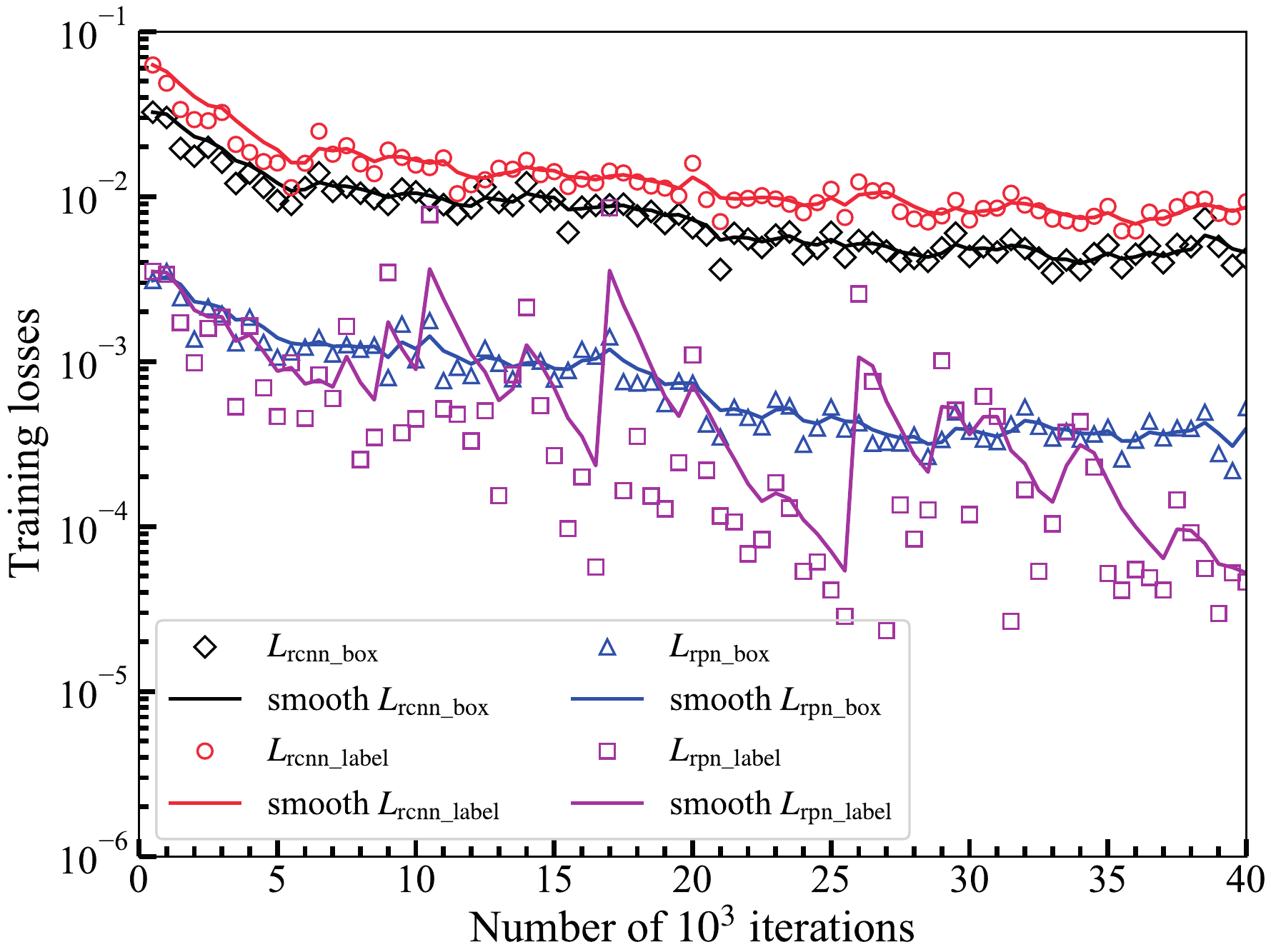}
\caption{RPN and Faster R-CNN loss ($Y$-axis) curves of using \textsc{HeTu}-101 network for training re-labelled augmented datasets. $x$-axis representes the number of iterations. The red line and circle markers are the smoothed and measured Faster R-CNN label classification loss ($L_{\rm rcnn\_label}$), the black line and diamond markers are the smoothed and measured Faster R-CNN bounding box loss ($L_{\rm rcnn\_box}$), the blue line and upper triangle markers are the smoothed and measured RPN label classification loss ($L_{\rm rpn\_label}$), the magenta line and squares marks are the smoothed and measured RPN bounding box loss ($L_{\rm rpn\_box}$).}
\label{fig:rpn_rcnn_loss_relabelld}
\end{figure}

In Fig.~\ref{fig:rpn_rcnn_loss_relabelld}, four additional loss curves of the \textsc{HeTu}-101 model on the re-labelled augmented datasets are shown for $L_{\rm rpn\_label}$, $L_{\rm rcnn\_label}$, $L_{\rm rpn\_box}$ and $L_{\rm rcnn\_box}$. These four losses are very small compared to the weight decay cost shown in Fig.~\ref{fig:total_loss_D1_relabelled}. $L_{\rm rcnn\_box}$ loss decreases from 0.033 to 0.0034, and $L_{\rm rcnn\_label}$ decreases from 0.063 to 0.0062, indicating that the loss is reduced by a factor of $\sim$10 after Faster R-CNN training. $L_{\rm rpn\_box}$ loss decreases from 0.0035 to 0.00022, and $L_{\rm rpn\_label}$ from 0.0085 to 0.000023, indicating a 16-fold reduction in  $L_{\rm rcnn\_box}$ and a 370-fold reduction in $L_{\rm rpn\_label}$ when the training converges. 

Using the trained \textsc{HeTu}-50 and \textsc{HeTu}-101 models,  the testing speeds per image are 4.6 ms and 5.4 ms, respectively. Both $AP$ and $mAP$ results are presented in Table~\ref{tab:AP_D1new1}. 
For the re-labelled augmented dataset,  $mAP$ increased to 94.2\% using the trained \textsc{HeTu}-101 model and to 92.6\% using the trained ResNet-101 model only, compared to 87.2\% using the \textsc{ClaRAN}. In particular,  $AP$s of all classes have been greatly improved using the trained ResNet-FPN model compare to only using the trained ResNet model. Especially the $AP$ of compact radio source has been greatly improved with the maximum of 0.995, which is close to 1.  For the re-labelled datasets, $mAP$ trained on un-augmented datasets  using \textsc{HeTu}-50 and \textsc{HeTu}-101 model are 88.6\% and 89.9\%, respectively. The $AP$s of CS and FRII were close to 1. The maximum increase in $AP$ for FRI is 0.066, and the maximum increase in $AP$ for CJ is 0.131. The $mAP$ of the trained model on the augmented datasets increases by 5.4 percentage and 4.2 percentage in the same backbone network due to the large increase in $AP$ for both FRI and CJ. This indicates that the data augmentation method can effectively solve the data imbalance problem. 

   \begin{table*}
   \centering
      \caption[]{$AP$ and $mAP$ results of the re-labelled dataset}
      \label{tab:AP_D1new1}
        \begin{tabular}{lllll}
        \hline\hline
         Class name & \textsc{HeTu}-50	& \textsc{HeTu}-101 & \textsc{HeTu}-50 	& \textsc{HeTu}-101 \\
         &&&(augmented) & (augmented) \\
         \hline
            CS	&0.9849 & 0.9860 & 0.9950&	0.9940 \\
            FRI	& 0.8167 & 0.8507 &0.8824&	0.8962 \\
            FRII &0.9810 & 0.9643& 0.9889&	0.9806 \\
            Core-jet& 0.7625 & 0.7951&0.8934&	0.8961 \\ \hline
            $mAP$ & 88.6\% & 89.9\% & 94.0\% &94.2\% \\
            \hline
        \end{tabular} 
   \end{table*}

\begin{figure}
\centering
\includegraphics[width=0.45\textwidth]{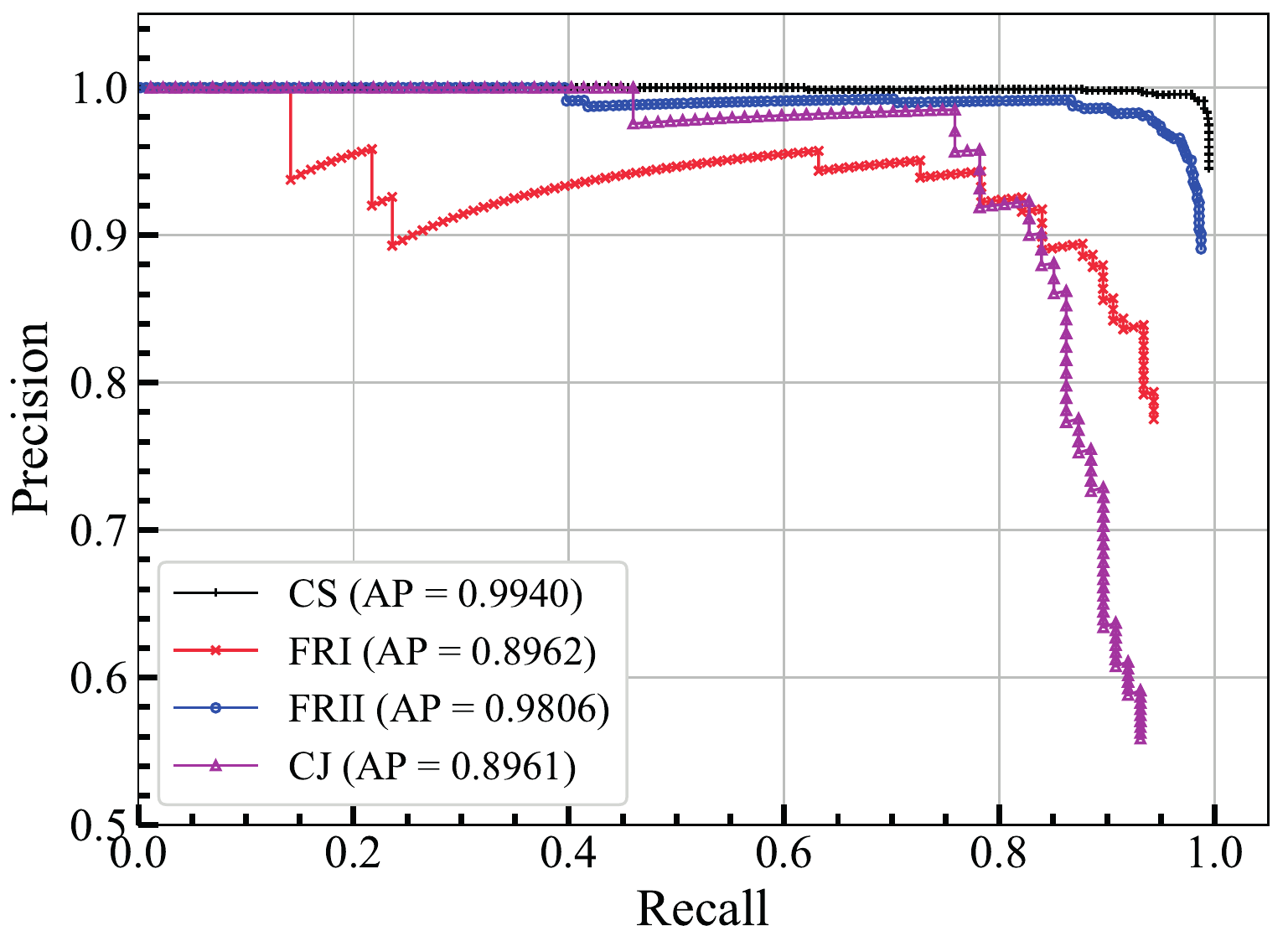}
\caption{(Color online) $P_{rec}-R$ of \textsc{HeTu}-101 model for different classes tested on the re-labelled augmented dataset. The black line with plus markers is the $P_{rec}-R$ curve of CS, the red line with $\times$ markers is the $P_{rec}-R$ curve of FRI, the blue line with open circle markers is the FRII $P_{rec}-R$ curve, and the magenta line with open upper triangle markers is the CJ $P_{rec}-R$ curve.}
\label{fig:P-R}
\end{figure}

\begin{figure}
\centering
\includegraphics[scale=0.3]{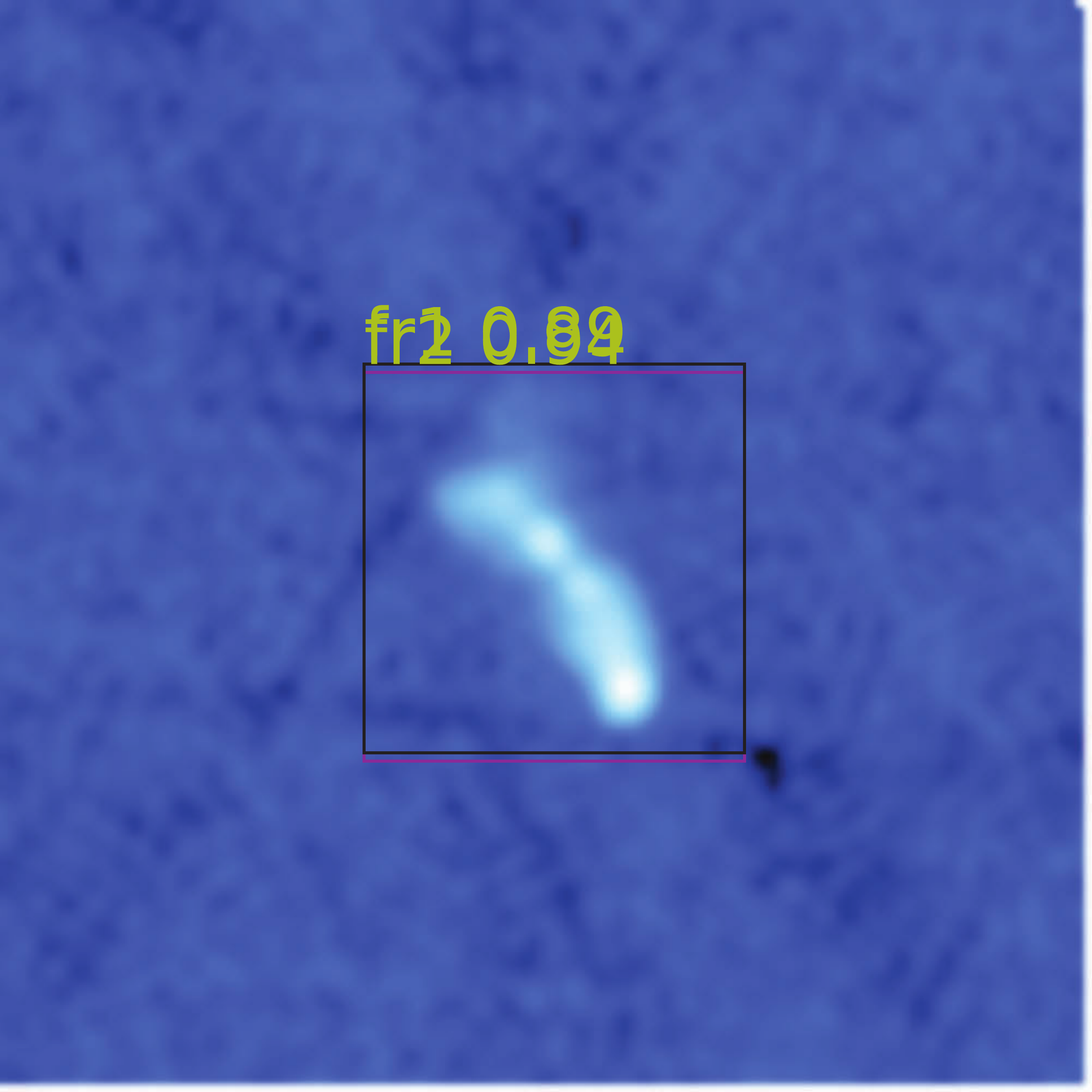}
\includegraphics[scale=0.3]{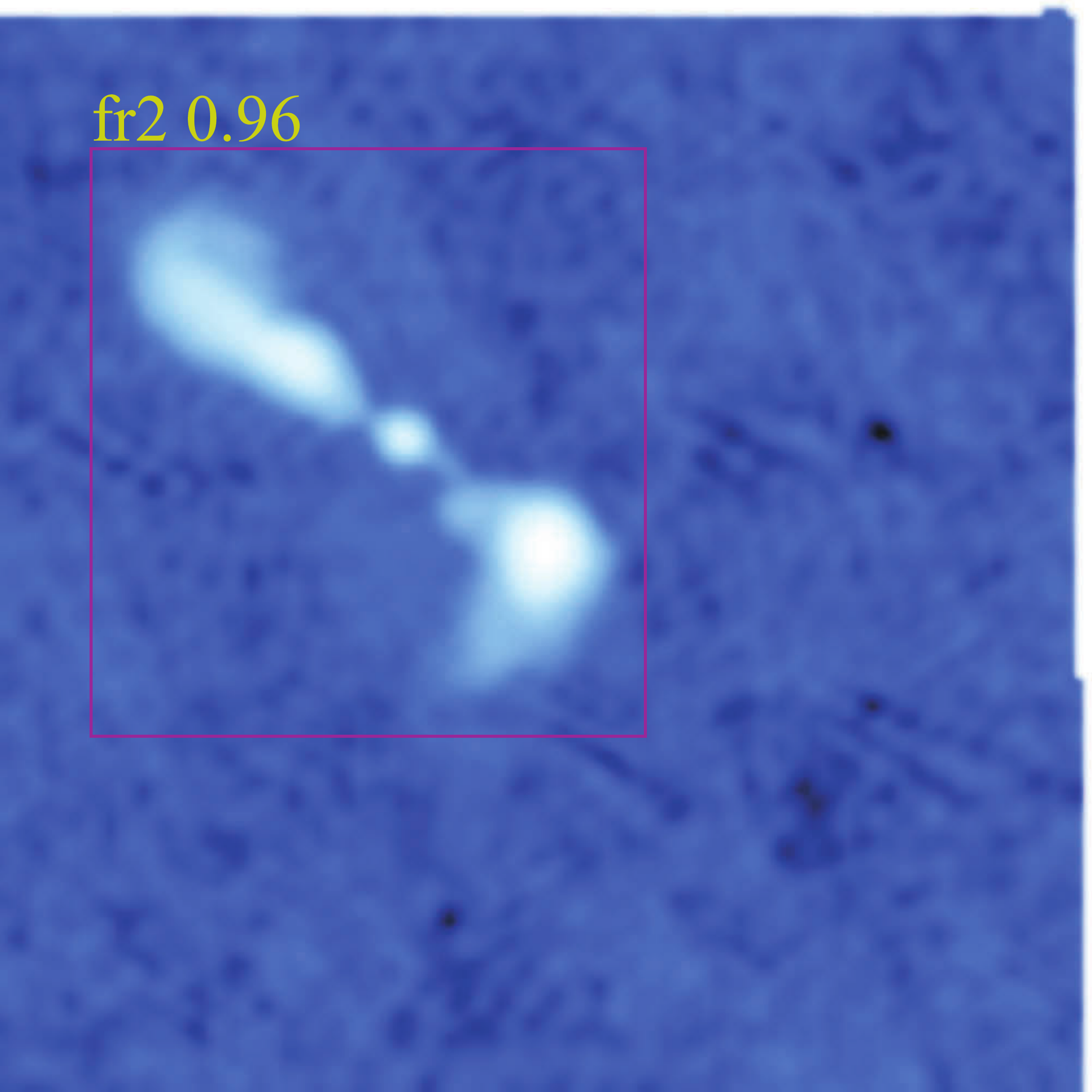}
\caption{An example of mis-identification between FRI and FRII. This may cause the fluctuations in the precision-recall curve in Fig.~\ref{fig:P-R}. }
\label{fig:fr1_fr2}
\end{figure}

The $P_{rec}-R$ curves of the \textsc{HeTu}-101 model tested on re-labelled and augmented datasets are plotted in Fig.~\ref{fig:P-R}. In general, the mAPs near the top-right corner of the $P_{\rm rec}-R$ curve have better precision than those far away. The $P_{\rm rec}-R$ curve for CS remained  a horizontal line at the $P_{\rm rec}$ level of $\sim$1.0 until $R$ reaches a level close to 1.0. This suggests that almost all detected sources are truly positive, and these sources account for near 100\% of the total true CS sources in the test datasets. This is consistent with the  high $mAP$ results shown in Table~\ref{tab:AP_D1new1}. The FRII $P_{\rm rec}-R$ curve has two horizontal lines at $P_{\rm rec}$ levels of 1.0 and 0.98, with a step at $R=0.4$. It shows that up to 98\% of the detected sources are truly positive. 
The CJ $P_{\rm rec}-R$ curve also has two horizontal lines similar to FRII, but its second line is at a lower level, and starts to decrease after $R=0.75$. The entire curve of CJ is almost above the $P_{\rm rec}$ level of 0.8. The FRI $P_{\rm rec}-R$ curve starts with a shorter horizontal line from $R=0.0$ to $R=0.12$, drops at $R=0.12$ and fluctuates between $R=0.12$ and $0.22$. The fluctuates are due to some sources in other classes being mistakenly identified as FRIs. The morphology of FRI sources is rather complex and some FRIs are difficult to distinguish from other categories (also discussed in Sec.~\ref{sec:predicting}). An example is shown in the left panel of Fig.~\ref{fig:fr1_fr2}, where an FRII source is mistakenly identified as an FRI, but it is not missed in the FRII class. This is why the P-R curve of FRI fluctuates between $R=0.12$ and $R=0.22$, while the P-R curve of FRII remains a flat horizontal line at $P=1.0$.  The right panel of Fig.~\ref{fig:fr1_fr2}  illustrates another situation: an FRI source is misidentified as an FRII. 
Seven missed FRI sources are re-identified in the training process, corresponding to the recall from $R=0.22$ to $R=0.62$, where the P-R curve shows an upward arch.

\begin{figure*}
\centering
\includegraphics[width=0.9\textwidth]{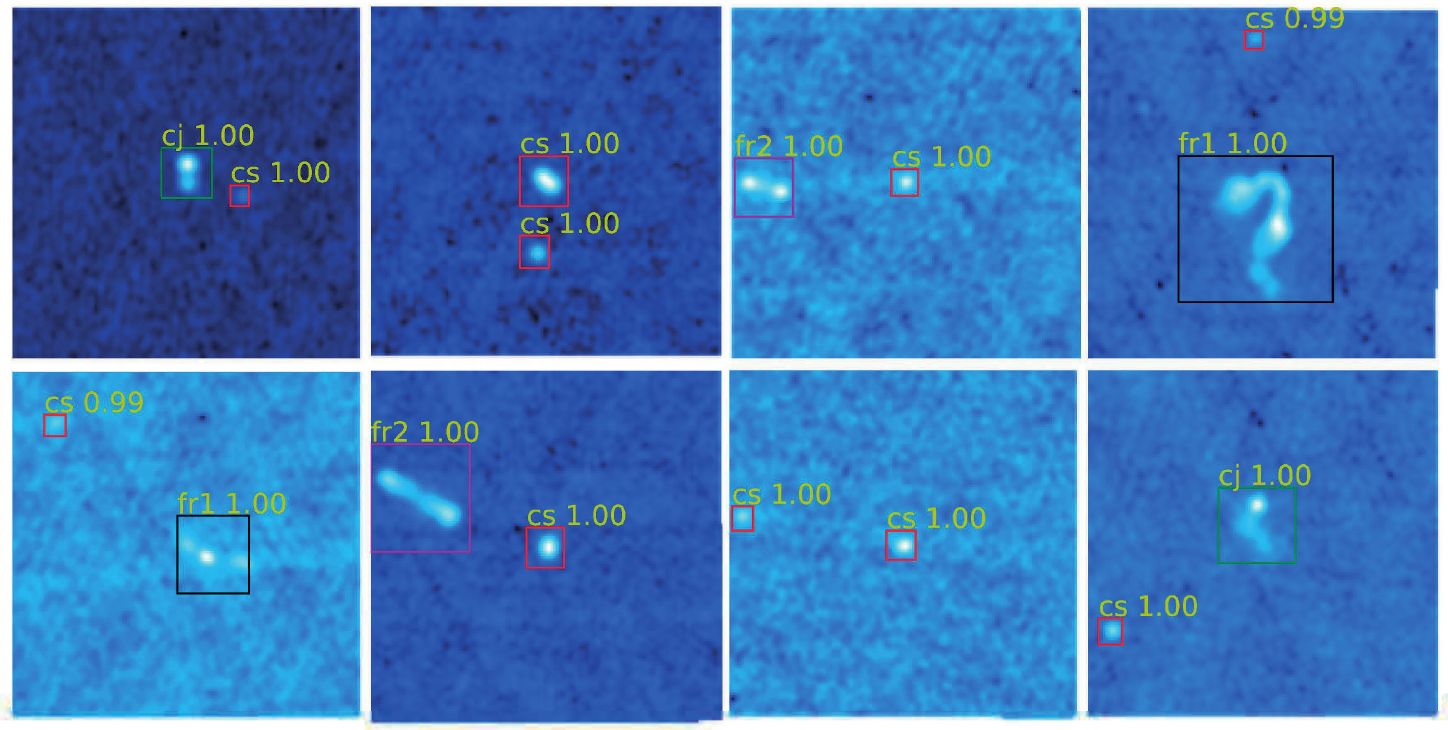}
    \caption{Examples of images detected by \textsc{HeTu}-101 model on re-labelled test datasets. The different colored rectangular boxes represent the identified sources: red for CS (cs), black for FRI (fr1), magenta for FRII (fr2) and green for core-jet (cj). Each source is labelled with a class name and a score between 0 and 1 on the top left of the rectangular box. 
    }
    \label{fig:result_D1new}
\end{figure*}

Examples of the images detected by the \textsc{HeTu}-101 model for re-labelled augmented datasets are depicted in Table~\ref{tab:relabelled} and shown in Fig.~\ref{fig:result_D1new}. The localization and size of the detected sources are labelled with four colored rectangular boxes: red for CS, black for FRI, magenta for FRII and green for CJ. Each source is labelled with the associated class or label name with a score between 0 and 1 on the top  of a rectangular box. In Fig.~\ref{fig:result_D1new}, all detected radio sources are identified with a higher score. 

In summary, \textsc{HeTu} using ResNet-FPN as the backbone network achieves higher recognition precision than \textsc{ClaRAN} and ResNet alone. The network depth affects the recognition precision to some extent: before augmentation, the precision of network depth is relatively low for classes (\textit{e.g.}, CJ and FRI) with less data, and the data augmentation significantly improves their recognition precision. After augmentation, increasing the network depth does not greatly improve the recognition performance. However, the training time does not increase significantly with increasing network depth, so the \textsc{HeTu}-101 model will be used as the final model of \textsc{HeTu}.
   
\subsection{Predicting}\label{sec:predicting}

\begin{figure*}
\centering
\includegraphics[width=0.90\textwidth]{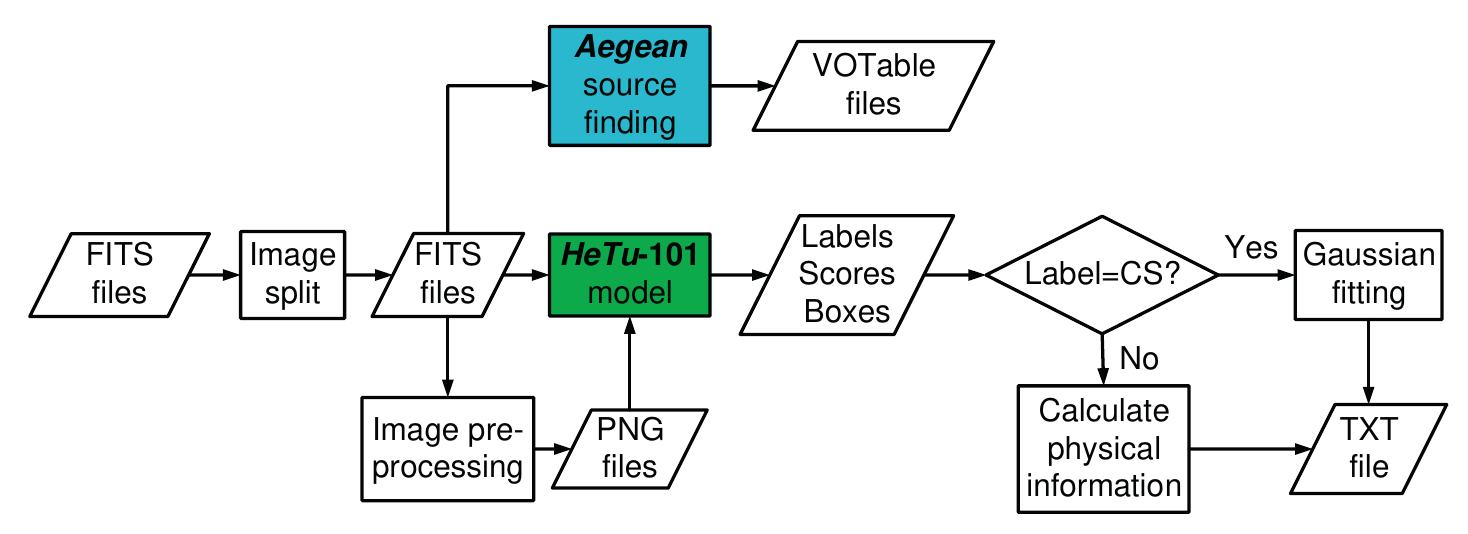}
\caption{The flowchart of \textsc{HeTu} predicting and \textsc{Aegean} source finding. First, the input images (FITS files) are split into many small-sized images, and each image is saved in FITS format. Second, source finding is performed for each FITS file using \textsc{Aegean} and a catalog is generated in VOTable file. Meanwhile, the small FITS files are pre-processed to create PNG-format files which  are used for inputs to \textsc{HeTu}-101 model for batch predicting. The labels, scores and boxes of detected sources are exported. If the detected source is a CS source, Gaussian fitting is performed on the source according to the box information, and the fitted parameters are output. Otherwise, the main physical information of the detected source is calculated and output to a TXT file.}
\label{fig:pred_hetu}
\end{figure*}

In this section, we apply \textsc{HeTu} to the MWA GLEAM images for radio source finding and classification, and compare the results with those obtained by the conventional source finding software \textsc{Aegean}. Fig.~\ref{fig:pred_hetu} shows the flowchart of the predicting experiment. First, the GLEAM images (in FITS format) are segmented into small images (see Sec.~\ref{sec:3.3.1} for the detailed image segmentation process). Then, the source finding of the segmented FITS files is performed by using \textsc{Aegean}, and the corresponding catalog of each image is generated in the VOTable file. At the same time, all the small-sized FITS files are pre-processed and saved in PNG-format images. To make the output information more practical and to create the same catalog results comparable with \textsc{Aegean}, the predict function in \textsc{HeTu} calculates the main physical information of the detected extended sources and also integrates a Gaussian fitting function to the detected CS sources by using the task \textsc{imfit}\footnote{\url{https://www.atnf.csiro.au/computing/software/miriad/doc/imfit.html}} in \textsc{miriad} software package. Finally, all resulting information is exported into a TXT file. 

\subsubsection{Dataset used for predicting experiment}\label{sec:3.3.1}

For the predicting data set, we selected the imaging data in the frequency range $200.32$ MHz--$231.04$ MHz obtained from the GLEAM observations on 2013 August 10. The sky center of this observation is at RA = 23h, DEC = $-17^\circ$. The frequency band is the highest of the GLEAM project, so the image resolution is the highest. The size of each snapshot image is 4000 pixel $\times$ 4000 pixel, and 1 pixel corresponds to 23.43$^{\prime\prime}$.
After data processing includes radio frequency interference mitigation, calibration, and imaging, we obtained $46$ total intensity (Stokes \textit{I}) snapshot images in FITS file format. For detailed data processing and results, please refer to Refs. \cite{2020LAO50e9501T,2017MNRAS.464.1146H}. 

Due to the higher noise level at the edges of each snapshot image,  the image quality in these areas is lower. The new snapshot images have been considered in this predicting experiment, which are generated by cropping 2000 pixel $\times$2000 pixel around the center of the original snapshot image. To ensure that \textsc{HeTu} detects the radio sources correctly, we further divide each new snapshot image into multiple smaller images of size 132 pixel$\times$132 pixel. To identify radio sources that are not to be missed due to image edge corruption, $20$ pixels are retained at the edge of each partition. Images with a width less than $132$ pixels are discarded, and finally a total of $13,294$ images are used for the predicting experiment in both FITS and PNG formats.  The field of view of each image is approximately $0.86^\circ\times0.86^\circ$. In order to make the morphology of all the radio sources in each image visually distinguishable, all images are displayed with `cool' colormaps. 60\% of the images have color bars ranging from $-10\sigma_m$ to $+10\sigma_m$, where the $\sigma_m$ is the local standard deviation in the image, and the others 40\% images have color bars ranging from $-5\sigma_m$ to $+5\sigma_m$.

\subsubsection{Results and evaluation}\label{sec:3.3.2}

\textsc{Aegean} is a source finder widely used for data processing in the Murchison Widefield Array (MWA) project and in particular the GLEAM project \cite{2018PASA...35...11H}. \textsc{Aegean} is performed on FITS files, while \textsc{HeTu} is performed on PNG images. This is the reason why we save each image file in both FITS and PNG formats. \textsc{Aegean} has two key parameters: seed threshold (${\sigma}_s$) and flood threshold (${\sigma}_f$). Pixel values above ${\sigma}_s$ are used to define an island, while pixel values above ${\sigma}_f$ are used to compose an island, then all pixels of this island will be considered as coming from the same source. We set ${{\sigma}_s}=6\sigma$ and use ${\sigma}_f=4\sigma$ (the default values in the GLEAM data processing) to find sources, where $\sigma$ is the local Root Mean Square (RMS) noise level of the image. The average RMS noise is $14.7$ mJy beam$^{-1}$ for all images. We have tried lower thresholds that resulted in more sources being detected, but also introduced more noise-induced fake sources, which impacts negatively on building the catalog. 
\textsc{Aegean} outputs a VOTable file after completing the source finding. The VOTable file is a table recording the properties of the identified sources, which include peak position (Right Ascension (RA) and Declination (DEC) in degrees), peak flux density in Jy beam$^{-1}$, local RMS in Jy beam$^{-1}$, and others parameters\footnote{https://github.com/PaulHancock/Aegean/wiki/Output-Formats}. We run the source finding task in \textsc{Aegean} on the GPU node of the CSRC-P described in  Sec.~\ref{sec:3} using its 20 CPU cores. 

The information of the detected sources is recorded into a TXT-format file (examples are shown in \ref{sec:3.3.3}), which for extended sources includes image name, label name, score, $4$ points coordinate values of the bounding box, peak flux density in Jy beam$^{-1}$, peak position in pixels and (RA, DEC) coordinates in degrees. For CS sources, in addition to the above parameter values, \textsc{HeTu} also performs an additional Gaussian fitting and exports the fitted parameter values. More details on the output format of \textsc{HeTu} are described in \textcolor{red}{\bf Sec.~\ref{sec:3.3.3}}. To ensure the reliability of the results, only information of sources with higher scores than 0.7 is recorded. 

\begin{figure}
\centering
\includegraphics[width=0.45\textwidth]{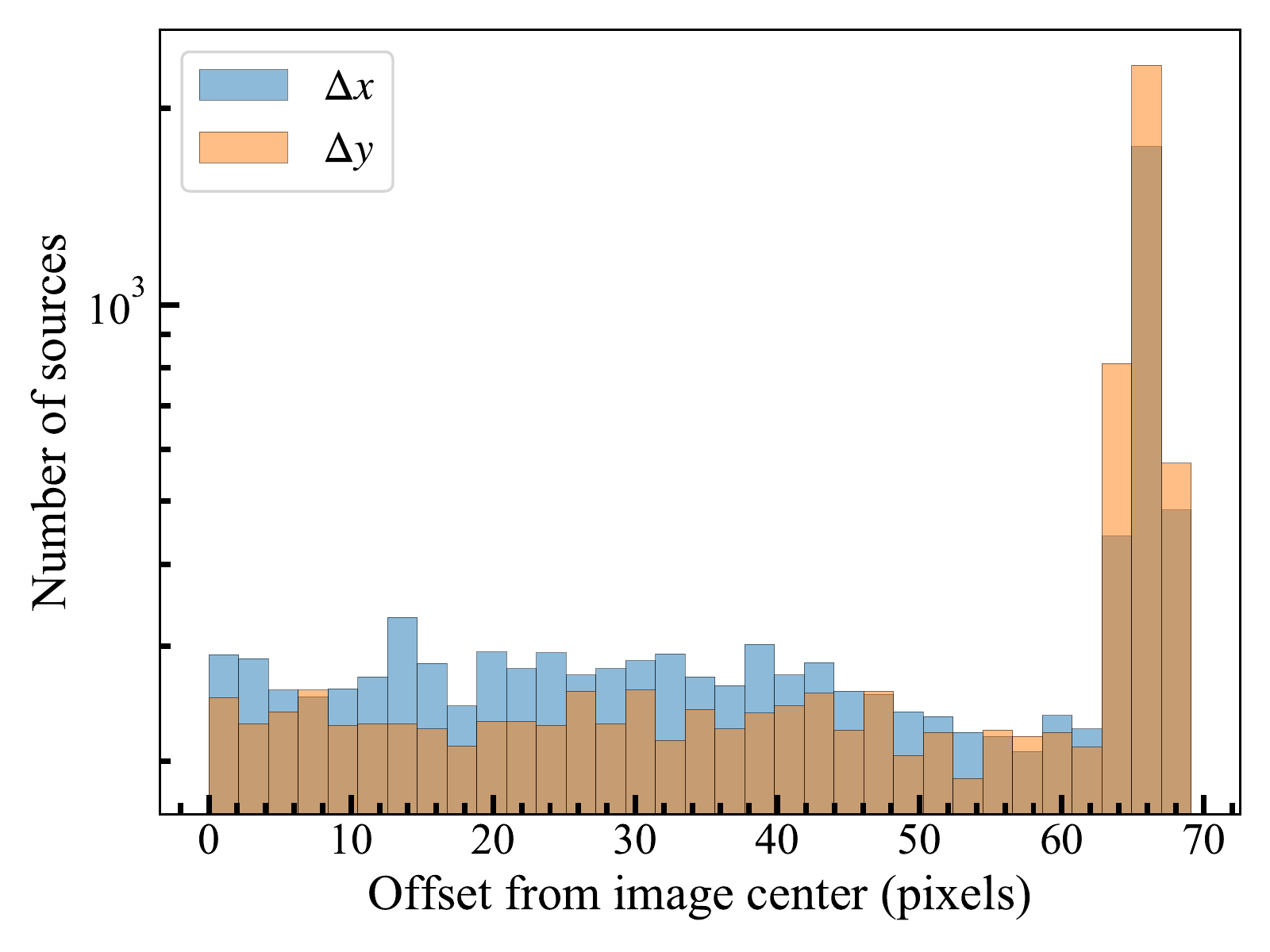}
\caption{Histogram of the source distance away from the image center. The $y$-axis is the number of sources not detected by \textsc{HeTu} but detected by \textsc{Aegean} above $5\sigma$. The $x$-axis is the offset of the $x$ and $y$ coordinates of the sources ($\Delta x$: royal blue color; and $\Delta y$: orange color) from the image center, \textit{i.e.}, the coordinate (66,66) pixel. The histogram shows an even distribution of the offsets, indicating that the sources are randomly distributed in the images. $5,686$ sources are located within 5 pixels offset from their local image boundaries, and have been discarded by \textsc{HeTu}.}
\label{fig:nomatch_all_aegean}
\end{figure}

\begin{figure*}
\centering
\includegraphics[width=0.9\textwidth]{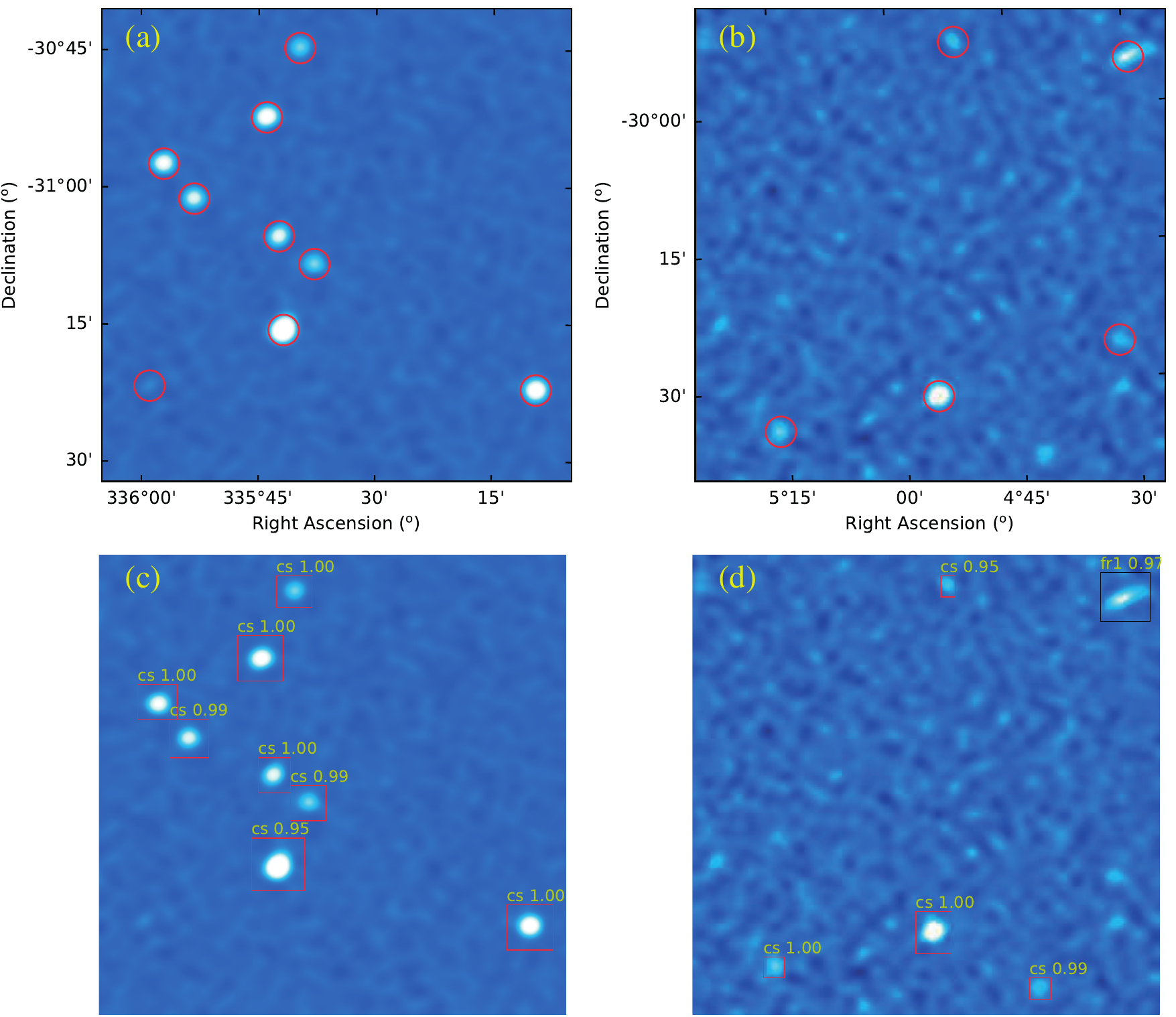}
\caption{Examples of detected results for \textsc{Aegean} with ${{\sigma}_s}=5\sigma$ (\textit{top}) and \textsc{HeTu} (\textit{bottom}). The detected sources are labelled with red cycles by \textsc{Aegean} and with red rectangles by \textsc{HeTu}. \textsc{HeTu} also labels the class name and associated score. 
The RMS noises of the (b) and (d) images are 15.6 mJy beam$^{-1}$. The peak flux density of CS source (score=0.99) in the bottom right corner of the (d) image is 82.2 mJy beam$^{-1}$.}
\label{fig:agean_resnet_fpn_cs}
\end{figure*}

\begin{figure*}
\centering
\includegraphics[width=0.9\textwidth]{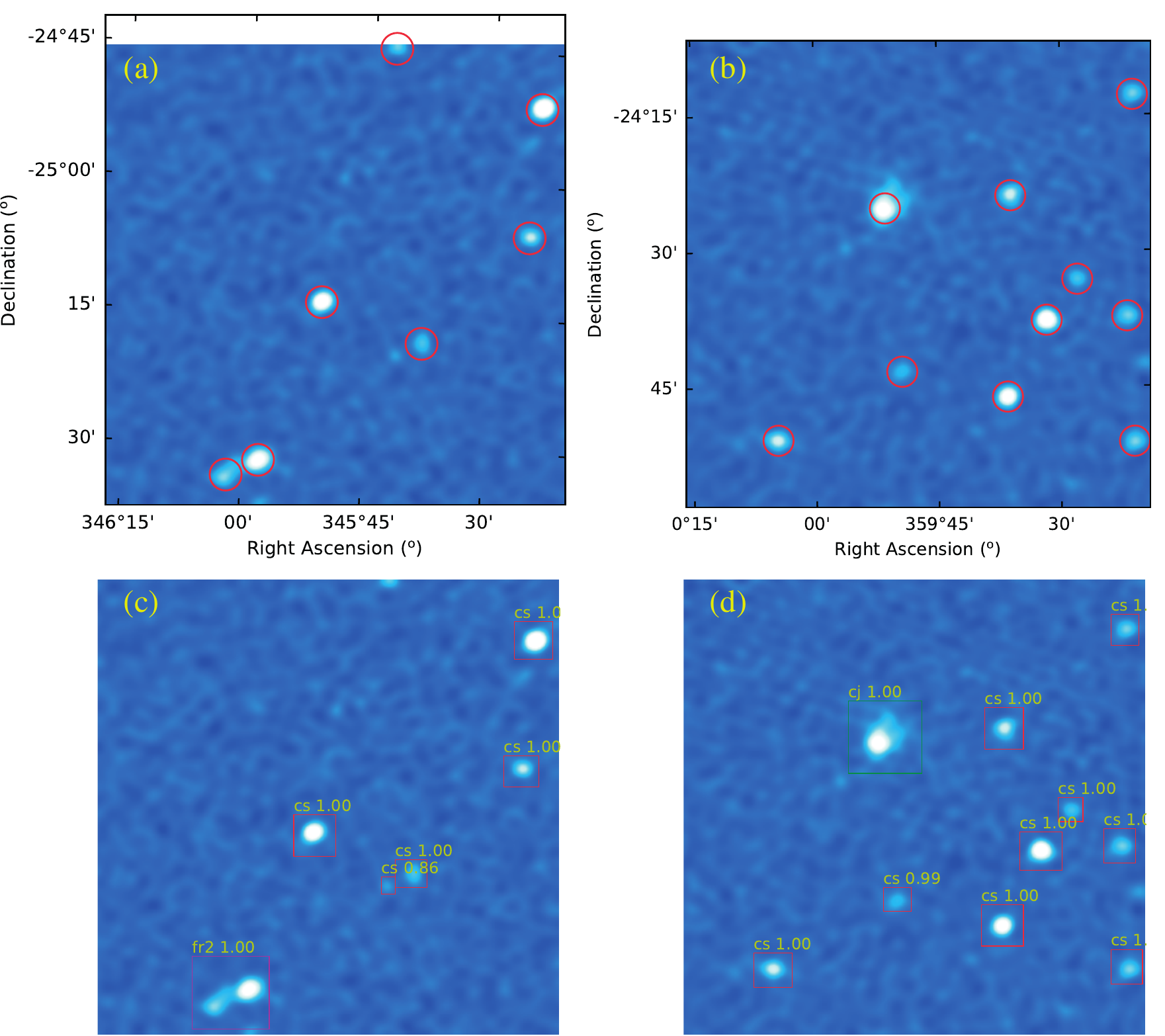}
\caption{Examples of detection results for \textsc{Aegean} with ${{\sigma}_s}=5\sigma$ and \textsc{HeTu}. Panels (a) and (b) show the results from \textsc{Aegean}. The detected sources of \textsc{Aegean} are labelled as localization with red circle. Panels (c) and (d) show the results corresponding to panels (a) and (b) from \textsc{HeTu}. The RMS noise in the (a) and (c) images are 11.1 mJy beam$^{-1}$. The peak flux density of CS source (score=0.86) in the (c) image is 56.4 mJy beam$^{-1}$.}
\label{fig:agean_resnet_fpn_fr2}
\end{figure*}

Table~\ref{tab:resnet+fpn_aegean} shows the comparison of the detection results of \textsc{HeTu} with those of \textsc{Aegean}.  Fig.~\ref{fig:agean_resnet_fpn_cs} shows example images of the detection results for \textsc{Aegean} and \textsc{HeTu}. One point to note is that the size of all the radio sources that can be detected within the GLEAM image is smaller than $1.63^\circ$ (or 250 pixels), while some extended sources may be missed if their source sizes are larger than 250 pixels. Whereas, such large-scale radio sources are very rare and have little impact on the total source detection rate. A total of $77,919$ sources are identified by \textsc{Aegean} above $6\sigma$, including $73,999$ CS sources. $8,523$ CS sources are detected by \textsc{Aegean}, but not identified by \textsc{HeTu}. We find that a large proportion of these sources are located close to the image edges. Fig.~\ref{fig:nomatch_all_aegean} shows the distribution of the source position offset from the image center. We find $4,594$ (or 54\%) sources are located within 5 pixels from the image boundary. \textsc{HeTu} identifies sources from their morphology and considers the sources close to the image edge incomplete and discards them. An example is shown in Fig.~\ref{fig:agean_resnet_fpn_fr2}: panel (a) contains an identified CS source in the upper edge of the image; however, this source is not identified by \textsc{HeTu} in panel (c). By subtracting these image edge sources, the number of the CS sources detected by \textsc{Aegean} is reduced to $69,405$. 

\textsc{HeTu} identifies in total $85,290$ sources using the trained \textsc{HeTu}-101 model: $82,821$ CS sources, 770 FRI sources, 685 FRII sources and 1,014 core-jet sources. The total number of \textsc{HeTu}-detected CS sources, $82,821$, is higher than the number of \textsc{Aegean} CS sources (\textit{i.e.}, $69,405$), because the number of \textsc{Aegean} detections is related to the threshold value while \textsc{HeTu} does not set flux density threshold.  For example, a CS source with a score of 0.99 at the bottom-right corner in panel (d) of Fig.~\ref{fig:agean_resnet_fpn_fr2} is detected by \textsc{HeTu} but not by \textsc{Aegean} in panel (b). The peak flux density of this CS source is 82.2 mJy beam$^{-1}$, while the RMS of this image is 15.6 mJy beam$^{-1}$, corresponding to a peak-to-noise ratio of $5.3\sigma$. Another example is seen in panels (a) and (c) of Fig.~\ref{fig:agean_resnet_fpn_fr2}, in which a CS source with a score of 0.86 is detected by \textsc{HeTu}, but not by \textsc{Aegean}. The ratio of the peak flux density to the noise level of this source is 5.1.
If we consider the same peak flux density threshold of $\geq6\sigma$, then we find $69,263$ CS sources from the \textsc{HeTu} catalog (see Table~\ref{tab:resnet+fpn_aegean}). We use the \textsc{TopCat} software~\cite{2005ASPC..347...29T} to cross match the sources detected by \textsc{HeTu} and \textsc{Aegean} with a searching radius of 30 arcsec and obtain 65,476 CS sources detected by both, accounting for 94.5\% of the total populations for \textsc{HeTu}-detected CS sources and 94.3\% for \textsc{Aegean}-detected CS sources. 

\begin{figure}
\centering
\includegraphics[width=0.45\textwidth]{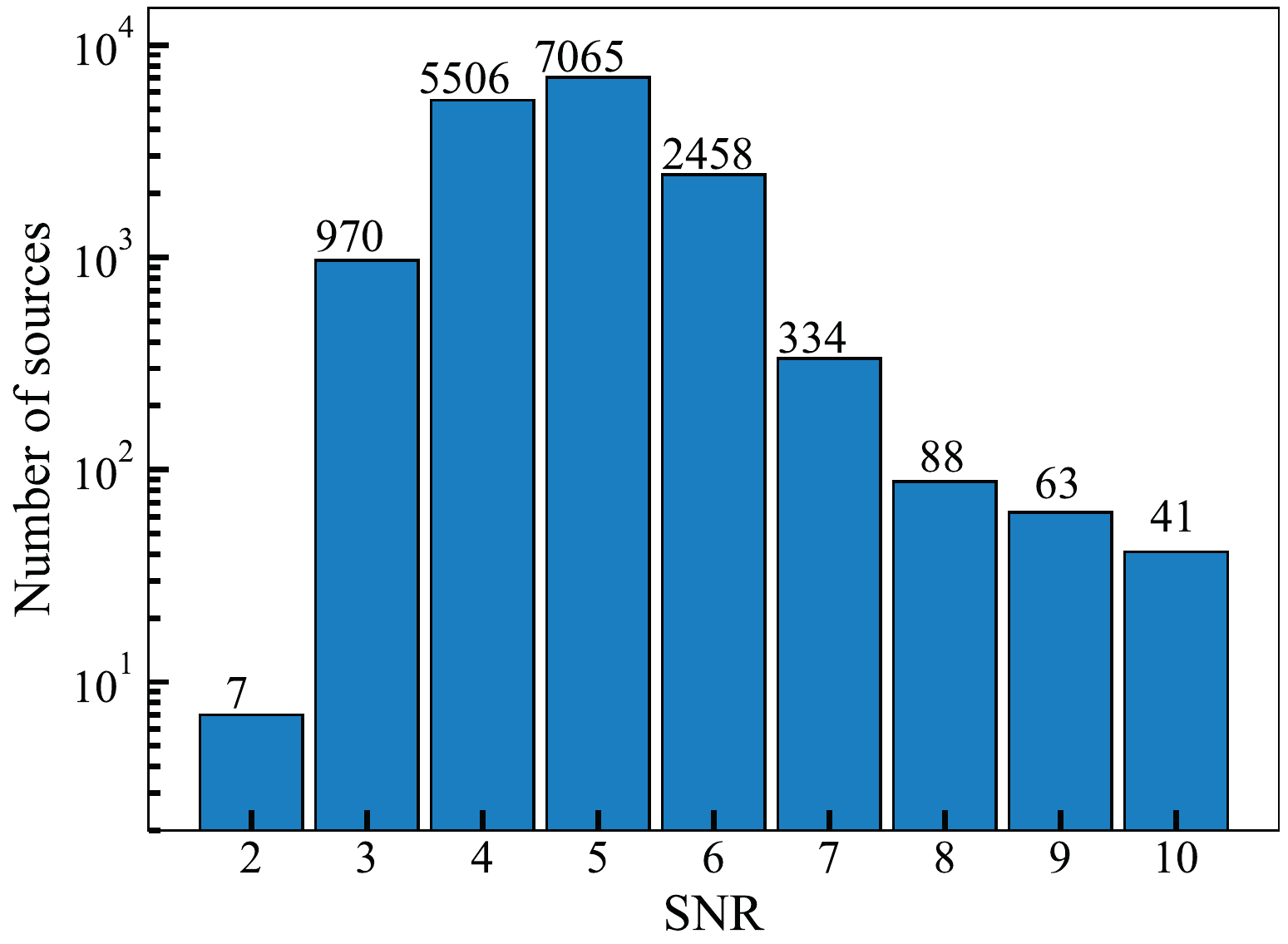}
\caption{Distribution of the number of the sources detected by \textsc{HeTu} but not by \textsc{Aegean} versus the signal to noise (SNR) ratio of the source's peak flux density. 
$x$-axis is the SNR of the source's peak flux density.
$y$-axis is the number of the source's peak flux density.
The numbers at the top of the histogram are the source numbers detected in the corresponding SNR bin.
Ten sources with $\mathrm{SNR}<2$ and 800 sources with $\mathrm{SNR}>10$ are out of the display zone. 
Because these data points are distributed over a wide range of SNRs, they are not shown in the plot for clarity purposes.
}
\label{fig:nomatch_all_hetu}
\end{figure}

If the \textsc{Aegean} detection threshold is set to a lower value, weaker sources can be detected, but at the cost of introducing a number of fake sources. 
For example, if the detection threshold is set to $5\sigma$, \textsc{Aegean} can detect 82,894 CS sources (5,786 image edge sources are not counted). 73,967 sources are cross-matched, accounting for 89.2\% and 96.9\% of the \textsc{Aegean}-detected and \textsc{HeTu}-detected CS sources, respectively. 
8,491 new sources are cross-matched compared to the $6\sigma$ values.  Fig.~\ref{fig:nomatch_all_hetu} shows the signal-to-noise ratio of \textsc{HeTu}-detected sources.
$7,065$ sources with $6>{\rm SNR}\geq5$ are detected by \textsc{HeTu} and they form the bulk of the new cross-matched $8,491$ sources. 
This means that lowering the threshold value does increase the detection number, but at the same time, more fake sources are introduced. There are 8,927 sources detected by \textsc{Aegean} but not by \textsc{HeTu}. We inspect these sources visually and find that they are mostly associated with the sidelobes of very bright sources. As shown in Table \ref{tab:resnet+fpn_aegean}, when the detection threshold is set as $4\sigma$, \textsc{Aegean} can detect $\sim100,000$ CS sources, but also results in a total of 17,646 un-matched CS sources. The un-matched number is doubled compared to the $5\sigma$ case, and 4.5 times the $6\sigma$ case. On the other hand, the ratio of the cross-matched CS sources to the \textsc{HeTu}-detected sources slightly changes from 94.5\% to 96.9\% and 96.2\% for the three thresholds.
Our experiments suggest that the \textsc{Aegean} detection using a threshold of $5\sigma$ gives an optimal compromise between the cross-matching rate and fidelity rate. The comparison between \textsc{HeTu} and \textsc{Aegean} catalogs below will be based on the $5\sigma$ threshold limit.

\textsc{Aegean} employs a component-based source finding algorithm which can identify an extended source with multiple discrete components (\textit{i.e.}, multiple sources). 
Diffuse emission between close components cannot be identified as Gaussian components, so \textsc{Aegean} cannot directly determine whether there is a connection between adjacent components, causing it to lack  the ability to automatically classify extended sources. The classification needs to be manually done in an offline manner at a later stage.
We associate the \textsc{HeTu}-detected extended sources with the brightest component in the corresponding \textsc{Aegean} source. 

All FRII sources ($N=685$) identified by \textsc{HeTu} are also detected by \textsc{Aegean}, corresponding to a cross-matching rate of 100\%.
\textsc{Aegean} identifies two radio lobes of an FRII as two separated sources; if an FRII has a bright central core, \textsc{Aegean} identifies it as an additional CS source. Instead, \textsc{HeTu} identifies an FRII as a single source, regardless of how many sub-components it contains. As a result, the detection component number of the FRII class by \textsc{Aegean} is 2.06 times that by \textsc{HeTu}.

\textsc{Aegean} identifies an FRI as at least one CS source, with the brightest one corresponding to the radio core. If the jet of an FRI contains bright hotspots, \textsc{Aegean} would identify it as one or more CS sources. On the other hand, \textsc{HeTu} identifies an FRI as a single source. The number of FRI sources identified by  \textsc{HeTu} is 770, and the number of FRI components identified by \textsc{Aegean} is about 1.7 times that by \textsc{HeTu}. 
For each FRI, we associate the brightest \textsc{Aegean} FRI component with the corresponding \textsc{HeTu} source.
Taking the $5\sigma$ \textsc{Aegean} results as an example, we get 747 cross-match sources. 
The cross-matched sources account for 97.4\% of the total \textsc{HeTu} FRI population.
The 20 FRI sources not recognized by \textsc{Aegean} are  mainly associated with some weaker emission features. 
That is, when the FRI jet is weak and diffuse, \textsc{Aegean} is unable to identify and fails to fit it as a Gaussian component. As shown in  Table~\ref{tab:resnet+fpn_aegean}, the FRI number increases as the detection threshold  decreases from $6\sigma$ to $4\sigma$, indicating that more weak FRIs are detected and the matching rate with the results of \textsc{HeTu} gradually increases.

\textsc{Aegean} identifies 1293 components associated with CJ sources above $5\sigma$, in which 851 have counterparts in \textsc{HeTu} CJ catalog. The cross-matching rate of the CJ class varies from 95.5\% to 98.9\%, depending on the detection threshold.

   \begin{table*}
   \centering
      \caption[]{Results of Detection in GLEAM Images }\label{tab:resnet+fpn_aegean}
       \begin{tabular}{lllllll}
        \hline\hline
        Threshold & Methods & CS & FRI & FRII & CJ & Runtime\\
        \hline
${{\sigma}_s}=6\sigma$ & \textsc{HeTu}   & 69,263 (94.5\%)	& 751 (94.9\%)	&685 (100\%)	&801 (98.9\%)  & 23.87 mins\\
                       & \textsc{Aegean} & 	69,405  (94.3\%) & $1,298$ components$^*$	& $1,415$ components	& $1,207$ components	& 499.35 mins\\
                       & Cross match     & 65,476& 713& 685& 799 & $-$ \\ \hline
${{\sigma}_s}=5\sigma$ &\textsc{HeTu}    & 76,328 (96.9\%) & 767 (97.4\%)	&685 (100\%)	&872 (97.6\%)  &  ... \\
                       & \textsc{Aegean} & 	82,894 (89.2\%) & $1,347$ components& $1,415$ components	& $1,293$ components	& 501.50 mins\\
                       & Cross match     & 73,967& 747& 685& 851 & $-$ \\ \hline
${{\sigma}_s}=4\sigma$ &\textsc{HeTu}    & 81,834 (96.2\%) & 770 (97.9\%)	&685 (100\%)	&961 (95.5\%)  &  ... \\
                       & \textsc{Aegean} & 	96,340 (81.7\%) & $1,367$	components& $1,415$ components	& $1,395$ components	& 504.52 mins\\
                       & Cross match     & 78,694& 754& 685& 918 & $-$ \\
        \hline
        \end{tabular} 
        \\[2mm]
        Note -- The trained model used in this experiment is \textsc{HeTu}-101. The figures in the brackets are the ratio of the cross-matched number to the detected number. \\
        $^*$: \textsc{Aegean} detected multiple components for FRI, FRII and CJ sources but did not directly classified them. The most prominent component detected by \textsc{Aegean} is associated with the source detected by \textsc{HeTu}. 
   \end{table*}

\begin{figure}
\centering
\includegraphics[scale=0.4]{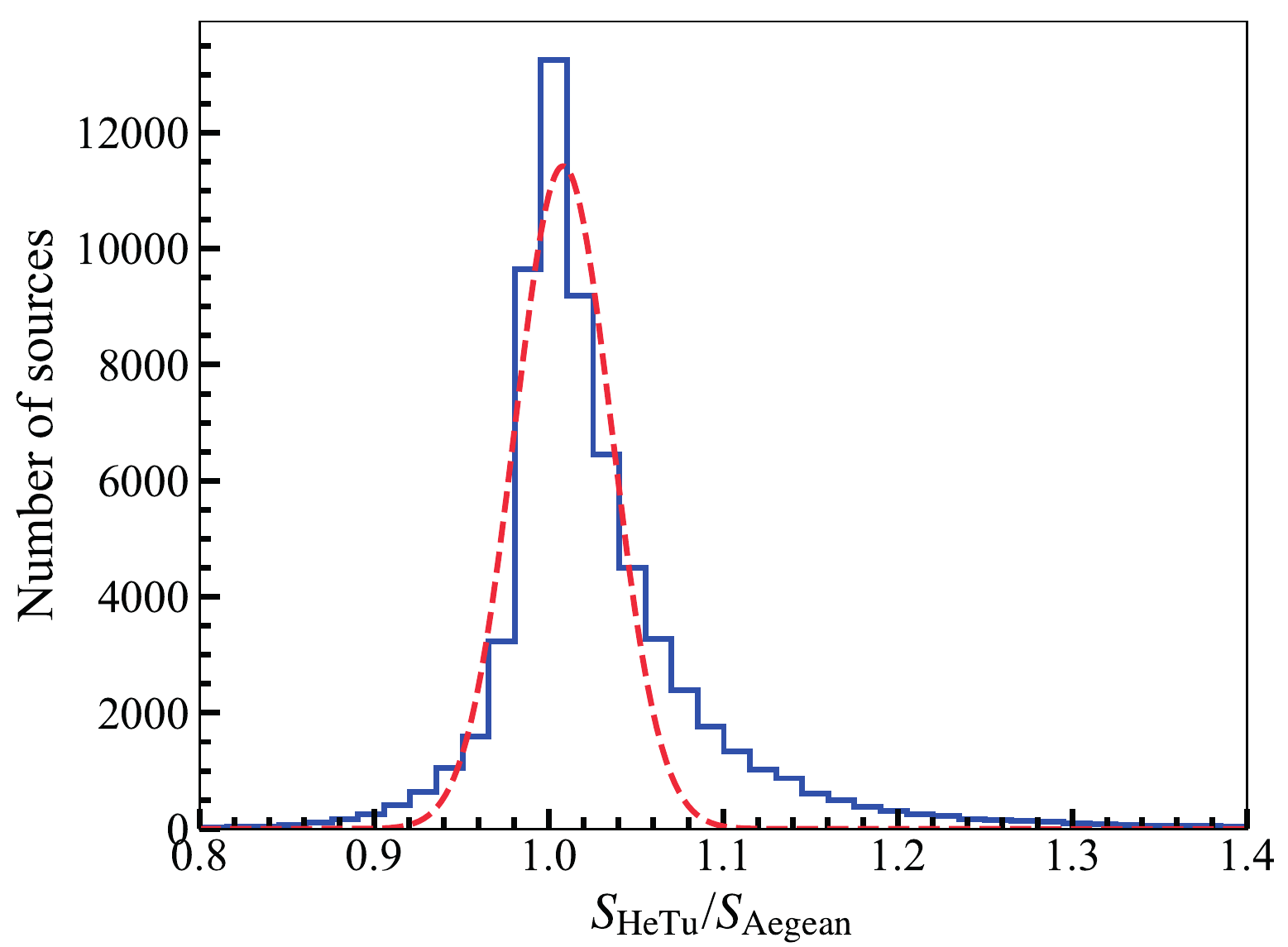}
\includegraphics[scale=0.4]{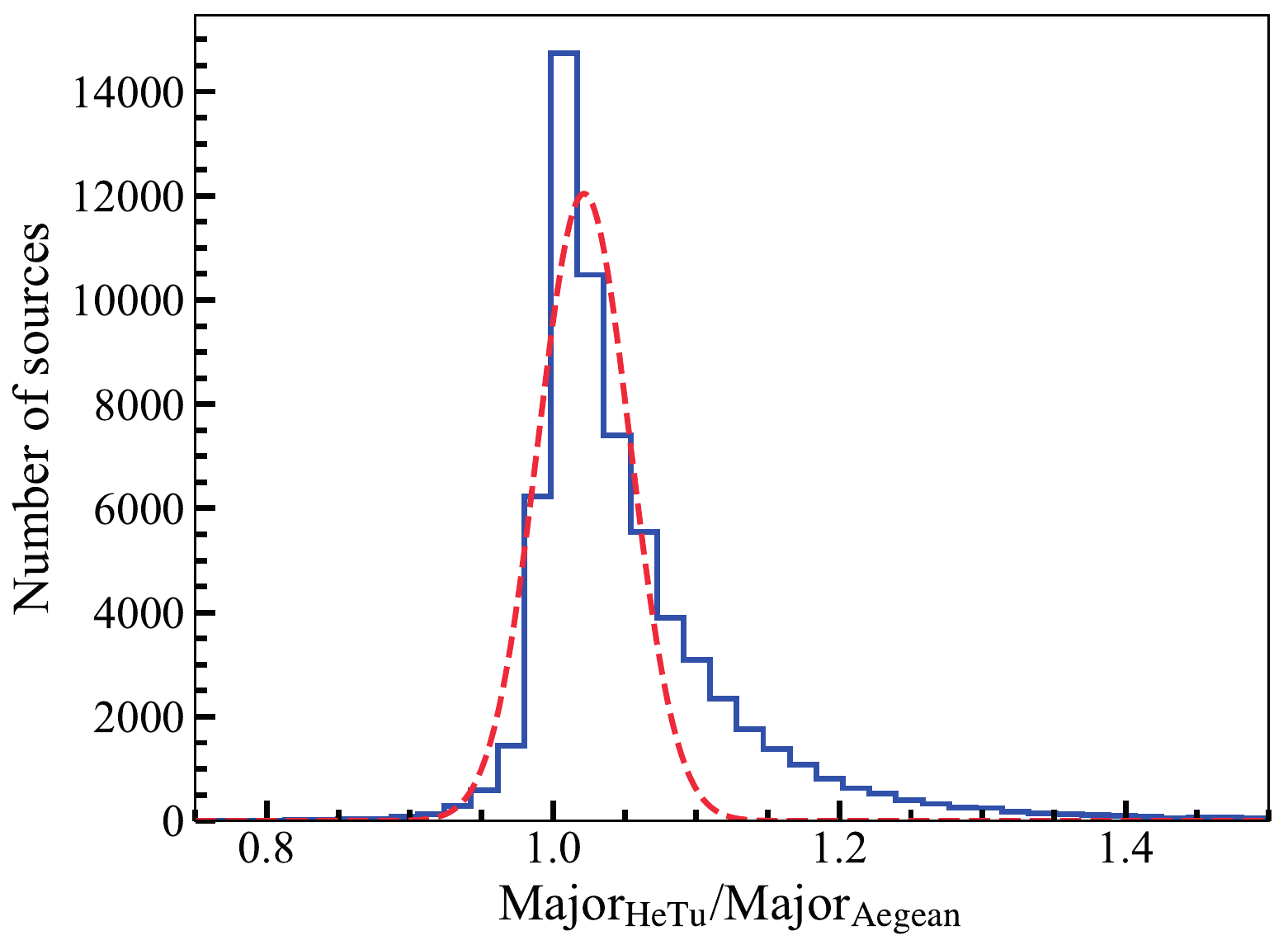}
\includegraphics[scale=0.4]{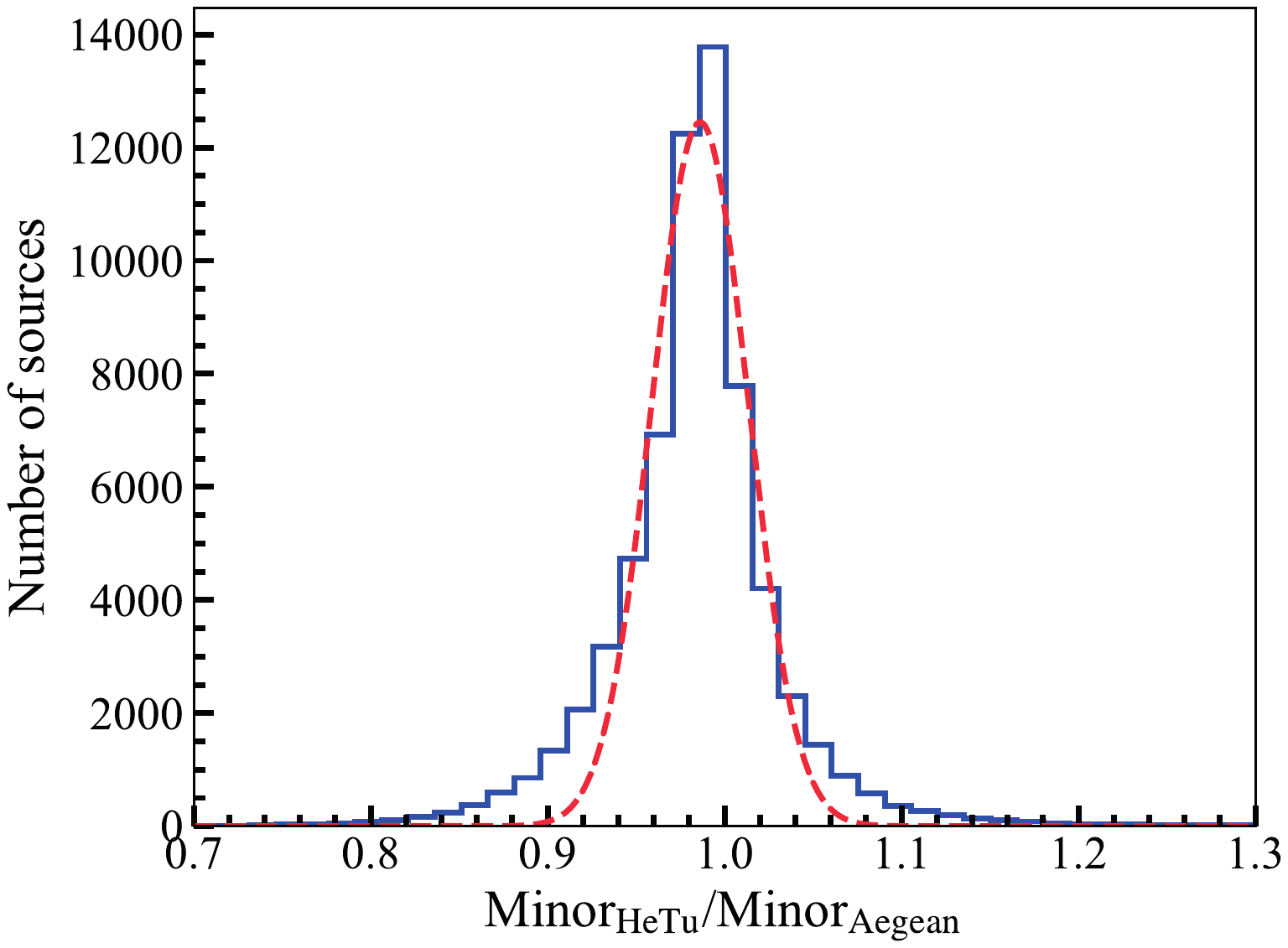}
\caption{Distribution of the fitted parameter values. 
\textit{Top}: histogram of the ratio of the fitted source flux densities ${\rm S_{HeTu}/S_{Aegean}}$. 
\textit{Middle}: histogram of the ratio of the fitted Gaussian major axis length  ${\rm Major_{HeTu}}/{\rm Major_{Aegean}}$. 
\textit{Bottom}: histogram of the ratio of the fitted Gaussian minor axis length  ${\rm Minor_{HeTu}}/{\rm Minor_{Aegean}}$. 
The dashed lines represent the Gaussian fits to the histograms. }
\label{fig:match_all_aegean_hetu_major}
\end{figure}

Fig.~\ref{fig:match_all_aegean_hetu_major} shows the distribution of the comparison of the fitted parameters, in forms of the ratio of the fitted source flux densities ${\rm S_{HeTu}/S_{Aegean}}$,  the ratio of the fitted Gaussian major axis length  ${\rm Major_{HeTu}}/{\rm Major_{Aegean}}$ and  the ratio of the fitted Gaussian minor axis length  ${\rm Minor_{HeTu}}/{\rm Minor_{Aegean}}$. The distribution of ${\rm S_{HeTu}/S_{Aegean}}$ is concentrated around 1.0, indicating that the flux densities fitted by two methods are in excellent agreement. In the region with ratio $>$1.0, the flux density and major axis length fitted by \textsc{HeTu} show an elevated shoulder (systematically larger than that fitted by \textsc{Aegean}). \textsc{Aegean} sets a $4\sigma$ limit on the minimal source flux density, while \textsc{HeTu} defines the scope of the source emission based on its morphology and does not set a limit on the flux density base. Therefore, for sources with somewhat resolved structures, the scopes of the sources used for fitting by \textsc{HeTu} are slightly larger than those used in the \textsc{Aegean} fitting. Since this extended structure is mainly along the major axis direction of the fitted elliptical Gaussian component, the fitted minor axis length (\textit{right} panel) is not affected by the above different flux density base adopted in \textsc{HeTu} and \textsc{Aegean} (see the bottom panel of Fig.~\ref{fig:match_all_aegean_hetu_major}).

\begin{figure}
\centering
\includegraphics[width=0.45\textwidth]{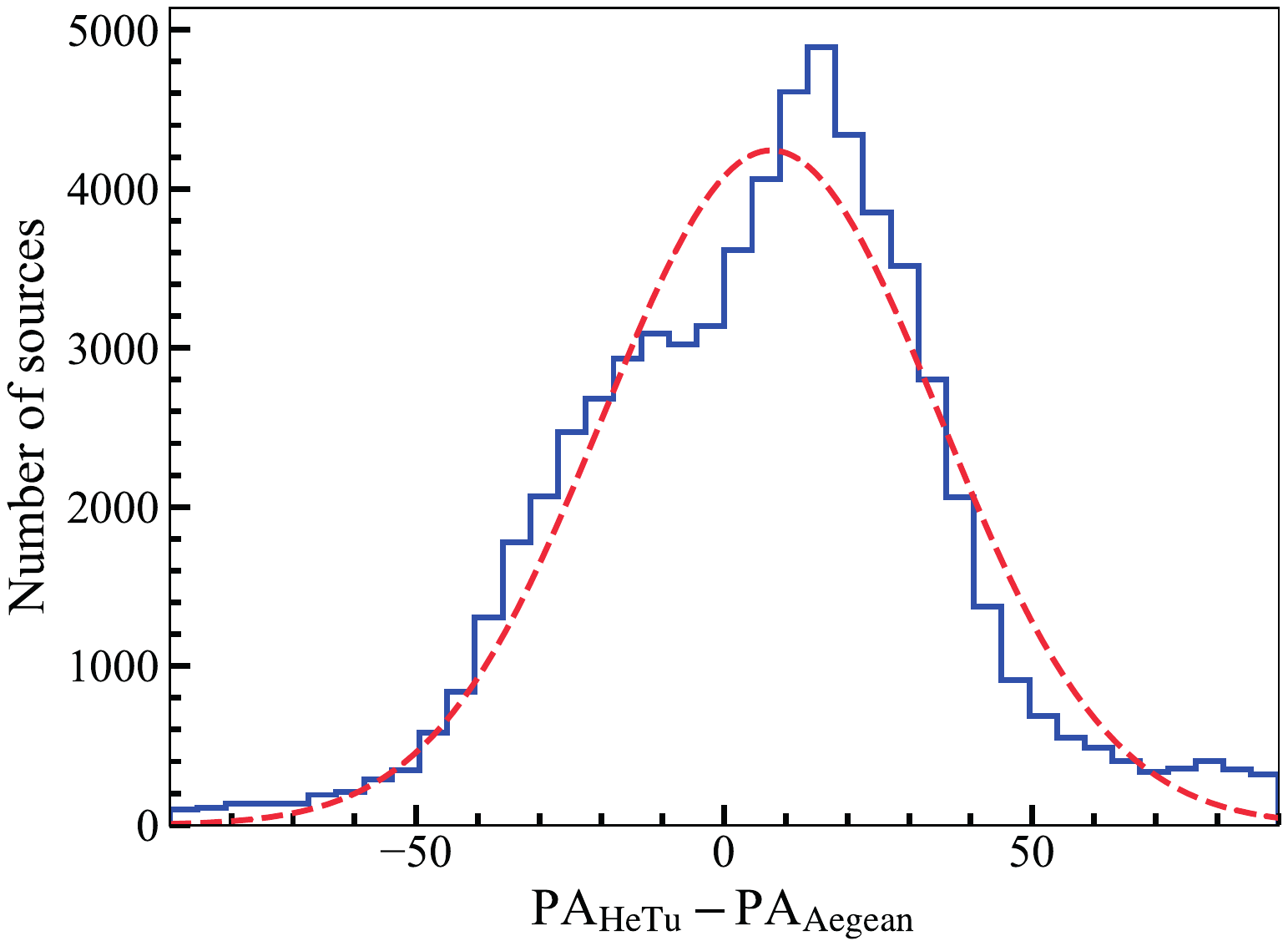}
\caption{Histogram of the difference between the fitted source position angles ${\rm PA_{HeTu}/PA_{Aegean}}$. The  dashed line is the Gaussian fits to the histograms. 
}
\label{fig:match_all_aegean_hetu_pa}
\end{figure}

Fig.~\ref{fig:match_all_aegean_hetu_pa} shows the histogram of the difference between the fitted source position angles ${\rm PA_{HeTu}/PA_{Aegean}}$. The distribution is concentrated around $0^\circ$, indicating that the morphology of the fitted model by \textsc{HeTu} and \textsc{Aegean} is in good agreement.

\textsc{HeTu} supports multiple-GPUs parallel batch execution. The predicting experiment of \textsc{HeTu} was carried out on a GPU node described in Sec.~\ref{sec:3} using 8 GPU devices. 
The batch prediction of all images by \textsc{HeTu} takes 23.87 minutes, which translates into an average runtime of about 100 ms per image. In comparison, the total runtime of \textsc{Aegean} is 501.50 minutes for the detection threshold of $5\sigma$, indicating that \textsc{HeTu} is about $21$ times faster than \textsc{Aegean}. If only identification is considered and no Gaussian fitting is performed, the runtime of \textsc{HeTu} is only 9.52 minutes. Thanks to the GPU acceleration capability and the deep learning framework, the running time of source finding and classification using \text{HeTu} is greatly improved  while maintaining high precision. Moreover, different from \textsc{Aegean} and other traditional component-based source finding software, \textsc{HeTu} can identify and classify radio sources simultaneously and obtains excellent performance especially in the classification of extended sources.

\subsubsection{Output format of \textsc{HeTu}}  \label{sec:3.3.3}

The information of the detected sources generated by \textsc{HeTu} is
recorded into a TXT file, and the columns of CS sources in the output files are as follows: 
\begin{itemize}
\setlength{\itemsep}{0.1mm}
\item imagename: image file name
\item classname: class name of the detected source
\item score: probability of being detected as a specific `classname' source within the bounding box
\item box: the border coordinate values of the four points in pixels, in the format xmin-ymin-xmax-ymax
\item local\_rms: local rms of the detected image, in Jy$\cdot {\rm beam}^{-1}$
\item peak: peak flux density of the detected source, in Jy$\cdot {\rm beam}^{-1}$
\item peak\_err: fitting error of the peak flux density, in Jy$\cdot {\rm beam}^{-1}$
\item int\_flux: integrated flux density of the fitted component, in Jy$\cdot {\rm beam}^{-1}$
\item int\_flux\_err: fitting error of the integrated flux density, in Jy$\cdot {\rm beam}^{-1}$
\item RA: Right Ascension of the emission peak of the fitted Gaussian model component, in hh:mm:ss
\item DEC: Declination of the emission peak of the fitted Gaussian model component, in dd:mm:ss
\item major: semi-major axis length of the fitted Gaussian model component, in arcsec 
\item major\_err: fitting error of the fitted semi-major axis length, in arcsec
\item minor: semi-minor axis length of the fitted Guassian model component, in arcsec
\item minor\_err: fitting error of the fitted semi-minor axis length, in arcsec
\item pa: position angle of the fitted Guassian model component, in degrees
\item pa\_err: fitting error of the Guassian model component's position angle, in degrees
\end{itemize}

The first 10 rows of the  predicting results of the GLEAM images for the CS class are shown in Tables~\ref{tab:example-cs-I} and \ref{tab:example-cs-II}.

Extended sources in output files have the following columns: 
\begin{itemize}
\setlength{\itemsep}{0.1mm}
\item imagename: image file name
\item classname: class name of the detected source
\item score: probability of being detected as a specific `classname' source within the bounding box
\item box: the border coordinate values of the four points in pixels, in the format xmin-ymin-xmax-ymax
\item peak: peak flux density in the bounding box, in Jy$\cdot {\rm beam}^{-1}$
\item peak\_x: x-coordinate of the peak position, in pexels
\item peak\_y: y-coordinate of the peak position, in pexels
\item RA: Right Ascension of the peak position, in degrees
\item DEC: Declination of the peak position, in degrees
\end{itemize}

The extended sources examples of output information are shown in Table~\ref{tab:example-extended}.

\begin{sidewaystable}
\caption{The first 10 rows of the output information for the detected CS sources -I}
\label{tab:example-cs-I}
\begin{tabular}{ccccccccc}
\hline \hline
imagename                             &classname  &score  &box                                      &local\_rms       &peak    &peak\_err   &int\_flux     &int\_flux\_err \\
\hline
1060195176\_200-231MHz\_1344\_1232.png  &cs           &1.00   &23.52101-51.55266-37.39066-65.50194      &0.0093           &0.57    &0.024       &0.68          &0.041          \\
1060195176\_200-231MHz\_1344\_1232.png  &cs           &1.00   &62.29266-26.80662-72.92683-37.71486      &0.0093           &0.22    &0.014       &0.22          &0.020          \\        
1060195176\_200-231MHz\_1344\_1232.png  &cs           &1.00   &48.67455-104.37875-59.72747-115.36621    &0.0093           &0.27    &0.012       &0.28          &0.019          \\         
1060195176\_200-231MHz\_1344\_1232.png  &cs           &1.00   &47.93584-69.70512-57.05483-78.76936      &0.0093           &0.17    &0.008       &0.19          &0.013          \\
1060195176\_200-231MHz\_1344\_1232.png  &cs           &1.00   &57.16669-77.96612-71.88255-92.78156      &0.0093           &1.02    &0.018       &1.05          &0.027          \\
1060195176\_200-231MHz\_1344\_1232.png  &cs           &1.00   &90.45789-120.70591-97.45632-128.00775    &0.0093           &0.10    &0.011       &0.12          &0.023          \\
1060189176\_200-231MHz\_784\_1904.png   &cs           &1.00   &12.42670-41.87839-26.41370-56.01190      &0.0301           &2.57    &0.042       &2.83          &0.067          \\
1060189176\_200-231MHz\_784\_1904.png   &cs           &1.00   &83.77074-76.54671-91.29866-84.01656      &0.0301           &0.28    &0.022       &0.37          &0.048          \\
1060189176\_200-231MHz\_784\_1904.png   &cs           &1.00   &33.40170-50.41240-39.44162-56.24206      &0.0301           &0.22    &0.018       &0.19          &0.028          \\
1060189176\_200-231MHz\_784\_1904.png   &cs           &1.00   &65.57304-88.52624-72.09946-95.49663      &0.0301           &0.28    &0.019       &0.34          &0.037          \\

\hline \hline

\end{tabular}

\end{sidewaystable}

\begin{sidewaystable}
\caption{The first 10 rows of output information for the detected CS sources -II}
\label{tab:example-cs-II}
\begin{tabular}{ccccccccc}
\hline \hline

imagename                                 &RA                  &DEC                     &major      &major\_err        &minor       &minor\_err       &pa           &pa\_err    \\
\hline
1060195176\_200-231MHz\_1344\_1232.png      &23:37:20.415        &\-24:29:22.259          &116.1      &4.90              &106.8       &4.51             &31.7         &0.0        \\         
1060195176\_200-231MHz\_1344\_1232.png      &23:36:15.468        &\-24:18:40.355          &110.7      &7.33              &93.4        &5.91             &\-65.0       &15.1       \\          
1060195176\_200-231MHz\_1344\_1232.png      &23:36:37.116        &\-24:49:20.989          &116.2      &5.59              &94.7        &4.53             &\-69.5       &9.4        \\         
1060195176\_200-231MHz\_1344\_1232.png      &23:36:40.842        &\-24:35:15.831          &115.0      &5.50              &98.9        &4.67             &88.8         &12.9       \\           
1060195176\_200-231MHz\_1344\_1232.png      &23:36:20.481        &\-24:39:27.576          &111.1      &1.99              &96.1        &1.72             &\-69.0       &5.1        \\         
1060195176\_200-231MHz\_1344\_1232.png      &23:35:28.137        &\-24:54:44.304          &117.7      &14.83             &110.7       &16.48            &34.9         &0.0        \\      
1060189176\_200-231MHz\_784\_1904.png       &21:37:13.466        &\-27:54:56.387          &111.3      &1.85              &103.1       &1.71             &\-5.8        &0.0        \\
1060189176\_200-231MHz\_784\_1904.png       &21:35:09.509        &\-28:05:51.049          &129.6      &11.71             &108.2       &9.86             &\-5.3        &0.0        \\         
1060189176\_200-231MHz\_784\_1904.png       &21:36:43.068        &\-27:56:24.400          &112.9      &12.59             &79.2        &7.81             &\-82.7       &10.4       \\      
1060189176\_200-231MHz\_784\_1904.png       &21:35:41.524        &\-28:11:00.263          &122.8      &9.84              &102.2       &7.64             &\-31.8       &0.0        \\
   
\hline \hline

\end{tabular}
\end{sidewaystable}

\begin{sidewaystable}
\caption{The first 2 rows of output information for the each extended sources (CJ, FRI and FRII) -II}
\label{tab:example-extended}
\begin{tabular}{ccccccccc}
\hline \hline

imagename                                 &classname                &score &box                   &peak      &peak\_x        &peak\_y       &RA       &DEC         \\
\hline
1060191576\_200-231MHz\_1456\_1120.png&core\_jet&0.73&49.10925,$-$110.58087,$-$64.83427,$-$124.15668&0.10856&53&15&339.92427&$-$24.14490\\
1060198776\_200-231MHz\_1680\_336.png&core\_jet&1.00&19.59369,$-$0.00000,$-$47.90993,$-$24.31009&5.62358&35&128&15.67706&$-$21.87859\\
1060181376\_200-231MHz\_1904\_1008.png&fr1&0.94&87.21804,$-$44.25601,$-$128.81941,$-$81.02393&0.99791&108&69&297.70891&$-$20.83345\\
1060183176\_200-231MHz\_1680\_1232.png&fr1&0.98&83.68005,$-$50.83970,$-$112.32713,$-$89.04667&0.61012&98&48&303.71708&$-$22.42738\\
1060181376\_200-231MHz\_1904\_1008.png&fr2&1.00&6.34757,$-$109.81445,$-$29.37298,$-$131.41068&2.61121&21&12&298.31460&$-$21.20801\\
1060187376\_200-231MHz\_1904\_784.png&fr2&0.98&57.28113,$-$55.46975,$-$73.90620,$-$72.06968&0.40528&64&65&324.65532&$-$20.88213 \\
   
\hline \hline

\end{tabular}
\end{sidewaystable}

\section{Conclusion}\label{sec:conclusion}

We developed an improved Faster R-CNN based radio source detector named \textsc{HeTu} 
that automatically identifies and classifies radio sources with associated morphological information (label name, score and bounding box). \textsc{HeTu} is built on the pioneering work of \textsc{ClaRAN}  
and further employs a combined ResNet and FPN network as the backbone network. A new morphology classification scheme more suitable for radio astronomy study (compact sources, Fanaroff-Riley type I, Fanaroff-Riley type II and core-jet sources) is adopted. 
\textsc{HeTu} supports training and testing in parallel using multiple GPU devices in a node as well as multiple GPU nodes. In our experiment using 8 GPU devices in a single node, the training speed is 2.5 times faster than without parallelism. Even faster execution can be achieved by using more GPU devices and more nodes. The trained  \textsc{HeTu} models achieve excellent performance with $mAP$ (mean Average Precision) of 87.6\%, 88.2\%, 88.0\% on the D1, D3, and D4 datasets, which are 4.4, 5.6 and 9.2 percentage higher than \textsc{ClaRAN} v0.1 using the same datasets.
On our re-labelled augmented dataset, $mAP$ is as high as 94.2\%. Some source classes have $AP$ values as high as 0.994 (compact sources) and 0.981 (Fanaroff-Riley type II galaxies). 
The augmented datasets can be used for training other CNN models.
Image testing using the \textsc{HeTu}-101 model takes 5.4 milliseconds per image, two orders of magnitude faster than the visual recognition.
The predicting experiment on the MWA GLEAM images shows that \textsc{HeTu}'s detection and classification speed is 43 ms per image, 21 times faster than the standard software package. The overall cross-matching rate between \textsc{HeTu} and \textsc{Aegean} (the GLEAM source finder) is $\sim97\%$, and the rate is 100\% for FRII sources. Note that we can not exclude that some of the identified FRII sources might be CJ-class sources, or vice versa, so further inspection with higher resolution images is needed to determine whether they are indeed FRII-type galaxies. 
Some weaker sources are detected by \textsc{HeTu} but not by \textsc{Aegean}, suggesting that \textsc{HeTu} not only has high recognition precision but also has excellent ability in identifying weak sources. Compared with traditional component-based software, \textsc{HeTu} classifies the detected sources into relevant classes in addition to identification, and outputs the results in a table to facilitate astronomers' research.

The ongoing and upcoming large radio continuum survey projects using the SKA pathfinder telescopes\footnote{see latest advances of the SKA pathfinder telescopes at the 2021 SKA Science Conference, ``A precursor view of the SKA Sky": \url{https://www.skatelescope.org/skascicon21/}} and SKA itself will produce a tremendous amount of images. Automated and accurate source finding and classification tools are particularly important to support these large sky surveys and to mine the data archive. 
Additional predicting experiments will be conducted to further improve \textsc{HeTu}'s recognition performance and speed. These studies will be of practical value in extending \textsc{HeTu} to other radio continuum survey projects.
Future work will focus on improving the current version of the detector so that it can handle even larger images. The development will also focus on upgrading the detector to support Mask R-CNN based instance segmentation, and on improving the measure statistics of the detected sources.
Another effort is to increase the numbers of Fanaroff-Riley type I galaxies and core-jet source images; this will contribute to reducing the imbalance between the different classes. Deep surveys may also discover irregular-shaped sources (e.g., circular radio objects found in the EMU pilot survey \cite{2021PASA...38....3N}) and enrich the radio source classification.   

Neural networks have been shown to understand data more deeply than imagined, but require large data sets for training (learning). The vastness of the Universe provides a naturally large amount of data for training neural networks, and, in turn, the applications of neural networks in astronomy will facilitate the development of AI. On the other hand, AI is actually a form of supervised machine learning, which means that a computer can perform certain tasks well only if it has a correctly labelled dataset to learn from, and the dataset should be large enough to train a good model; moreover, a well-trained model performs a single (fixed) type of task. Successful AI applications have proven that once the dataset is large enough, it can get better and better at specific tasks; for example, as we have seen, the best human chess players are helpless in front of the AlphaGo. In some specific areas, AI has far surpassed human beings, but it is important to note that AI is not an ``all-around champion". However even so, the speed and efficiency of AI is increasingly shaping our understanding of the world we live in. The network framework of \textsc{HeTu} developed in this study can be used not only for astronomical source identification and classification but also be potentially used in other fields such as medical CT image analysis (e.g., automated tumor detection), target recognition (e.g., autonomous driving), etc.

\section*{Acknowledgements}
This work made use of resources provided by China SKA Regional Centre prototype (An, Wu, Hong, Nat Astron, 2019, 3, 1030) which is funded by the National Key R\&D Programme of China (2018YFA0404603) and Chinese Academy of Sciences (114231KYSB20170003). We acknowledge the contribution and the participation of
more than twelve thousand volunteers in the Radio Galaxy Zoo project. Their contributions are individually acknowledged at \url{http://rgzauthors.galaxyzoo.org}. BL and TA thank Chen Wu for helpful discussions and insightful suggestions on the motivation and the methodology of the paper. BL thanks Qi Dang for advice on the experiments and helpful discussion on the deep learning models, thanks Xiaofeng Li for his help with visual identification of re-labeled data and Xiaolong Yang and Sumit Jaiswal for the help on making Figures C14 and C17. TA thanks Natasha Hurley-Walker and Ivy Wong for helpful discussion. This project is implemented based on \textsc{ClaRAN} and Tensorpack\footnote{\url{https://github.com/tensorpack/tensorpack}}.









 \bibliographystyle{elsarticle-num}
 \bibliography{main_final_comb}

\begin{thebibliography}{10}
\expandafter\ifx\csname url\endcsname\relax
  \def\url#1{\texttt{#1}}\fi
\expandafter\ifx\csname urlprefix\endcsname\relax\def\urlprefix{URL }\fi
\expandafter\ifx\csname href\endcsname\relax
  \def\href#1#2{#2} \def\path#1{#1}\fi

\bibitem{2019SCPMA..6289531A}
T.~{An}, {Science opportunities and challenges associated with SKA big data},
  Science China Physics, Mechanics, and Astronomy 62~(8) (2019) 989531.
\newblock \href {http://arxiv.org/abs/1901.07756} {\path{arXiv:1901.07756}},
  \href {https://doi.org/10.1007/s11433-018-9360-x}
  {\path{doi:10.1007/s11433-018-9360-x}}.

\bibitem{2020SRCWP}
P.~{Quinn}, M.~{van Haarlem}, T.~{An}, D.~{Barbosa}, R.~{Bolton},
  A.~{Chrysostomou}, J.~{Conway}, S.~{Gaudet}, H.-R. {Kl\"{o}ckner},
  A.~{Possenti}, S.~{Ratcliffe}, A.~{Scaife}, L.~{Verdes-Montenegro}, J.-P.
  {Vilotte}, Y.~{Wadadekar}, {SKA Regional Centres White Paper v1.0} (May
  2020).

\bibitem{2011PASA...28..215N}
R.~P. {Norris}, A.~M. {Hopkins}, J.~{Afonso}, {et al.}, {EMU: Evolutionary map
  of the universe}, \pasa 28~(3) (2011) 215--248.
\newblock \href {http://arxiv.org/abs/1106.3219} {\path{arXiv:1106.3219}},
  \href {https://doi.org/10.1071/AS11021} {\path{doi:10.1071/AS11021}}.

\bibitem{2021MNRAS.500.3821B}
A.~{Bonaldi}, T.~{An}, M.~{Br{\"u}ggen}, {et al.}, {Square kilometre array
  science data challenge 1: analysis and results}, \mnras 500~(3) (2021)
  3821--3837.
\newblock \href {http://arxiv.org/abs/2009.13346} {\path{arXiv:2009.13346}},
  \href {https://doi.org/10.1093/mnras/staa3023}
  {\path{doi:10.1093/mnras/staa3023}}.

\bibitem{2019MNRAS.482.1211W}
C.~{Wu}, O.~I. {Wong}, L.~{Rudnick}, {et al.}, {Radio Galaxy Zoo: CLARAN - a
  deep learning classifier for radio morphologies}, \mnras 482~(1) (2019)
  1211--1230.
\newblock \href {http://arxiv.org/abs/1805.12008} {\path{arXiv:1805.12008}},
  \href {https://doi.org/10.1093/mnras/sty2646}
  {\path{doi:10.1093/mnras/sty2646}}.

\bibitem{2021MNRAS.tmp..358B}
B.~{Becker}, M.~{Vaccari}, M.~{Prescott}, {et al.}, {CNN architecture
  comparison for radio galaxy classification}, \mnras (Feb. 2021).
\newblock \href {http://arxiv.org/abs/2102.03780} {\path{arXiv:2102.03780}},
  \href {https://doi.org/10.1093/mnras/stab325}
  {\path{doi:10.1093/mnras/stab325}}.

\bibitem{2019NatAs...3.1030A}
T.~{An}, X.-P. {Wu}, X.~{Hong}, {SKA data take centre stage in China}, Nature
  Astronomy 3 (2019) 1030--1030.
\newblock \href {https://doi.org/10.1038/s41550-019-0943-4}
  {\path{doi:10.1038/s41550-019-0943-4}}.

\bibitem{2016arXiv161203144L}
T.-Y. {Lin}, P.~{Doll{\'a}r}, R.~{Girshick}, {et al.}, {Feature Pyramid
  Networks for Object Detection} (2017) 936--944,\href
  {https://doi.org/10.1109/CVPR.2017.106} {\path{doi:10.1109/CVPR.2017.106}}.

\bibitem{2017MNRAS.464.1146H}
N.~{Hurley-Walker}, J.~R. {Callingham}, P.~{Hancock}, {et al.}, {GaLactic and
  Extragalactic All-sky Murchison Widefield Array (GLEAM) survey - I. A
  low-frequency extragalactic catalogue}, \mnras 464~(1) (2017) 1146--1167.
\newblock \href {http://arxiv.org/abs/1610.08318} {\path{arXiv:1610.08318}},
  \href {https://doi.org/10.1093/mnras/stw2337}
  {\path{doi:10.1093/mnras/stw2337}}.

\bibitem{2013PASA...30....7T}
S.~J. {Tingay}, R.~{Goeke}, J.~D. {Bowman}, D.~{Emrich}, S.~M. {Ord}, D.~A.
  {Mitchell}, M.~F. {Morales}, T.~{Booler}, B.~{Crosse}, R.~B. {Wayth}, C.~J.
  {Lonsdale}, S.~{Tremblay}, D.~{Pallot}, T.~{Colegate}, A.~{Wicenec},
  N.~{Kudryavtseva}, W.~{Arcus}, D.~{Barnes}, G.~{Bernardi}, F.~{Briggs},
  S.~{Burns}, J.~D. {Bunton}, R.~J. {Cappallo}, B.~E. {Corey}, A.~{Deshpande},
  L.~{Desouza}, B.~M. {Gaensler}, L.~J. {Greenhill}, P.~J. {Hall}, B.~J.
  {Hazelton}, D.~{Herne}, J.~N. {Hewitt}, M.~{Johnston-Hollitt}, D.~L.
  {Kaplan}, J.~C. {Kasper}, B.~B. {Kincaid}, R.~{Koenig}, E.~{Kratzenberg},
  M.~J. {Lynch}, B.~{Mckinley}, S.~R. {Mcwhirter}, E.~{Morgan}, D.~{Oberoi},
  J.~{Pathikulangara}, T.~{Prabu}, R.~A. {Remillard}, A.~E.~E. {Rogers},
  A.~{Roshi}, J.~E. {Salah}, R.~J. {Sault}, N.~{Udaya-Shankar},
  F.~{Schlagenhaufer}, K.~S. {Srivani}, J.~{Stevens}, R.~{Subrahmanyan},
  M.~{Waterson}, R.~L. {Webster}, A.~R. {Whitney}, A.~{Williams}, C.~L.
  {Williams}, J.~S.~B. {Wyithe}, {The Murchison Widefield Array: The Square
  Kilometre Array Precursor at Low Radio Frequencies}, \pasa 30 (2013) e007.
\newblock \href {http://arxiv.org/abs/1206.6945} {\path{arXiv:1206.6945}},
  \href {https://doi.org/10.1017/pasa.2012.007}
  {\path{doi:10.1017/pasa.2012.007}}.

\bibitem{2018PASA...35...11H}
P.~J. {Hancock}, C.~M. {Trott}, N.~{Hurley-Walker}, {Source Finding in the Era
  of the SKA (Precursors): Aegean 2.0}, \pasa 35 (2018) e011.
\newblock \href {http://arxiv.org/abs/1801.05548} {\path{arXiv:1801.05548}},
  \href {https://doi.org/10.1017/pasa.2018.3} {\path{doi:10.1017/pasa.2018.3}}.

\bibitem{2019ApJ...873..111I}
{\v{Z}}.~{Ivezi{\'c}}, S.~M. {Kahn}, J.~A. {Tyson}, B.~{Abel}, E.~{Acosta},
  R.~{Allsman}, D.~{Alonso}, Y.~{AlSayyad}, S.~F. {Anderson}, J.~{Andrew},
  J.~R.~P. {Angel}, G.~Z. {Angeli}, R.~{Ansari}, P.~{Antilogus}, C.~{Araujo},
  R.~{Armstrong}, K.~T. {Arndt}, P.~{Astier}, {\'E}.~{Aubourg}, N.~{Auza},
  T.~S. {Axelrod}, D.~J. {Bard}, J.~D. {Barr}, A.~{Barrau}, J.~G. {Bartlett},
  A.~E. {Bauer}, B.~J. {Bauman}, S.~{Baumont}, E.~{Bechtol}, K.~{Bechtol},
  A.~C. {Becker}, J.~{Becla}, C.~{Beldica}, S.~{Bellavia}, F.~B. {Bianco},
  R.~{Biswas}, G.~{Blanc}, J.~{Blazek}, R.~D. {Blandford}, J.~S. {Bloom},
  J.~{Bogart}, T.~W. {Bond}, M.~T. {Booth}, A.~W. {Borgland}, K.~{Borne}, J.~F.
  {Bosch}, D.~{Boutigny}, C.~A. {Brackett}, A.~{Bradshaw}, W.~N. {Brandt},
  M.~E. {Brown}, J.~S. {Bullock}, P.~{Burchat}, D.~L. {Burke}, G.~{Cagnoli},
  D.~{Calabrese}, S.~{Callahan}, A.~L. {Callen}, J.~L. {Carlin}, E.~L.
  {Carlson}, S.~{Chandrasekharan}, G.~{Charles-Emerson}, S.~{Chesley}, E.~C.
  {Cheu}, H.-F. {Chiang}, J.~{Chiang}, C.~{Chirino}, D.~{Chow}, D.~R. {Ciardi},
  C.~F. {Claver}, J.~{Cohen-Tanugi}, J.~J. {Cockrum}, R.~{Coles}, A.~J.
  {Connolly}, K.~H. {Cook}, A.~{Cooray}, K.~R. {Covey}, C.~{Cribbs}, W.~{Cui},
  R.~{Cutri}, P.~N. {Daly}, S.~F. {Daniel}, F.~{Daruich}, G.~{Daubard},
  G.~{Daues}, W.~{Dawson}, F.~{Delgado}, A.~{Dellapenna}, R.~{de Peyster},
  M.~{de Val-Borro}, S.~W. {Digel}, P.~{Doherty}, R.~{Dubois}, G.~P.
  {Dubois-Felsmann}, J.~{Durech}, F.~{Economou}, T.~{Eifler}, M.~{Eracleous},
  B.~L. {Emmons}, A.~{Fausti Neto}, H.~{Ferguson}, E.~{Figueroa},
  M.~{Fisher-Levine}, W.~{Focke}, M.~D. {Foss}, J.~{Frank}, M.~D. {Freemon},
  E.~{Gangler}, E.~{Gawiser}, J.~C. {Geary}, P.~{Gee}, M.~{Geha}, C.~J.~B.
  {Gessner}, R.~R. {Gibson}, D.~K. {Gilmore}, T.~{Glanzman}, W.~{Glick},
  T.~{Goldina}, D.~A. {Goldstein}, I.~{Goodenow}, M.~L. {Graham}, W.~J.
  {Gressler}, P.~{Gris}, L.~P. {Guy}, A.~{Guyonnet}, G.~{Haller}, R.~{Harris},
  P.~A. {Hascall}, J.~{Haupt}, F.~{Hernandez}, S.~{Herrmann}, E.~{Hileman},
  J.~{Hoblitt}, J.~A. {Hodgson}, C.~{Hogan}, J.~D. {Howard}, D.~{Huang}, M.~E.
  {Huffer}, P.~{Ingraham}, W.~R. {Innes}, S.~H. {Jacoby}, B.~{Jain},
  F.~{Jammes}, M.~J. {Jee}, T.~{Jenness}, G.~{Jernigan}, D.~{Jevremovi{\'c}},
  K.~{Johns}, A.~S. {Johnson}, M.~W.~G. {Johnson}, R.~L. {Jones},
  C.~{Juramy-Gilles}, M.~{Juri{\'c}}, J.~S. {Kalirai}, N.~J. {Kallivayalil},
  B.~{Kalmbach}, J.~P. {Kantor}, P.~{Karst}, M.~M. {Kasliwal}, H.~{Kelly},
  R.~{Kessler}, V.~{Kinnison}, D.~{Kirkby}, L.~{Knox}, I.~V. {Kotov}, V.~L.
  {Krabbendam}, K.~S. {Krughoff}, P.~{Kub{\'a}nek}, J.~{Kuczewski},
  S.~{Kulkarni}, J.~{Ku}, N.~R. {Kurita}, C.~S. {Lage}, R.~{Lambert},
  T.~{Lange}, J.~B. {Langton}, L.~{Le Guillou}, D.~{Levine}, M.~{Liang}, K.-T.
  {Lim}, C.~J. {Lintott}, K.~E. {Long}, M.~{Lopez}, P.~J. {Lotz}, R.~H.
  {Lupton}, N.~B. {Lust}, L.~A. {MacArthur}, A.~{Mahabal}, R.~{Mandelbaum},
  T.~W. {Markiewicz}, D.~S. {Marsh}, P.~J. {Marshall}, S.~{Marshall}, M.~{May},
  R.~{McKercher}, M.~{McQueen}, J.~{Meyers}, M.~{Migliore}, M.~{Miller}, D.~J.
  {Mills}, C.~{Miraval}, J.~{Moeyens}, F.~E. {Moolekamp}, D.~G. {Monet},
  M.~{Moniez}, S.~{Monkewitz}, C.~{Montgomery}, C.~B. {Morrison}, F.~{Mueller},
  G.~P. {Muller}, F.~{Mu{\~n}oz Arancibia}, D.~R. {Neill}, S.~P. {Newbry},
  J.-Y. {Nief}, A.~{Nomerotski}, M.~{Nordby}, P.~{O'Connor}, J.~{Oliver}, S.~S.
  {Olivier}, K.~{Olsen}, W.~{O'Mullane}, S.~{Ortiz}, S.~{Osier}, R.~E. {Owen},
  R.~{Pain}, P.~E. {Palecek}, J.~K. {Parejko}, J.~B. {Parsons}, N.~M. {Pease},
  J.~M. {Peterson}, J.~R. {Peterson}, D.~L. {Petravick}, M.~E. {Libby Petrick},
  C.~E. {Petry}, F.~{Pierfederici}, S.~{Pietrowicz}, R.~{Pike}, P.~A. {Pinto},
  R.~{Plante}, S.~{Plate}, J.~P. {Plutchak}, P.~A. {Price}, M.~{Prouza},
  V.~{Radeka}, J.~{Rajagopal}, A.~P. {Rasmussen}, N.~{Regnault}, K.~A. {Reil},
  D.~J. {Reiss}, M.~A. {Reuter}, S.~T. {Ridgway}, V.~J. {Riot}, S.~{Ritz},
  S.~{Robinson}, W.~{Roby}, A.~{Roodman}, W.~{Rosing}, C.~{Roucelle}, M.~R.
  {Rumore}, S.~{Russo}, A.~{Saha}, B.~{Sassolas}, T.~L. {Schalk},
  P.~{Schellart}, R.~H. {Schindler}, S.~{Schmidt}, D.~P. {Schneider}, M.~D.
  {Schneider}, W.~{Schoening}, G.~{Schumacher}, M.~E. {Schwamb}, J.~{Sebag},
  B.~{Selvy}, G.~H. {Sembroski}, L.~G. {Seppala}, A.~{Serio}, E.~{Serrano},
  R.~A. {Shaw}, I.~{Shipsey}, J.~{Sick}, N.~{Silvestri}, C.~T. {Slater}, J.~A.
  {Smith}, R.~C. {Smith}, S.~{Sobhani}, C.~{Soldahl}, L.~{Storrie-Lombardi},
  E.~{Stover}, M.~A. {Strauss}, R.~A. {Street}, C.~W. {Stubbs}, I.~S.
  {Sullivan}, D.~{Sweeney}, J.~D. {Swinbank}, A.~{Szalay}, P.~{Takacs}, S.~A.
  {Tether}, J.~J. {Thaler}, J.~G. {Thayer}, S.~{Thomas}, A.~J. {Thornton},
  V.~{Thukral}, J.~{Tice}, D.~E. {Trilling}, M.~{Turri}, R.~{Van Berg},
  D.~{Vanden Berk}, K.~{Vetter}, F.~{Virieux}, T.~{Vucina}, W.~{Wahl},
  L.~{Walkowicz}, B.~{Walsh}, C.~W. {Walter}, D.~L. {Wang}, S.-Y. {Wang},
  M.~{Warner}, O.~{Wiecha}, B.~{Willman}, S.~E. {Winters}, D.~{Wittman}, S.~C.
  {Wolff}, W.~M. {Wood-Vasey}, X.~{Wu}, B.~{Xin}, P.~{Yoachim}, H.~{Zhan},
  {LSST: From Science Drivers to Reference Design and Anticipated Data
  Products}, \apj 873~(2) (2019) 111.
\newblock \href {http://arxiv.org/abs/0805.2366} {\path{arXiv:0805.2366}},
  \href {https://doi.org/10.3847/1538-4357/ab042c}
  {\path{doi:10.3847/1538-4357/ab042c}}.

\bibitem{2000AJ....120.1579Y}
D.~G. {York}, J.~{Adelman}, J.~{Anderson}, John~E., S.~F. {Anderson},
  J.~{Annis}, N.~A. {Bahcall}, J.~A. {Bakken}, R.~{Barkhouser}, S.~{Bastian},
  E.~{Berman}, W.~N. {Boroski}, S.~{Bracker}, C.~{Briegel}, J.~W. {Briggs},
  J.~{Brinkmann}, R.~{Brunner}, S.~{Burles}, L.~{Carey}, M.~A. {Carr}, F.~J.
  {Castander}, B.~{Chen}, P.~L. {Colestock}, A.~J. {Connolly}, J.~H. {Crocker},
  I.~{Csabai}, P.~C. {Czarapata}, J.~E. {Davis}, M.~{Doi}, T.~{Dombeck},
  D.~{Eisenstein}, N.~{Ellman}, B.~R. {Elms}, M.~L. {Evans}, X.~{Fan}, G.~R.
  {Federwitz}, L.~{Fiscelli}, S.~{Friedman}, J.~A. {Frieman}, M.~{Fukugita},
  B.~{Gillespie}, J.~E. {Gunn}, V.~K. {Gurbani}, E.~{de Haas}, M.~{Haldeman},
  F.~H. {Harris}, J.~{Hayes}, T.~M. {Heckman}, G.~S. {Hennessy}, R.~B.
  {Hindsley}, S.~{Holm}, D.~J. {Holmgren}, C.-h. {Huang}, C.~{Hull},
  D.~{Husby}, S.-I. {Ichikawa}, T.~{Ichikawa}, {\v{Z}}.~{Ivezi{\'c}},
  S.~{Kent}, R.~S.~J. {Kim}, E.~{Kinney}, M.~{Klaene}, A.~N. {Kleinman},
  S.~{Kleinman}, G.~R. {Knapp}, J.~{Korienek}, R.~G. {Kron}, P.~Z. {Kunszt},
  D.~Q. {Lamb}, B.~{Lee}, R.~F. {Leger}, S.~{Limmongkol}, C.~{Lindenmeyer},
  D.~C. {Long}, C.~{Loomis}, J.~{Loveday}, R.~{Lucinio}, R.~H. {Lupton},
  B.~{MacKinnon}, E.~J. {Mannery}, P.~M. {Mantsch}, B.~{Margon}, P.~{McGehee},
  T.~A. {McKay}, A.~{Meiksin}, A.~{Merelli}, D.~G. {Monet}, J.~A. {Munn}, V.~K.
  {Narayanan}, T.~{Nash}, E.~{Neilsen}, R.~{Neswold}, H.~J. {Newberg}, R.~C.
  {Nichol}, T.~{Nicinski}, M.~{Nonino}, N.~{Okada}, S.~{Okamura}, J.~P.
  {Ostriker}, R.~{Owen}, A.~G. {Pauls}, J.~{Peoples}, R.~L. {Peterson},
  D.~{Petravick}, J.~R. {Pier}, A.~{Pope}, R.~{Pordes}, A.~{Prosapio},
  R.~{Rechenmacher}, T.~R. {Quinn}, G.~T. {Richards}, M.~W. {Richmond}, C.~H.
  {Rivetta}, C.~M. {Rockosi}, K.~{Ruthmansdorfer}, D.~{Sandford}, D.~J.
  {Schlegel}, D.~P. {Schneider}, M.~{Sekiguchi}, G.~{Sergey}, K.~{Shimasaku},
  W.~A. {Siegmund}, S.~{Smee}, J.~A. {Smith}, S.~{Snedden}, R.~{Stone},
  C.~{Stoughton}, M.~A. {Strauss}, C.~{Stubbs}, M.~{SubbaRao}, A.~S. {Szalay},
  I.~{Szapudi}, G.~P. {Szokoly}, A.~R. {Thakar}, C.~{Tremonti}, D.~L. {Tucker},
  A.~{Uomoto}, D.~{Vanden Berk}, M.~S. {Vogeley}, P.~{Waddell}, S.-i. {Wang},
  M.~{Watanabe}, D.~H. {Weinberg}, B.~{Yanny}, N.~{Yasuda}, {SDSS
  Collaboration}, {The Sloan Digital Sky Survey: Technical Summary}, \aj
  120~(3) (2000) 1579--1587.
\newblock \href {http://arxiv.org/abs/astro-ph/0006396}
  {\path{arXiv:astro-ph/0006396}}, \href {https://doi.org/10.1086/301513}
  {\path{doi:10.1086/301513}}.

\bibitem{2009IEEEP..97.1482D}
P.~E. {Dewdney}, P.~J. {Hall}, R.~T. {Schilizzi}, T.~J.~L.~W. {Lazio}, {The
  Square Kilometre Array}, IEEE Proceedings 97~(8) (2009) 1482--1496.
\newblock \href {https://doi.org/10.1109/JPROC.2009.2021005}
  {\path{doi:10.1109/JPROC.2009.2021005}}.

\bibitem{2015aska.confE..81M}
S.~{Makhathini}, M.~{Jarvis}, O.~{Smirnov}, I.~{Heywood}, {Morphological
  classification of radio sources for galaxy evolution and cosmology with the
  SKA}, in: Advancing Astrophysics with the Square Kilometre Array (AASKA14),
  2015, p.~81.
\newblock \href {http://arxiv.org/abs/1412.5990} {\path{arXiv:1412.5990}}.

\bibitem{2003ASSL..285..109G}
E.~W. {Greisen}, {AIPS, the VLA, and the VLBA}, Vol. 285, 2003, p. 109.
\newblock \href {https://doi.org/10.1007/0-306-48080-8_7}
  {\path{doi:10.1007/0-306-48080-8_7}}.

\bibitem{1997ApJ...475..479W}
R.~L. {White}, R.~H. {Becker}, D.~J. {Helfand}, M.~D. {Gregg}, {A Catalog of
  1.4 GHz Radio Sources from the FIRST Survey}, \apj 475~(2) (1997) 479--493.
\newblock \href {https://doi.org/10.1086/303564} {\path{doi:10.1086/303564}}.

\bibitem{1998AJ....115.1693C}
J.~J. {Condon}, W.~D. {Cotton}, E.~W. {Greisen}, Q.~F. {Yin}, R.~A. {Perley},
  G.~B. {Taylor}, J.~J. {Broderick}, {The NRAO VLA Sky Survey}, \aj 115~(5)
  (1998) 1693--1716.
\newblock \href {https://doi.org/10.1086/300337} {\path{doi:10.1086/300337}}.

\bibitem{1995ASPC...77..433S}
R.~J. {Sault}, P.~J. {Teuben}, M.~C.~H. {Wright}, {A Retrospective View of
  MIRIAD}, in: R.~A. {Shaw}, H.~E. {Payne}, J.~J.~E. {Hayes} (Eds.),
  Astronomical Data Analysis Software and Systems IV, Vol.~77 of Astronomical
  Society of the Pacific Conference Series, 1995, p. 433.
\newblock \href {http://arxiv.org/abs/astro-ph/0612759}
  {\path{arXiv:astro-ph/0612759}}.

\bibitem{2002AJ....123.1086H}
A.~M. {Hopkins}, C.~J. {Miller}, A.~J. {Connolly}, C.~{Genovese}, R.~C.
  {Nichol}, L.~{Wasserman}, {A New Source Detection Algorithm Using the
  False-Discovery Rate}, \aj 123~(2) (2002) 1086--1094.
\newblock \href {http://arxiv.org/abs/astro-ph/0110570}
  {\path{arXiv:astro-ph/0110570}}, \href {https://doi.org/10.1086/338316}
  {\path{doi:10.1086/338316}}.

\bibitem{2000A&AS..146...41P}
I.~{Prandoni}, L.~{Gregorini}, P.~{Parma}, H.~R. {de Ruiter}, G.~{Vettolani},
  M.~H. {Wieringa}, R.~D. {Ekers}, {The ATESP radio survey. II. The source
  catalogue}, \aaps 146 (2000) 41--55.
\newblock \href {http://arxiv.org/abs/astro-ph/0007398}
  {\path{arXiv:astro-ph/0007398}}, \href {https://doi.org/10.1051/aas:2000361}
  {\path{doi:10.1051/aas:2000361}}.

\bibitem{2015MNRAS.448.3731R}
M.~{Regis}, L.~{Richter}, S.~{Colafrancesco}, M.~{Massardi}, W.~J.~G. {de
  Blok}, S.~{Profumo}, N.~{Orford}, {Local Group dSph radio survey with ATCA
  (I): observations and background sources}, \mnras 448~(4) (2015) 3731--3746.
\newblock \href {http://arxiv.org/abs/1407.5479} {\path{arXiv:1407.5479}},
  \href {https://doi.org/10.1093/mnras/stu2747}
  {\path{doi:10.1093/mnras/stu2747}}.

\bibitem{2015PASA...32...37H}
A.~M. {Hopkins}, M.~T. {Whiting}, N.~{Seymour}, K.~E. {Chow}, R.~P. {Norris},
  L.~{Bonavera}, R.~{Breton}, D.~{Carbone}, C.~{Ferrari}, T.~M.~O. {Franzen},
  H.~{Garsden}, J.~{Gonz{\'a}lez-Nuevo}, C.~A. {Hales}, P.~J. {Hancock},
  G.~{Heald}, D.~{Herranz}, M.~{Huynh}, R.~J. {Jurek}, M.~{L{\'o}pez-Caniego},
  M.~{Massardi}, N.~{Mohan}, S.~{Molinari}, E.~{Orr{\`u}}, R.~{Paladino},
  M.~{Pestalozzi}, R.~{Pizzo}, D.~{Rafferty}, H.~J.~A. {R{\"o}ttgering},
  L.~{Rudnick}, E.~{Schisano}, A.~{Shulevski}, J.~{Swinbank}, R.~{Taylor},
  A.~J. {van der Horst}, {The ASKAP/EMU Source Finding Data Challenge}, \pasa
  32 (2015) e037.
\newblock \href {http://arxiv.org/abs/1509.03931} {\path{arXiv:1509.03931}},
  \href {https://doi.org/10.1017/pasa.2015.37}
  {\path{doi:10.1017/pasa.2015.37}}.

\bibitem{1996A&AS..117..393B}
E.~{Bertin}, S.~{Arnouts}, {SExtractor: Software for source extraction.}, \aaps
  117 (1996) 393--404.
\newblock \href {https://doi.org/10.1051/aas:1996164}
  {\path{doi:10.1051/aas:1996164}}.

\bibitem{2012MNRAS.421.3242W}
M.~T. {Whiting}, {DUCHAMP: a 3D source finder for spectral-line data}, \mnras
  421~(4) (2012) 3242--3256.
\newblock \href {http://arxiv.org/abs/1201.2710} {\path{arXiv:1201.2710}},
  \href {https://doi.org/10.1111/j.1365-2966.2012.20548.x}
  {\path{doi:10.1111/j.1365-2966.2012.20548.x}}.

\bibitem{2012MNRAS.425..979H}
C.~A. {Hales}, T.~{Murphy}, J.~R. {Curran}, E.~{Middelberg}, B.~M. {Gaensler},
  R.~P. {Norris}, {BLOBCAT: software to catalogue flood-filled blobs in radio
  images of total intensity and linear polarization}, \mnras 425~(2) (2012)
  979--996.
\newblock \href {http://arxiv.org/abs/1205.5313} {\path{arXiv:1205.5313}},
  \href {https://doi.org/10.1111/j.1365-2966.2012.21373.x}
  {\path{doi:10.1111/j.1365-2966.2012.21373.x}}.

\bibitem{2012MNRAS.422.1812H}
P.~J. {Hancock}, T.~{Murphy}, B.~M. {Gaensler}, A.~{Hopkins}, J.~R. {Curran},
  {Compact continuum source finding for next generation radio surveys}, \mnras
  422~(2) (2012) 1812--1824.
\newblock \href {http://arxiv.org/abs/1202.4500} {\path{arXiv:1202.4500}},
  \href {https://doi.org/10.1111/j.1365-2966.2012.20768.x}
  {\path{doi:10.1111/j.1365-2966.2012.20768.x}}.

\bibitem{2015ascl.soft02007M}
N.~{Mohan}, D.~{Rafferty}, {PyBDSF: Python Blob Detection and Source Finder}
  (Feb. 2015).
\newblock \href {http://arxiv.org/abs/1502.007} {\path{arXiv:1502.007}}.

\bibitem{2015PASA...32...25W}
R.~B. {Wayth}, E.~{Lenc}, M.~E. {Bell}, J.~R. {Callingham}, K.~S.
  {Dwarakanath}, T.~M.~O. {Franzen}, B.~Q. {For}, B.~{Gaensler}, P.~{Hancock},
  L.~{Hindson}, N.~{Hurley-Walker}, C.~A. {Jackson}, M.~{Johnston-Hollitt},
  A.~D. {Kapi{\'n}ska}, B.~{McKinley}, J.~{Morgan}, A.~R. {Offringa},
  P.~{Procopio}, L.~{Staveley-Smith}, C.~{Wu}, Q.~{Zheng}, C.~M. {Trott},
  G.~{Bernardi}, J.~D. {Bowman}, F.~{Briggs}, R.~J. {Cappallo}, B.~E. {Corey},
  A.~A. {Deshpande}, D.~{Emrich}, R.~{Goeke}, L.~J. {Greenhill}, B.~J.
  {Hazelton}, D.~L. {Kaplan}, J.~C. {Kasper}, E.~{Kratzenberg}, C.~J.
  {Lonsdale}, M.~J. {Lynch}, S.~R. {McWhirter}, D.~A. {Mitchell}, M.~F.
  {Morales}, E.~{Morgan}, D.~{Oberoi}, S.~M. {Ord}, T.~{Prabu}, A.~E.~E.
  {Rogers}, A.~{Roshi}, N.~U. {Shankar}, K.~S. {Srivani}, R.~{Subrahmanyan},
  S.~J. {Tingay}, M.~{Waterson}, R.~L. {Webster}, A.~R. {Whitney},
  A.~{Williams}, C.~L. {Williams}, {GLEAM: The GaLactic and Extragalactic
  All-Sky MWA Survey}, \pasa 32 (2015) e025.
\newblock \href {http://arxiv.org/abs/1505.06041} {\path{arXiv:1505.06041}},
  \href {https://doi.org/10.1017/pasa.2015.26}
  {\path{doi:10.1017/pasa.2015.26}}.

\bibitem{2019PASA...36...37R}
S.~{Riggi}, F.~{Vitello}, U.~{Becciani}, C.~{Buemi}, F.~{Bufano},
  A.~{Calanducci}, F.~{Cavallaro}, A.~{Costa}, A.~{Ingallinera}, P.~{Leto},
  S.~{Loru}, R.~P. {Norris}, F.~{Schillir{\`o}}, E.~{Sciacca}, C.~{Trigilio},
  G.~{Umana}, {Cuc(aesar) source finder: Recent developments and testing},
  \pasa 36 (2019) e037.
\newblock \href {http://arxiv.org/abs/1909.06116} {\path{arXiv:1909.06116}},
  \href {https://doi.org/10.1017/pasa.2019.29}
  {\path{doi:10.1017/pasa.2019.29}}.

\bibitem{2019MNRAS.487.3971H}
C.~L. {Hale}, A.~S.~G. {Robotham}, L.~J.~M. {Davies}, M.~J. {Jarvis}, S.~P.
  {Driver}, I.~{Heywood}, {Radio source extraction with PROFOUND}, \mnras
  487~(3) (2019) 3971--3989.
\newblock \href {http://arxiv.org/abs/1902.01440} {\path{arXiv:1902.01440}},
  \href {https://doi.org/10.1093/mnras/stz1462}
  {\path{doi:10.1093/mnras/stz1462}}.

\bibitem{2017ApJS..230...20A}
A.~K. {Aniyan}, K.~{Thorat}, {Classifying Radio Galaxies with the Convolutional
  Neural Network}, \apjs 230~(2) (2017) 20.
\newblock \href {http://arxiv.org/abs/1705.03413} {\path{arXiv:1705.03413}},
  \href {https://doi.org/10.3847/1538-4365/aa7333}
  {\path{doi:10.3847/1538-4365/aa7333}}.

\bibitem{2018MNRAS.480.2085A}
W.~{Alhassan}, A.~R. {Taylor}, M.~{Vaccari}, {The FIRST Classifier: compact and
  extended radio galaxy classification using deep Convolutional Neural
  Networks}, \mnras 480~(2) (2018) 2085--2093.
\newblock \href {http://arxiv.org/abs/1807.10380} {\path{arXiv:1807.10380}},
  \href {https://doi.org/10.1093/mnras/sty2038}
  {\path{doi:10.1093/mnras/sty2038}}.

\bibitem{2020MNRAS.498.5620S}
Y.~{Su}, Y.~{Zhang}, G.~{Liang}, J.~A. {ZuHone}, D.~J. {Barnes}, N.~B.
  {Jacobs}, M.~{Ntampaka}, W.~R. {Forman}, P.~E.~J. {Nulsen}, R.~P. {Kraft},
  C.~{Jones}, {A deep learning view of the census of galaxy clusters in
  IllustrisTNG}, \mnras 498~(4) (2020) 5620--5628.
\newblock \href {http://arxiv.org/abs/2007.05144} {\path{arXiv:2007.05144}},
  \href {https://doi.org/10.1093/mnras/staa2690}
  {\path{doi:10.1093/mnras/staa2690}}.

\bibitem{2019Galax...8....3L}
V.~{Lukic}, F.~{de Gasperin}, M.~{Br{\"u}ggen}, {ConvoSource:
  Radio-Astronomical Source-Finding with Convolutional Neural Networks},
  Galaxies 8~(1) (2019) 3.
\newblock \href {http://arxiv.org/abs/1910.03631} {\path{arXiv:1910.03631}},
  \href {https://doi.org/10.3390/galaxies8010003}
  {\path{doi:10.3390/galaxies8010003}}.

\bibitem{2021MNRAS.501.4579B}
M.~{Bowles}, A.~M.~M. {Scaife}, F.~{Porter}, H.~{Tang}, D.~J. {Bastien},
  {Attention-gating for improved radio galaxy classification}, \mnras 501~(3)
  (2021) 4579--4595.
\newblock \href {http://arxiv.org/abs/2012.01248} {\path{arXiv:2012.01248}},
  \href {https://doi.org/10.1093/mnras/staa3946}
  {\path{doi:10.1093/mnras/staa3946}}.

\bibitem{2021arXiv210208252S}
A.~M.~M. {Scaife}, F.~{Porter}, {Fanaroff-Riley classification of radio
  galaxies using group-equivariant convolutional neural networks}, arXiv
  e-prints (2021) arXiv:2102.08252\href {http://arxiv.org/abs/2102.08252}
  {\path{arXiv:2102.08252}}.

\bibitem{2021arXiv210201007B}
D.~J. {Bastien}, A.~M.~M. {Scaife}, H.~{Tang}, M.~{Bowles}, F.~{Porter},
  {Structured Variational Inference for Simulating Populations of Radio
  Galaxies}, arXiv e-prints (2021) arXiv:2102.01007\href
  {http://arxiv.org/abs/2102.01007} {\path{arXiv:2102.01007}}.

\bibitem{2019MNRAS.484.2793V}
A.~{Vafaei Sadr}, E.~E. {Vos}, B.~A. {Bassett}, Z.~{Hosenie}, N.~{Oozeer},
  M.~{Lochner}, {DEEPSOURCE: point source detection using deep learning},
  \mnras 484~(2) (2019) 2793--2806.
\newblock \href {http://arxiv.org/abs/1807.02701} {\path{arXiv:1807.02701}},
  \href {https://doi.org/10.1093/mnras/stz131}
  {\path{doi:10.1093/mnras/stz131}}.

\bibitem{2019arXiv190505055Z}
Z.~{Zou}, Z.~{Shi}, Y.~{Guo}, J.~{Ye}, {Object Detection in 20 Years: A
  Survey}, arXiv e-prints (2019) arXiv:1905.05055\href
  {http://arxiv.org/abs/1905.05055} {\path{arXiv:1905.05055}}.

\bibitem{2015arXiv151202325L}
W.~{Liu}, D.~{Anguelov}, D.~{Erhan}, C.~{Szegedy}, S.~{Reed}, C.-Y. {Fu}, A.~C.
  {Berg}, {SSD: Single Shot MultiBox Detector}, arXiv e-prints (2015)
  arXiv:1512.02325\href {http://arxiv.org/abs/1512.02325}
  {\path{arXiv:1512.02325}}.

\bibitem{2015arXiv150602640R}
J.~{Redmon}, S.~{Divvala}, R.~{Girshick}, A.~{Farhadi}, {You Only Look Once:
  Unified, Real-Time Object Detection}, arXiv e-prints (2015)
  arXiv:1506.02640\href {http://arxiv.org/abs/1506.02640}
  {\path{arXiv:1506.02640}}.

\bibitem{2016arXiv161208242R}
J.~{Redmon}, A.~{Farhadi}, {YOLO9000: Better, Faster, Stronger}, arXiv e-prints
  (2016) arXiv:1612.08242\href {http://arxiv.org/abs/1612.08242}
  {\path{arXiv:1612.08242}}.

\bibitem{2018arXiv180402767R}
J.~{Redmon}, A.~{Farhadi}, {YOLOv3: An Incremental Improvement}, arXiv e-prints
  (2018) arXiv:1804.02767\href {http://arxiv.org/abs/1804.02767}
  {\path{arXiv:1804.02767}}.

\bibitem{2020arXiv200410934B}
A.~{Bochkovskiy}, C.-Y. {Wang}, H.-Y.~M. {Liao}, {YOLOv4: Optimal Speed and
  Accuracy of Object Detection}, arXiv e-prints (2020) arXiv:2004.10934\href
  {http://arxiv.org/abs/2004.10934} {\path{arXiv:2004.10934}}.

\bibitem{2014arXiv1409.5403G}
R.~{Girshick}, F.~{Iandola}, T.~{Darrell}, J.~{Malik}, {Deformable Part Models
  are Convolutional Neural Networks}, arXiv e-prints (2014)
  arXiv:1409.5403\href {http://arxiv.org/abs/1409.5403}
  {\path{arXiv:1409.5403}}.

\bibitem{2015arXiv150201852H}
K.~{He}, X.~{Zhang}, S.~{Ren}, J.~{Sun}, {Delving Deep into Rectifiers:
  Surpassing Human-Level Performance on ImageNet Classification}, arXiv
  e-prints (2015) arXiv:1502.01852\href {http://arxiv.org/abs/1502.01852}
  {\path{arXiv:1502.01852}}.

\bibitem{2015arXiv150408083G}
R.~{Girshick}, {Fast R-CNN}, arXiv e-prints (2015) arXiv:1504.08083\href
  {http://arxiv.org/abs/1504.08083} {\path{arXiv:1504.08083}}.

\bibitem{2015arXiv150601497R}
S.~{Ren}, K.~{He}, R.~{Girshick}, J.~{Sun}, {Faster R-CNN: Towards Real-Time
  Object Detection with Region Proposal Networks}, arXiv e-prints (2015)
  arXiv:1506.01497\href {http://arxiv.org/abs/1506.01497}
  {\path{arXiv:1506.01497}}.

\bibitem{2017arXiv170306870H}
K.~{He}, G.~{Gkioxari}, P.~{Doll{\'a}r}, R.~{Girshick}, {Mask R-CNN}, arXiv
  e-prints (2017) arXiv:1703.06870\href {http://arxiv.org/abs/1703.06870}
  {\path{arXiv:1703.06870}}.

\bibitem{2014arXiv1409.1556S}
K.~{Simonyan}, A.~{Zisserman}, {Very Deep Convolutional Networks for
  Large-Scale Image Recognition}, arXiv e-prints (2014) arXiv:1409.1556\href
  {http://arxiv.org/abs/1409.1556} {\path{arXiv:1409.1556}}.

\bibitem{2015arXiv151203385H}
K.~{He}, X.~{Zhang}, S.~{Ren}, J.~{Sun}, {Deep Residual Learning for Image
  Recognition}, arXiv e-prints (2015) arXiv:1512.03385\href
  {http://arxiv.org/abs/1512.03385} {\path{arXiv:1512.03385}}.

\bibitem{2013arXiv1311.2901Z}
M.~D. {Zeiler}, R.~{Fergus}, {Visualizing and Understanding Convolutional
  Networks}, arXiv e-prints (2013) arXiv:1311.2901\href
  {http://arxiv.org/abs/1311.2901} {\path{arXiv:1311.2901}}.

\bibitem{hua2017mgupgma}
G.-J. Hua, C.-L. Hung, C.-Y. Lin, F.-C. Wu, Y.-W. Chan, C.~Y. Tang, Mgupgma: a
  fast upgma algorithm with multiple graphics processing units using nccl,
  Evolutionary Bioinformatics 13 (2017) 1176934317734220.

\bibitem{2015MNRAS.453.2326B}
J.~K. {Banfield}, O.~I. {Wong}, K.~W. {Willett}, R.~P. {Norris}, L.~{Rudnick},
  S.~S. {Shabala}, B.~D. {Simmons}, C.~{Snyder}, A.~{Garon}, N.~{Seymour},
  E.~{Middelberg}, H.~{Andernach}, C.~J. {Lintott}, K.~{Jacob}, A.~D.
  {Kapi{\'n}ska}, M.~Y. {Mao}, K.~L. {Masters}, M.~J. {Jarvis},
  K.~{Schawinski}, E.~{Paget}, R.~{Simpson}, H.~R. {Kl{\"o}ckner},
  S.~{Bamford}, T.~{Burchell}, K.~E. {Chow}, G.~{Cotter}, L.~{Fortson},
  I.~{Heywood}, T.~W. {Jones}, S.~{Kaviraj}, {\'A}.~R. {L{\'o}pez-S{\'a}nchez},
  W.~P. {Maksym}, K.~{Polsterer}, K.~{Borden}, R.~P. {Hollow}, L.~{Whyte},
  {Radio Galaxy Zoo: host galaxies and radio morphologies derived from visual
  inspection}, \mnras 453~(3) (2015) 2326--2340.
\newblock \href {http://arxiv.org/abs/1507.07272} {\path{arXiv:1507.07272}},
  \href {https://doi.org/10.1093/mnras/stv1688}
  {\path{doi:10.1093/mnras/stv1688}}.

\bibitem{1995ApJ...450..559B}
R.~H. {Becker}, R.~L. {White}, D.~J. {Helfand}, {The FIRST Survey: Faint Images
  of the Radio Sky at Twenty Centimeters}, \apj 450 (1995) 559.
\newblock \href {https://doi.org/10.1086/176166} {\path{doi:10.1086/176166}}.

\bibitem{2014yCat.2328....0C}
R.~M. {Cutri}, {et al.}, {VizieR Online Data Catalog: AllWISE Data Release
  (Cutri+ 2013)}, VizieR Online Data Catalog (2014) II/328.

\bibitem{Everingham15}
M.~Everingham, S.~M.~A. Eslami, L.~V. Gool, C.~K.~I. Williams, J.~Winn,
  A.~Zisserman, The pascal visual object classes challenge: A retrospective,
  International Journal of Computer Vision 111~(1) (2015) 98--136.
\newblock \href {https://doi.org/10.1007/s11263-014-0733-5}
  {\path{doi:10.1007/s11263-014-0733-5}}.

\bibitem{1974MNRAS.167P..31F}
B.~L. {Fanaroff}, J.~M. {Riley}, {The morphology of extragalactic radio sources
  of high and low luminosity}, \mnras 167 (1974) 31P--36P.
\newblock \href {https://doi.org/10.1093/mnras/167.1.31P}
  {\path{doi:10.1093/mnras/167.1.31P}}.

\bibitem{2019ApJS..240...34M}
Z.~{Ma}, H.~{Xu}, J.~{Zhu}, D.~{Hu}, W.~{Li}, C.~{Shan}, Z.~{Zhu}, L.~{Gu},
  J.~{Li}, C.~{Liu}, X.~{Wu}, {A Machine Learning Based Morphological
  Classification of 14,245 Radio AGNs Selected from the Best-Heckman Sample},
  \apjs 240~(2) (2019) 34.
\newblock \href {http://arxiv.org/abs/1812.07190} {\path{arXiv:1812.07190}},
  \href {https://doi.org/10.3847/1538-4365/aaf9a2}
  {\path{doi:10.3847/1538-4365/aaf9a2}}.

\bibitem{2019BasR...31.1017S}
C.~M. {Shorten}, P.~G. {Fitzgerald}, {Post-orogenic thermal history and
  exhumation of the northern Appalachian Basin: Low-temperature
  thermochronologic constraints}, Basin Research 31~(6) (2019) 1017--1039.
\newblock \href {https://doi.org/10.1111/bre.12354}
  {\path{doi:10.1111/bre.12354}}.

\bibitem{2020LAO50e9501T}
B.~{Lao}, T.~{An}, {Deployment of SKA low frequency imaging system in China SKA
  Regional Centre}, Scientia Sinica Physica, Mechanica \& Astronomica 50~(5)
  (2020) 059501.
\newblock \href {https://doi.org/10.1360/SSPMA-2019-0332}
  {\path{doi:10.1360/SSPMA-2019-0332}}.

\bibitem{2005ASPC..347...29T}
M.~B. {Taylor}, {TOPCAT \& STIL: Starlink Table/VOTable Processing Software},
  in: P.~{Shopbell}, M.~{Britton}, R.~{Ebert} (Eds.), Astronomical Data
  Analysis Software and Systems XIV, Vol. 347 of Astronomical Society of the
  Pacific Conference Series, 2005, p.~29.

\bibitem{2021PASA...38....3N}
R.~P. {Norris}, H.~T. {Intema}, A.~D. {Kapi{\'n}ska}, B.~S. {Koribalski},
  E.~{Lenc}, L.~{Rudnick}, R.~Z.~E. {Alsaberi}, C.~{Anderson}, G.~E.
  {Anderson}, E.~{Crawford}, R.~{Crocker}, J.~{English}, M.~D. {Filipovi{\'c}},
  T.~J. {Galvin}, A.~M. {Hopkins}, N.~{Hurley-Walker}, S.~{Inoue}, K.~{Luken},
  P.~J. {Macgregor}, P.~{Manojlovi{\'c}}, J.~{Marvil}, A.~N. {O'Brien},
  L.~{Park}, W.~{Raja}, D.~{Shobhana}, T.~{Venturi}, J.~D. {Collier},
  C.~{Hale}, A.~{Hotan}, V.~{Moss}, M.~{Whiting}, {Unexpected circular radio
  objects at high Galactic latitude}, \pasa 38 (2021) e003.
\newblock \href {http://arxiv.org/abs/2006.14805} {\path{arXiv:2006.14805}},
  \href {https://doi.org/10.1017/pasa.2020.52}
  {\path{doi:10.1017/pasa.2020.52}}.

\end{thebibliography}





\end{document}